\documentclass{article}
\usepackage{graphicx}
\usepackage[centertags]{amsmath}
\usepackage{amsfonts}
\usepackage{amssymb}
\usepackage{amsthm}

\title{Dynamical systems with internal degrees of freedom 
in non-Euclidean spaces}

\author{J. J. S\l awianowski, V. Kovalchuk,\\
B. Go\l ubowska, A. Martens, E. E. Ro\.zko\\
Institute of Fundamental Technological Research,\\
Polish Academy of Sciences,\\
21 \'{S}wi\c{e}tokrzyska str., 00-049 Warsaw, Poland\\
e-mails: jslawian@ippt.gov.pl, vkoval@ippt.gov.pl,\\
bgolub@ippt.gov.pl, amartens@ippt.gov.pl,\\ 
erozko@ippt.gov.pl}

\begin{document}

\maketitle

\begin{abstract}

Presented is description of kinematics and dynamics of material
points with internal degrees of freedom moving in a Riemannian 
manifold. The models of internal degrees of freedom we concentrate 
on are based on the orthogonal and affine groups. Roughly speaking, 
we consider infinitesimal gyroscopes and homogeneously deformable 
gyroscopes (affienly-rigid bodies) in curved manifolds. We follow 
our earlier models of extended rigid and affinely-rigid bodies 
moving in a flat space. It is well known that in curved spaces 
in general there is no well-defined concept of extended rigid 
or affinely-rigid body. Our infinitesimal models are mathematically 
well defined and physically they may be interpreted as an 
approximate description of "small" rigid and affinely-rigid bodies. 
We derive equations of motion and show how internal degrees of
freedom interact with spatial geometry, first of all with the 
curvature but also with the torsion. Integrability and degeneracy 
problems are discussed.

\end{abstract}


\section{Introduction}

The primary concept of Newton mechanics is that of the material
point moving in three-dimensional Euclidean space. A good deal of
the theory depends only on the affine sector of geometry. The
metric structure becomes essential when constructing particular
functional models of forces; the concepts of energy, work, and
power (time rate of work) also depend in an essential way on the
metric tensor. The Galilei relativity principle implies that, as a
matter of fact, it is not three-dimensional Euclidean space but
rather four-dimensional Galilean space-time that is a proper arena
of mechanics. This space-time has relatively complicated
structure, does not carry any natural four-dimensional metric
tensor and fails to be the Cartesian product of space and time.
There exists the absolute time, but the absolute space does not.
In the sequel we concentrate on the other kind of problems, so the
analysis of the subtle space-time aspects will be almost absent in
our treatment. Newton theory becomes essentially realistic and
viable when multiparticle systems are analyzed. It is just there
where metrical concepts become almost unavoidable, because it is
practically impossible to construct any realistic model of
interparticle forces without the explicit use of the metric
tensor. Extended bodies are described as discrete or continuous
systems of material points. Their motion consists of that of the
center of mass, i.e., translational motion and the relative motion
of constituents with respect to the center of mass. The total
configuration space may be identified with the Cartesian product
of the physical space (translational motion) and the configuration
space of relative motion. In many physical problems the structure
of mutual interactions leads to certain hierarchy of degrees of
freedom of the relative motion; in particular, some constraints
may appear. The effective configuration space becomes then the
Cartesian product of the physical space and some manifold of
additional degrees of freedom. There are situations when this
auxiliary manifold and the corresponding dynamics are postulated
as something rather primary then derived from the multiparticle
models. Usually the guiding hints are based on some symmetry
principles. In this way the concept of internal degrees of freedom
replaces that of relative motion. Sometimes it is a merely
convenient procedure, but one can also admit something like
essentially internal degrees of freedom not derivable from any
multiparticle model. After all, the very concept of the material
point is an abstraction of a small piece of matter. Why to reject
a priori an abstraction of a small and particularly shaped piece
of matter, thus, the material point with extra attached geometric
objects, as something primary for mechanics? In this way the
configuration space becomes as follows:
\[
Q=M\times Q^{\rm int},
\]
the Cartesian product of physical space $M$ (translational motion)
and some internal configuration space $Q_{\rm int}$. The usual
d'Alembert procedure of deriving equations of motion is not then
reliable, perhaps just essentially inadequate and should be
replaced by something else, probably based on appropriate
invariance assumptions.

In standard quantum theory there are quantities which have the
status of essentially internal variables, e.g., spin. If $M$ is
not the Euclidean space but some general manifold, then the
concept of essentially internal degrees of freedom becomes even
more justified, as we shall see.

And just in connection with this, the last step of generalization:
why $Q=M \times Q^{\rm int}$? Perhaps any spatial point $x\in M$
has its own manifold of internal variables $Q^{\rm int}_{x}$? Then
the total configuration space would be
\[
Q=\bigcup_{x\in M}Q^{\rm int}_{x}.
\]
In this way, the concept of fibre bundle appears as a most natural
mathematical framework for describing internal degrees of freedom.

As mentioned, such concepts are particularly natural when $M$ is a
general manifold endowed with some geometric structure based on
the affine connection, metric tensor, or both (interrelated or
not). In any case, the primary mathematical concept underlying any
model of internal degrees of freedom is a principal fibre bundle
$(Q,M,\pi)$, where $M$ denotes the base manifold (physical space),
$Q$ is the total bundle manifold (configuration space), and
$\pi:Q\rightarrow M$ is the bundle projection. We shall often use
the following abbreviation for fibres: $Q_{x}=\pi^{-1}(x)$.
Obviously, in the relativistic theory a more adequate formulation
is one using the space-time manifold as a basis of the fibre
bundle. The same is true in non-relativistic theory when the
problems of Galilean relativity principle are important. However,
in this treatise we do not touch such problems or do it merely
exceptionally.

As is well known, the general fibre bundle $(Q,M,\pi)$ does not
need to be diffeomorphic with the Cartesian product $M\times
Q^{\rm int}$. And even if such a diffeomorphism does exist it is
not in general unique in the sense that there is no canonically
distinguished choice (there are exceptions like, e.g., tangent and
cotangent bundles over Lie groups, bundles of linear frames or
co-frames over Lie groups, and so on). Therefore, in general, the
total motion $\varrho:\mathbb{R}\rightarrow Q$ does not split in a
well-defined way into translational and internal motion. More
precisely, only translational motion is well defined as a
projection
\[
\varrho_{\rm tr}:=\pi\circ\varrho:\mathbb{R}\rightarrow M.
\]
Similarly, generalized velocity $\dot{\varrho}(t)\in
T_{\varrho(t)}Q$ does not split into translational and internal
velocities. Only the first one is well defined as a
$T\pi$-projection of $\dot{\varrho}(t)$ to the tangent space
$T_{\pi(\varrho(t))}M$; obviously, it is identical with
$\dot{\varrho}_{\rm tr}(t)\in T_{\varrho_{\rm tr}(t)}M$. Without
additional geometric objects the time-rate of internal
configuration is not well defined.
\bigskip

\begin{center}
\includegraphics[scale=0.35]{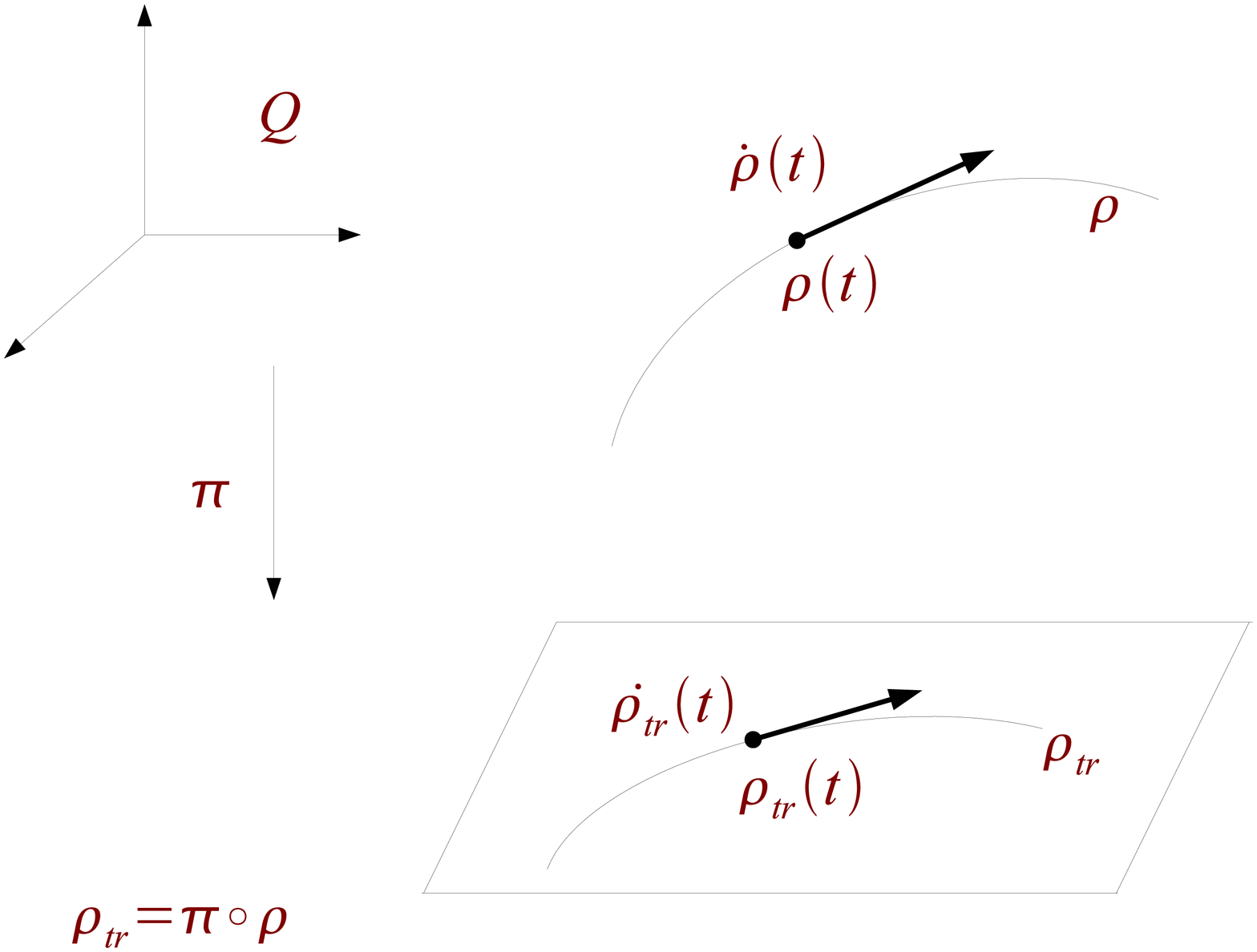}

{\rm Fig. 1}
\end{center}
\bigskip

The very concept of internal motion and internal trajectory is
meaningful only when $Q=M\times Q^{\rm int}$, i.e., when $Q$ is
the Cartesian product. In general, when the bundle $(Q,M,\pi)$ is
nontrivial (perhaps even not trivializable) and no additional
connection-like structures in $Q$ are defined, it is only the
total motion in $Q$ and the translational one in $M$ that are well
defined. The same concerns velocities.

In original Newton theory $M$ is the Euclidean three-dimensional
space. We shall consider the manifolds with a more general
geometric structure and the fibre bundles $(Q,M,\pi)$ over them.
Having in view both the mathematical clarity and certain
nonstandard physical applications we admit the general dimension,
i.e., $\dim M=n$.

\section{Kinematics and canonical formalism for affine models}

In a long series of earlier papers we have developed the theory of
extended affinely-rigid bodies in Euclidean spaces
\cite{Gol01,Gol02,Gol03,Gol04,Gol04S,Gol06,Mart02,Mart03,Mart04_1,Mart04_2,Roz05,JJS73_2,JJS74_2,JJS75_1,JJS75_2,JJS82_1,JJS82_2,JJS88,JJS02_1,JJS02_2,JJS03,JJS04,JJS04S,JJS-VK03,JJS-VK04,JJS-VK04S,all-book04,all04,all05}.
Strictly speaking, kinematics of such objects preassumes only
affine structure of the physical space. Affine constraints mean
that all affine relations between material points remain invariant
during admissible motions; material straight lines continue to be
straight lines, their parallelism is preserved, etc. Assuming that
all metrical relations are preserved one obtains the rigid body in
the usual sense.

The concepts of extended metrically- and affinely-rigid bodies
break down when the Euclidean or affine space is replaced by a
differential manifold with geometry given by the metric tensor,
affine connection, or both of them (interrelated or not). For
example, let $(M,g)$ be a Riemann space, where $M$ is a manifold
and $g$ is a metric tensor defined on it. An extended continuous
system of material points moving in $M$ is metrically rigid if all
infinitesimal distances are invariant during any admissible
motion. Of course, then also finite along-geodetic distances are
constant. If some standard reference configuration is fixed, the
configuration space becomes identified with Diff$(M,g)$, the
isometry group, i.e., the group of all diffeomorphisms
$\varphi:M\rightarrow M$ preserving the metric tensor,
$\varphi^{\ast}g=g$. But it is well known that for the generic
Riemann space $(M,g)$ with not vanishing curvature tensor
$R[g]\neq 0$ it is rather typical that the isometry group is
trivial, i.e., Diff$(M,g)=\{{\rm id}_M\}$. Obviously, the highest
possible dimension of Diff$(M,g)$ in an $n$-dimensional $M$ equals
$n(n+1)/2$. This highest dimension is attained only in
constant-curvature spaces, like, e.g., spheres and pseudospheres
in $\mathbb{R}^{n+1}$. And in fact, in such manifolds the concept
of a finite extended rigid body continues to be well defined. But
such manifolds, with $n(n+1)/2$ independent Killing vectors, are
extremely exceptional in category of all Riemann manifolds. Let us
also mention that even if $R[g]=0$, i.e., $(M,g)$ is locally
Euclidean, but $M$ is topologically not equivalent to
$\mathbb{R}^{n}$ (e.g., torus
$\mathbb{T}^{n}=\mathbb{R}^{n}/\mathbb{Z}^{n}$ with the metric
tensor inherited from $\mathbb{R}^{n}$), there may be obstacles
against the existence of the global $n(n+1)/2$-dimensional
isometry groups. So, in a generic Riemann manifold the concept of
extended rigid body fails to be well defined.

The same concerns extended affinely-rigid bodies. Let $(M,\Gamma)$
be an affine-connection space, i.e., a differential manifold $M$
endowed with an affine connection $\Gamma$. This connection may be
quite arbitrary, not necessarily one derived explicitly from some
metric tensor $g$ (at this stage the metric structure does not
need to exist at all). Let $\nabla$ denote the covariant
differentiation corresponding to $\Gamma$. We say that
$\varphi\in$ Diff$(M)$ is an affine transformation of
$(M,\Gamma)$, i.e., that $\varphi\in$ Diff$(M,\Gamma)$ if for any
pair of vector fields $X$, $Y$ on $M$ the following holds:
\[
\nabla_{\varphi_{\ast}X}\left(\varphi_{\ast}Y\right)=\varphi_{\ast}\left(\nabla_{X}Y\right),
\]
where, as usual, $\nabla_{X}$ denotes the covariant
differentiation along the vector field $X$. Roughly speaking, the
above property means that the $\nabla$-operation is ''transparent"
with respect to the transformation $\varphi$. In the usual flat
affine space this is just an equivalent definition of the
$n(n+1)$-dimensional affine group GAf$(M)$. However, when the
curvature tensor $\mathcal{R}[\Gamma]$ of the affine connection
$\Gamma$ is not vanishing, then, as a rule, the dimension of
Diff$(M,\Gamma)$ is smaller than $n(n+1)$ and the generic
situation is that Diff$(M,\Gamma)=\{{\rm id}_{M}\}$, i.e., the
only affine transformation is the identity mapping. Therefore, in
principle the concept of extended affinely-rigid body breaks down.

However, ``very small" regions of $(M,\Gamma)$ approximately look
like flat affine spaces; similarly, ``very small" regions of
Riemannian manifolds $(M,g)$ look like Euclidean spaces.
Therefore, approximately, one can consider ``very small" rigid and
affinely-rigid bodies in non-Euclidean manifolds. This is always
an approximation, the better one, the more body shrinks.
Everything becomes rigorous in the limit of vanishing size.
Roughly speaking, the body is not any longer injected in the
physical space $M$, but in a tangent space $T_{x}M$, where $x\in
M$ represents the spatial position of the body ``as a whole" and
is a remnant of the centre of mass position in the flat-space
theory. And configurations of affine bodies in the linear space
$T_{x}M$ may be identified with linear frames, i.e., ordered bases
in $T_{x}M$.

This is a natural analogue of some description used in mechanics
of extended affine bodies. Namely, let $V$ be the linear space of
translations in the physical affine space $M$. The configuration
space of extended affinely-rigid body in $M$ may be identified
with $M\times F(V)$, where $F(V)$ denotes the manifold of linear
frames in $V$. This is meant in the following sense: $x\in M$ is
an instantaneous position of the centre of mass, and the frame
$e=(e_{1},\ldots,e_{A},\ldots,e_{n})\in F(V)$ is materially frozen
into the body, i.e., co-moving with it. More precisely, if
$a=(a^{1},\ldots,a^{n})$ are Lagrange (reference) coordinates of
some material point (its identification labels), and configuration
is given by $q=(x,e)=(x;e_{1},\ldots,e_{n})\in Q=M\times F(V)$,
then the current spatial position $y\in M$ of this $a$-th point
satisfies
\[
\overrightarrow{xy}=a^{K}e_{K},
\]
where $\overrightarrow{xy}$ denotes the radius-vector of $y$ with respect to the instantaneous
position $x\in M$ of the centre of mass. Analytically,
\[
y^{i}=x^{i}+e^{i}{}_{K}a^{K}
\]
with respect to some Cartesian coordinates.

Essentially the same remains true for infinitesimal bodies injected in tangent spaces
$T_{x}M$. They are somewhere placed in space, $x\in M$, and have the extra attached internal
variables $e_{K}\in T_{x}M$, $K=\overline{1,n}$. Let us describe it in geometric terms.

The configuration space of infinitesimal affinely-rigid body moving in the physical space $M$
is given by the manifold $FM$ of linear frames in $M$,
\[
Q=FM=\bigcup_{x\in M}F_{x}M,
\]
where $F_{x}M$ denotes the manifold of linear frames in the
tangent space $T_{x}M$. Obviously, just as $F(V)$ is an open
submanifold of
\[
V^{n}= \underbrace{V\times V\times\cdots\times V}_{n \
\text{terms}},
\]
so $FM$ is an open submanifold in the Whitney sum of fibre bundles
\[
\bigoplus_{n}TM=\bigcup_{x\in M} \underbrace{T_{x}M\times
T_{x}M\times\cdots\times T_{x}M}_{n\ \text{terms}}.
\]
These open subsets consist of linearly independent linear
$n$-tuples.

To be more precise, in the theory of the flat-space extended
continuous bodies one should not use the total $F(V)$, but rather
one of its connected components, e.g., one positively oriented
with respect to some fixed standard orientation. And so is in
mechanics of structured material points; by the way, it is tacitly
assumed that $M$ is orientable. Although, one can argue that if
once the framework of classically imaginable extended bodies is
left and affine degrees of freedom are essentially internal, the
exotic orientation-changing motions might be perhaps admissible
and even (at least mathematically) interesting. The problem
becomes particularly interesting, perhaps just mathematically
exciting, when $M$ is not orientable, like, e.g., M\"obius band.

The manifold $FM$ carries a natural structure of the principal
fibre bundle over the base $M$ \cite{Kob-Nom63,Ster64}. The
projection $\pi:FM\rightarrow M$ assigns to any linear frame the
point at which it is attached, thus $\pi\left(F_{x}M\right)=x$.
The structural group GL$(n,\mathbb{R})$ acts on $FM$ according to
the standard rule; thus, for any $L\in$ GL$(n,\mathbb{R})$ we have
that
\begin{equation}\label{a11}
FM\ni e=(\ldots,e_{A},\ldots)\mapsto eL:=(\ldots,e_{B}L^{B}{}_{A},\ldots).
\end{equation}
Obviously, dim $FM=n(n+1)$; this is the number of degrees of freedom ($n$ translational and
$n^{2}$ internal ones). For any linear frame $e$ there exists a unique dual co-frame
$\widetilde{e}=(\ldots,e^{A},\ldots)$, where
\[
\left\langle e^{A},e_{B}\right\rangle=\delta^{A}{}_{B}
\]
and for any $x\in M$, $v\in T_{x}M$ and $p\in T_{x}^{\ast}M$ the
symbol $\langle p,v\rangle$ denotes the evaluation of the linear
function $p$ on the vector $v$.

The manifold of all co-frames, i.e.,
\[
Q^{\ast}=F^{\ast}M=\bigcup_{x\in M}F_{x}^{\ast}M,
\]
is canonically diffeomorphic with $FM$ just in the sense of duality. Therefore, the
configuration space of infinitesimal affine body may be represented either as $FM$ or
$F^{\ast}M$; it depends on the particular problem which representation is more convenient.
Just as previously, $F^{\ast}M$ is an open subset of the Whitney sum
\[
\bigoplus_{n}T^{\ast}M=\bigcup_{x\in M}
\underbrace{T_{x}^{\ast}M\times T_{x}^{\ast}M\times\cdots\times
T_{x}^{\ast}M}_{n\ \text{terms}};
\]
it consists of linearly independent $n$-tuples. The natural
projection onto the base manifold will be denoted by
$\pi^{\ast}:F^{\ast}M\rightarrow M$, and then
$\pi^{\ast}\left(F_{x}^{\ast}M\right)=x$. Obviously, $F^{\ast}M$
also carries the structure of the principal fibre bundle with
GL$(n,\mathbb{R})$ as a structural group:
\begin{equation}\label{a12}
F^{\ast}M\ni\widetilde{e}=\left(\ldots,e^{A},\ldots\right)\mapsto\widetilde{e}L:=
\left(\ldots,L^{-1A}{}_{B}e^{B},\ldots\right)
\end{equation}
for any $L\in$ GL$(n,\mathbb{R})$.

There is one subtle point, namely, GL$(n,\mathbb{R})$ is not
connected. It consists of two connected components, i.e., the
subgroup GL$^{+}(n,\mathbb{R})$ of positive-determinant matrices
and GL$^{-}(n,\mathbb{R})$ is the coset (not a subgroup, of
course) of negative-determinant matrices. In theory of Lie groups,
fibre bundles and connections one usually deals with connected
groups. Thus, it seems rather natural to deal with reductions to
the subgroup GL$^{+}(n,\mathbb{R})$. But this has again to do with
orientability of $M$. Then the unconnected manifold $FM$ is
replaced by $F^{+}M$, i.e., the manifold of linear frames
positively oriented with respect to some fixed orientation on $M$.

Any system of local coordinates $x^{i}$, $i=\overline{1,n}$, on
$M$ induces local coordinates $(x^{i},e^{i}{}_{A})$,
$A=\overline{1,n}$, on $FM$ and $(x^{i},e^{A}{}_{i})$ on
$F^{\ast}M$, where $e^{i}{}_{A}$, $e^{A}{}_{i}$ are respectively
components of $e_{A}$, $e^{A}$ with respect to coordinates
$x^{i}$, and
\[
e^{A}{}_{i}e^{i}{}_{B}=\delta^{A}{}_{B},\qquad e^{i}{}_{A}e^{A}{}_{j}=\delta^{i}{}_{j}.
\]
To avoid the crowd of symbols we do not distinguish graphically between $x^{i}$ and their
pull-backs to $FM$ and $F^{\ast}M$.

In certain problems it is convenient to admit the local, $x$-dependent action of the
structural group, like in gauge theories. This means that we consider (sufficiently smooth)
fields $L:M\rightarrow$ GL$(n,\mathbb{R})$; they form an infinite-dimensional group under the
pointwise multiplication,
\[
(L_{1}L_{2})(x)=L_{1}(x)L_{2}(x).
\]
And this group acts, also in a pointwise way, on $FM$ and $F^{\ast}M$. Thus, if $e\in F_{x}M$,
$\widetilde{e}\in F_{x}^{\ast}M$, then the action of $L$ is given by
\begin{equation}\label{a13}
e\mapsto eL(x),\qquad \widetilde{e}\mapsto\widetilde{e}L(x).
\end{equation}
From the mechanical point of view the structural action of
GL$(n,\mathbb{R})$ on $Q=FM$ and $Q^{\ast}=F^{\ast}M$ corresponds
to material transformations. We shall use the term ``micromaterial
transformations". This is exactly the infinitesimal limit of the
usual material transformations. In fact, material points with
affine internal degrees of freedom may be interpreted as
affinely-rigid bodies injected into instantaneous tangent spaces
$T_{x}M$, where $x\in M$ is the spatial position of the body. But
it is obvious that $e\in F_{x}M$ is canonically identical with
some linear isomorphism of $\mathbb{R}^{n}$ onto $T_{x}M$
(similarly, $\widetilde{e}\in F_{x}^{\ast}M$ is a linear
isomorphism of $T_{x}M$ onto $\mathbb{R}^{n}$). In this way,
$\mathbb{R}^{n}$ plays the role of the ``micromaterial space"
(Lagrange variables) and $T_{x}M$ is the ``microphysical space"
(Euler variables) of infinitesimal affinely-rigid body. The frame
$e\in F_{x}M$, i.e., the ``placement" is then exactly the
counterpart of what was denoted by $\varphi\in$ LI$(U,V)$ in our
papers about extended flat-space bodies
\cite{JJS73_2,JJS74_2,JJS75_1,JJS75_2,JJS82_1,JJS82_2,JJS88,JJS02_1,JJS02_2,JJS04,JJS04S,JJS-VK03,JJS-VK04,JJS-VK04S,all-book04,all04,all05}.

And what does correspond to spatial or physical transformations,
i.e., what are ``microspatial" kinematical symmetries? Here the
situation is more complicated. Obviously, microspatial
transformations must act on the left in tangent spaces $T_{x}M$.
But for different points $x\in M$ the tangent spaces $T_{x}M$ are
logically different linear spaces and their linear groups
GL$(T_{x}M)$ are also logically different sets. Without additional
strong geometric structures there is no natural isomorphism
between different GL$(T_{x}M)$. So, there is only one possibility,
in an essential way infinite-dimensional. Namely, the fields $T$
of mixed non-degenerate second-order tensors on $M$,
\[
M\ni x\mapsto T_{x}\in{\rm GL}(T_{x}M)\subset{\rm L}(T_{x}M)\simeq
T{}^{1}_{1}(T_{x}M),
\]
give rise to the following transformations of $FM$, $F^{\ast}M$:
\begin{eqnarray}
F_{x}M\ni e=(\ldots,e_{A},\ldots)&\mapsto&\left(\ldots,T_{x}\circ e_{A},\ldots\right),
\label{a14a}\\ F^{\ast}_{x}M\ni \widetilde{e}=(\ldots,e^{A},\ldots)&\mapsto&\left(\ldots,
e^{A}\circ T^{-1}_{x},\ldots\right).\label{a14b}
\end{eqnarray}
In a structure-less manifold $M$ and in a connection manifold
$(M,\Gamma)$ with not vanishing curvature tensor
$\mathcal{R}(\Gamma)$, there are no natural isomorphisms between
different $T_{x}M$, therefore, nothing like the ``constancy" of
$T$ may be defined. If the components field $T^{i}{}_{j}$ is
accidentally constant in some coordinates $x^{i}$, then in other
coordinates it is no longer true. Therefore, by its very nature
the above group of left-acting transformations is
infinite-dimensional, flexible. Of course, in the right-hand side
action of GL$(n,\mathbb{R})$ (\ref{a12}) $L$ may be put
$x$-dependent like in gauge theories of fields and continua, but
need not to be so; its constancy is well defined. $L\in$
GL$(n,\mathbb{R})$ are just matrices by their very nature, not
matrix representants of linear mappings in $T_{x}M$ with respect
to some coordinates.

When some coordinates $x^{i}$ are fixed in $M$, the second-order mixed tensor field $T$ is
represented by a system of functions, i.e., components $T^{i}{}_{j}(x^{a})$,
\[
T=T^{i}{}_{j}(x)\frac{\partial}{\partial x^{i}}\otimes dx^{j}.
\]
Analytically, the action of $T$ on the configuration space $Q$ is described by
\begin{equation}\label{a15}
\left(\ldots,x^{a},\ldots;\ldots,e^{i}{}_{A},\ldots\right)\mapsto
\left(\ldots,x^{a},\ldots;\ldots,T^{i}{}_{j}(x)e^{j}{}_{A},\ldots\right).
\end{equation}
Similarly, $x$-dependent $L\in$ GL$(n,\mathbb{R})$ act as follows:
\begin{equation}\label{b15}
\left(\ldots,x^{a},\ldots;\ldots,e^{i}{}_{A},\ldots\right)\mapsto
\left(\ldots,x^{a},\ldots;\ldots,e^{i}{}_{B}L^{B}{}_{A}(x),\ldots\right).
\end{equation}
In particular, this formula holds for global, $x$-independent
$L\in$ GL$(n,\mathbb{R})$. It is seen that when $x\in M$ is kept
fixed, these are just the well-known formulas for an
affinely-rigid body in a flat space $T_{x}M$. The difference is
that there is no counterpart of translations and general affine
mappings in $M$ and in the total $FM$. The reason is that, as
mentioned, even if $M$ is endowed with some affine connection
$\Gamma$ with not vanishing curvature tensor, the group of affine
transformations is generically trivial or its dimension is smaller
than $n(n+1)$. And if $M$ is completely amorphous, i.e., even if
any affine connection is not fixed, then it is only the total
diffeomorphism group Diff$(M)$ that may replace the group of
spatial affine transformations in mechanics of extended affine
bodies in a flat space.

In mechanics of affine extended bodies
\cite{JJS73_2,JJS74_2,JJS75_1,JJS75_2,JJS82_1,JJS82_2,JJS88,JJS02_1,JJS02_2,JJS04,JJS04S,JJS-VK03,JJS-VK04,JJS-VK04S,all-book04,all04,all05}
we considered various kinds of additional constraints; for obvious
reasons the most important of them were metrical constraints,
i.e., rigid body in the literal sense. Not only affine relations
between material points are then preserved but also metrical ones,
i.e., distances and angles. In flat-space theory one is dealing
then, also on the kinematical level, with the Euclidean structure
$(M,V,g)$, where $V$ is the linear space of translations in $M$,
and $g\in V^{\ast}\otimes V^{\ast}$ is the metric tensor of $M$.
The configuration space of the rigid body may be identified with
$Q=M\times F(V,g)$, where $F(V,g)\subset F(V)$ is the manifold of
$g$-orthonormal frames $e=(e_{1},\ldots,e_{A},\ldots,e_{n})$,
i.e.,
\[
g(e_{A},e_{B})=g_{ij}e^{i}{}_{A}e^{j}{}_{B}=\delta_{AB}.
\]
More precisely, one should use rather some connected component of $F(V,g)$, e.g., the manifold
$F^{+}(V,g)$ of $g$-orthonormal frames positively oriented with respect to some fixed
orientation of $M$.

When $(M,g)$ is a general Riemann manifold, then, as a rule, there
are no extended rigid bodies with the usual number of $n(n+1)/2$
degrees of freedom; the only exception are constant-curvature
spaces. In a generic case the not trivial isometries do not exist
at all. But just as in affine theory we can speak about
infinitesimal rigid bodies. They describe in a good approximation
the behaviour of very small ``almost rigid bodies", in which the
changes of distances between neighbouring particles are
higher-order small in comparison with the initial distances
themselves.

The configuration space of infinitesimal rigid body in $(M,g)$ may
be identified with $F(M,g)$, i.e., the manifold of all
$g$-orthonormal frames in all tangent spaces of $M$. Again, to be
more precise, if $M$ is orientable, we should restrict ourselves
to $F^{+}(M,g)$, i.e., the connected manifold of $g$-orthonormal
frames positively oriented with respect to some fixed orientation
in $M$. Obviously, $F(M,g)$, $F^{+}(M,g)$ are
$n(n+1)/2$-dimensional manifolds; there are $n$ translational
degrees of freedom ($M$) and $n(n-1)/2$ rotational ones (fibres
$(F_{x}M,g_{x})$). Just as in affine model, it does not matter
whether we use $F(M,g)$ or the manifold $F^{\ast}(M,g)$ of
$g$-orthonormal co-frames. Projections onto the base manifold $M$
are just the restrictions of the previous $\pi$, $\pi^{\ast}$ to
submanifolds $F(M,g)$, $F^{\ast}(M,g)$; for brevity we denote them
by the same symbols. The structure groups of $F(M,g)$,
$F^{+}(M,g)$ are respectively O$(n,\mathbb{R})\subset$
GL$(n,\mathbb{R})$, SO$(n,\mathbb{R})\subset$ GL$(n,\mathbb{R})$;
they act on the bundle manifolds on the right, just in the sense
of formulas (\ref{a11}), (\ref{a12}). Similarly, the
left-hand-side spatial transformations are given by (\ref{a14a})
and (\ref{a14b}), where now the field $T$ takes values in
orthogonal groups of $(T_{x}M,g_{x})$, i.e.,
\begin{eqnarray}
M\ni x&\mapsto& T_{x}\in{\rm O}\left(T_{x}M,g_{x}\right)\subset{\rm GL}(T_{x}M),\nonumber\\
M\ni x&\mapsto& T_{x}\in{\rm SO}\left(T_{x}M,g_{x}\right)\subset{\rm
O}\left(T_{x}M,g_{x}\right);\nonumber
\end{eqnarray}
obviously, $g_{x}\in T^{\ast}_{x}M\otimes T^{\ast}_{x}M$ is the metric tensor of $T_{x}M$,
i.e., the value of the field $g$ at $x\in M$.

What concerns the very description of degrees of freedom,
infinitesimal affinely-rigid body is well defined in any quite
amorphous differential manifold. Obviously, the usual, i.e.,
metrically-rigid body is meaningful only when $M$ is endowed with
some metric tensor $g$.

But, as mentioned, even on the purely kinematical level there are
some difficulties with systems with internal degrees of freedom.
Namely, for any motion $\varrho:\mathbb{R}\rightarrow FM$ it is
only the tangent vector $\dot{\varrho}(t)\in T_{\varrho(t)}FM$ and
its projection $\dot{\varrho}_{\rm tr}(t)\in T_{\pi(\varrho(t))}M$
(see Fig. $2$) that are well defined velocities, i.e.,
respectively the total generalized and translational velocities.

There is no well-defined time rate of internal variables
evolution. The minimal geometric structure necessary to define it
is a connection on the fibre bundle $(Q,M,\pi)$. In mechanics of
infinitesimal affine bodies and infinitesimal gyroscopes this is
simply the affine connection. Therefore, from now on we assume
that some affine (more precisely linear) connection $\Gamma$ is
fixed in $M$. Analytically it is given by the system of components
$\Gamma^{i}{}_{jk}$ transforming under the change of coordinates
$x^{i}$ on $M$ according to the well-known linear-inhomogeneous
rule. In modern differential geometry \cite{Kob-Nom63,Ster64}
linear connection is described by some differential one-form
$\omega$ on $FM$ (or $F(M,g)$) taking values in the Lie algebra of
the structural group GL$(n,\mathbb{R})^{\prime}\simeq$
L$(n,\mathbb{R})$ (or SO$(n,\mathbb{R})^{\prime}$, i.e., the space
of skew-symmetric matrices). It satisfies certain conditions that
are quoted in \cite{Kob-Nom63,Ster64}. Analytically $\omega$ is
represented by the system of differential one-forms
$\omega^{K}{}_{L}$ related in the following way to
coordinate-dependent quantities $\Gamma^{i}{}_{jk}$:
\[
\omega^{A}{}_{B}=e^{A}{}_{i}\left(de^{i}{}_{B}+\Gamma^{i}{}_{jk}e^{j}{}_{B}dx^{k}\right).
\]
\begin{center}
\includegraphics[scale=0.35]{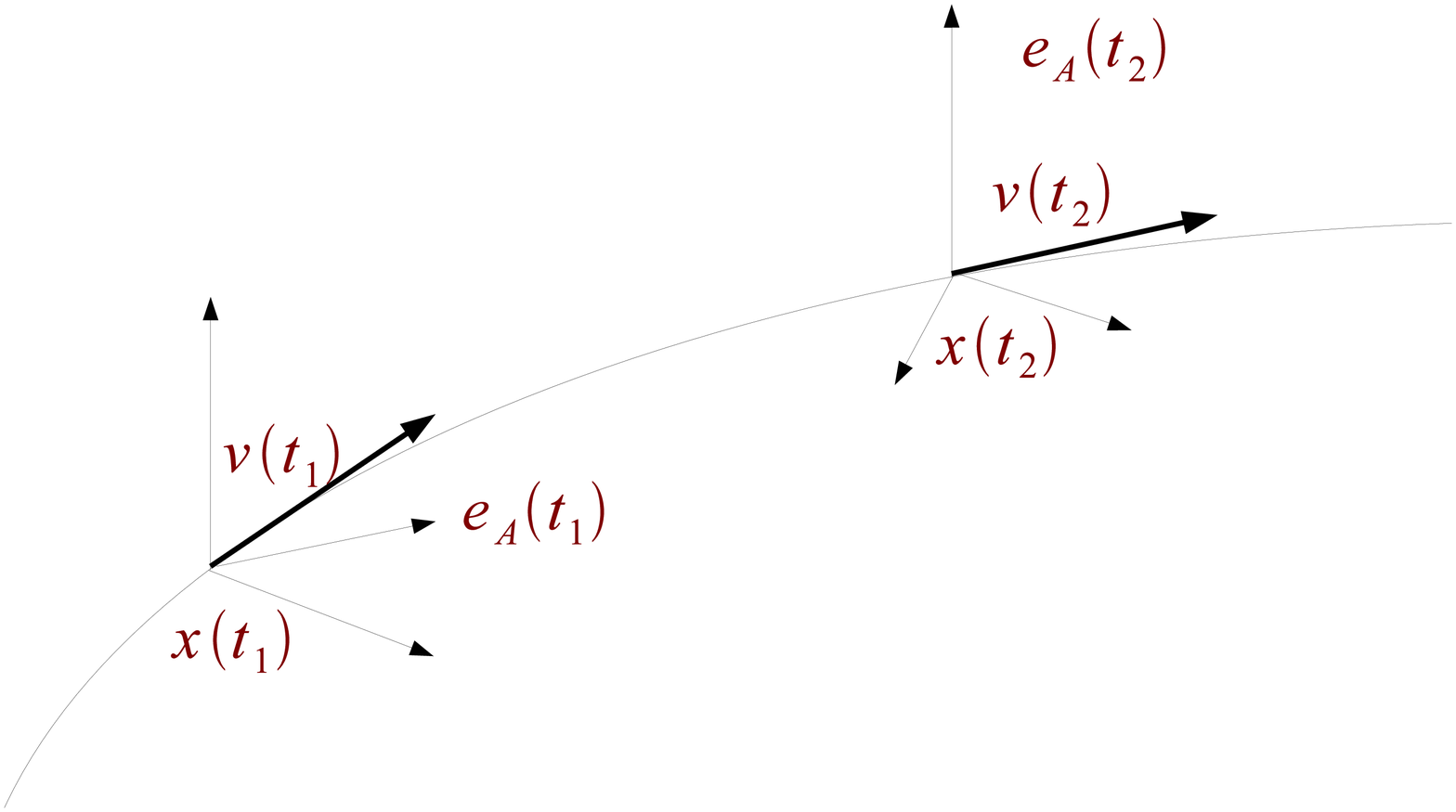}

{\rm Fig. 2}
\end{center}
\bigskip

One uses also the connection-independent canonical
$\mathbb{R}^{n}$-valued differential one-form $\theta$ on $FM$; it
is intrinsically defined and its coordinate representation is
given by
\[
\theta^{A}=e^{A}{}_{i}dx^{i}.
\]
It vanishes when evaluated on vectors tangent to the fibres
$F_{x}M$, $F_{x}(M,g)$, i.e., vertical ones. The kernel of
$\omega_{e}$ is transversal to the space of vertical vectors; its
elements are by definition ``horizontal" vectors in $T_{e}FM$,
$T_{e}F(M,g)$. The subspaces of vertical and horizontal vectors at
$e$ are denoted respectively by $V_{e}$, $H_{e}$. Obviously,
$T_{e}FM=V_{e}\oplus H_{e}$, and similarly for $F(M,g)$. Roughly
speaking, horizontal vectors establish isomorphisms between fibres
over infinitesimally remote points of $M$. Therefore,
infinitesimally, the concepts like the change of internal state
and the velocity of internal motion become meaningful.

At any $e$, the co-vectors $\omega_{e},\theta_{e}\in
T^{\ast}_{e}FM$ ($T^{\ast}_{e}F(M,g)$ in the gyroscopic case) form
a basis of $T^{\ast}_{e}FM$. The dual basis in $T_{e}FM$
($T_{e}F(M,g)$) consists of vectors denoted by $E^{A}{}_{B}$,
$H_{A}$ and
\begin{eqnarray}
\langle\omega^{K}{}_{L},E^{A}{}_{B}\rangle=\delta^{K}{}_{B}\delta^{A}{}_{L},&\quad&
\langle\omega^{K}{}_{L},H_{A}\rangle=0,\nonumber\\
\langle\theta^{K},E^{A}{}_{B}\rangle=0,&\quad&
\langle\theta^{K},H_{A}\rangle=\delta^{K}{}_{A}.\nonumber
\end{eqnarray}
One can easily show that, after identifying vector fields with
differential operators \cite{Kob-Nom63,Ster64}, we have that
\[
E^{K}{}_{L}=e^{i}{}_{L}\frac{\partial}{\partial e^{i}{}_{K}},\qquad
H_{L}=e^{i}{}_{L}\left(\frac{\partial}{\partial x^{i}}-
\Gamma^{k}{}_{ji}e^{j}{}_{A}\frac{\partial}{\partial e^{k}{}_{A}}\right).
\]
At any point $e$, the corresponding vectors
$\left(E^{K}{}_{L}\right)_{e}$ are vertical, i.e., tangent to the
fibres, $\left(E^{K}{}_{L}\right)_{e}\in V_{e}$, whereas
$\left(H_{K}\right)_{e}$ are horizontal,
$\left(H_{K}\right)_{e}\in H_{e}$. Dually to the situation with
co-vectors $\omega^{A}{}_{B}$, $\theta^{A}$, now $E^{K}{}_{L}$ are
connection-independent, whereas $H_{L}$ depend explicitly on the
connection. In the literature \cite{Kob-Nom63,Ster64} $H_{K}$ are
referred to as standard horizontal vector fields and $E^{K}{}_{L}$
as fundamental vector fields. $E^{K}{}_{L}$ are infinitesimal
generators of the action of the structural group
GL$(n,\mathbb{R})$ on $FM$. Roughly speaking, $H_{K}$ generate the
parallel transport in the sense of $\Gamma$.

The splitting $T_{e}FM=V_{e}\oplus H_{e}$ enables one to decompose
every vector at $e$ into vertical and horizontal parts. In
particular, this may be done for generalized velocities
$\dot{\varrho}(t)\in T_{\varrho(t)}FM$. Projecting the horizontal
component to $M$, we do not obtain anything new, just the
translational velocity $\dot{\varrho}_{\rm tr}(t)\in
T_{\pi(\varrho(t))}M$. The vertical part, however, is a new,
$\Gamma$-depending object. It is a measure of the internal
velocity, a kind of the time rate of internal configuration. And,
as expected, it simply coincides with the system of covariant
derivatives of the frame vectors $e_{A}$ along the curve
describing translational motion. If $x^{i}$, $e^{i}{}_{A}$ are
coordinates on $FM$ induced by those $x^{i}$ on $M$, and if
instead of sophisticated symbols $x^{i}\circ \varrho$,
$e^{i}{}_{A}\circ \varrho$ we simply write $x^{i}(t)$,
$e^{i}{}_{A}(t)$ for the time dependence of generalized
coordinates of the object, then translational velocity $v$ has the
components
\[
v^{i}=\frac{dx^{i}}{dt},
\]
whereas the internal velocity is given as follows:
\[
V^{i}{}_{A}=\frac{De^{i}{}_{A}}{Dt}=\frac{de^{i}{}_{A}}{dt}+
\Gamma^{i}{}_{jk}\left(x(t)\right)e^{j}{}_{A}\frac{dx^{k}}{dt}.
\]
It is very important to stress that if $(M,\Gamma)$ is
non-Euclidean, i.e., the curvature and torsion tensors do not
vanish:
\begin{eqnarray}
\mathcal{R}^{i}{}_{jkl}&=&\partial_{k}\Gamma^{i}{}_{jl}-\partial_{l}\Gamma^{i}{}_{jk}+
\Gamma^{i}{}_{ak}\Gamma^{a}{}_{jl}-\Gamma^{i}{}_{al}\Gamma^{a}{}_{jk}\neq 0,\nonumber\\
S^{i}{}_{jk}&=&\frac{1}{2}\left(\Gamma^{i}{}_{jk}-\Gamma^{i}{}_{kj}\right)\neq 0,\nonumber
\end{eqnarray}
then the system $\left(v^{i},V^{i}{}_{A}\right)$ is the aholonomic
velocity in the sense that there are no coordinates $q^{i}$,
$q^{i}{}_{A}$ in $FM$ for which the following could hold:
\[
v^{i}=\frac{dq^{i}}{dt},\qquad V^{i}{}_{A}=\frac{dq^{i}{}_{A}}{dt}.
\]
Generalized velocities at $e\in\pi^{-1}(x)$ are represented by the
$(n+1)$-tuples of vectors $\left(V;\ldots,V_{A},\ldots\right)\in
\left(T_{x}M\right)^{n+1}$ attached at $x=\pi(e)\in M$. Therefore,
the affine connection $\Gamma$ enables one to identify the tangent
bundle $TFM$ with the following fibre bundle over $M$:
\[
\mathcal{T}FM=\bigcup_{x\in M}F_{x}M\oplus\left(T_{x}M\right)^{n+1}\subset \bigcup_{x\in
M}\left(T_{x}M\right)^{2n+1}.
\]
It is an open subset of the last Whitney sum; namely, the first $n$-tuple is the submanifold
of $\left(T_{x}M\right)^{n}$ consisting of linearly independent systems. Obviously, for any
fixed affine connection $\Gamma$ the above-mentioned diffeomorphism of $TFM$ onto
$\mathcal{T}FM$ is canonical.

Similarly, canonical momenta at $e\in\pi^{-1}(x)$ are represented
by the $(n+1)$-tuples of co-vectors at $x=\pi(e)$, i.e.,
$\left(P;\ldots,P^{A},\ldots\right)\in
\left(T^{\ast}_{x}M\right)^{n+1}$.

It is seen again that $\Gamma$ establishes a distinguished diffeomorphism of the cotangent
bundle $T^{\ast}FM$ (phase space of the system) with the bundle
\[
\mathcal{T}^{\ast}FM=\bigcup_{x\in M}F_{x}M\oplus\left(T^{\ast}_{x}M\right)^{n+1}\subset
\bigcup_{x\in M}\left(T_{x}M\right)^{n}\oplus\left(T^{\ast}_{x}M\right)^{n+1}.
\]
Denoting generalized velocities by
\[
v^{i}=\frac{dx^{i}}{dt},\qquad v^{i}{}_{A}=\frac{de^{i}{}_{A}}{dt}
\]
and their conjugate canonical momenta by $p_{i}$, $p^{A}{}_{i}$,
we have that
\[
p_{i}v^{i}+p^{A}{}_{i}v^{i}{}_{A}=P_{i}V^{i}+P^{A}{}_{i}V^{i}{}_{A},
\]
and
\begin{eqnarray}
V^{i}=v^{i},&\quad& V^{i}{}_{A}=v^{i}{}_{A}+\Gamma^{i}{}_{jk}(x)e^{j}{}_{A}v^{k},\nonumber\\
P_{i}=p_{i}-e^{j}{}_{A}p^{A}{}_{k}\Gamma^{k}{}_{ji},&\quad& P^{A}{}_{i}=p^{A}{}_{i}.\nonumber
\end{eqnarray}
Let us observe the following anti-dualism: it is easily seen that
$V^{i}$ and $P^{A}{}_{i}$ are connection-independent, whereas
$P_{i}$ and $V^{i}{}_{A}$ depend explicitly on $\Gamma$. Geometric
reasons for this are that that $V$ is simply the $\pi$-projection
of generalized velocity $\dot{\varrho}(t)$ to $M$ and
$P^{A}{}_{i}$ are components of the linear functional on
$T_{\varrho(t)}F_{x(t)}M$ ($x(t)=\pi(\varrho(t))$) which is simply
the usual restriction of some functional on $T_{\varrho(t)}FM$ to
the subspace $T_{\varrho(t)}F_{x(t)}M$. The restriction procedure
is obviously connection-independent.

\noindent{\bf Remark:} $v^{i}{}_{A}$ are not components of vectors
in $T_{x}M$, and similarly, $p_{i}$ are not elements of covectors.
Only $v^{i}$ and $p^{A}{}_{i}$ are so. The total systems
$\left(v^{i},v^{i}{}_{A}\right)$, $\left(p_{i},p^{A}{}_{i}\right)$
represent the higher-floor vectors and covectors in $T_{e}FM$. The
con\-nection-dependent systems $V^{i}{}_{A}$, $P_{i}$ are
respectively components of vectors and covectors in $T_{x}M$
(elements of $T_{x}M$ itself and $T^{\ast}_{x}M$).

Quite a similar reasoning works for infinitesimal gyroscope, i.e.,
for the bundle $F(M,g)$. The situation there is more complicated
technically, because the quantities $e^{i}{}_{A}$ are no longer
independent, i.e., they satisfy the orthonormality conditions
\[
g_{ij}e^{i}{}_{A}e^{j}{}_{B}=\delta_{AB}.
\]
This fact creates some problems in equations of motion; to overcome them one uses an auxiliary
technical tool, namely, a field of linear orthonormal frames defined all over the manifold
$M$. We return to this question later on.

In mechanics of extended affinely- and metrically-rigid bodies one
uses the concepts of affine velocity (Eringen's ``gyration") and
affine spin (affine momentum, hypermomentum). They may be as well
defined for infinitesimal objects, i.e., for internal degrees of
freedom.

Every tensor object in the tangent space $T_{e}FM$ or
$T_{e}F(M,g)$ may be expressed in terms of its components with
respect to the frame $e$ itself. In particular, this concerns the
translational velocity $V$ and the system of internal velocities
$\left(\ldots,V_{A},\ldots\right)$. Similarly, the covariant
conjugate momenta $P$, $\left(\ldots,P^{A},\ldots\right)$ may be
expanded with respect to the dual co-frame $\widetilde{e}$. So, we
have that
\[
V=\widehat{V}^{A}e_{A},\qquad V_{A}=e_{B}\widehat{\Omega}^{B}{}_{A},
\]
i.e.,
\[
\widehat{V}^{A}=\langle e^{A},V\rangle=e^{A}{}_{i}V^{i},\qquad
\widehat{\Omega}^{A}{}_{B}=\langle e^{A},V_{B}\rangle=e^{A}{}_{i}V^{i}{}_{B},
\]
or explicitly, using differentiation symbols,
\[
\widehat{\Omega}^{A}{}_{B}=\left\langle
e^{A},\frac{De_{B}}{Dt}\right\rangle=e^{A}{}_{i}\frac{De^{i}{}_{B}}{Dt}.
\]
This expression is exactly the co-moving affine velocity known
from the theory of extended affine bodies in a flat space.
$\widehat{V}^{A}$ are co-moving components of the translational
velocity. Geometrically, these objects belong to the matrix
spaces, i.e.,
\[
\widehat{\Omega}\in{\rm L}(n,\mathbb{R})\simeq{\rm GL}(n,\mathbb{R})^{\prime},\qquad
\widehat{V}\in\mathbb{R}^{n}\simeq{\rm L}(1,n;\mathbb{R}).
\]
But having $e$ instantaneously fixed, we can associate with these numerical objects the
corresponding quantities in the tangent space $T_{x}M$, where $x=\pi(e)$ ($e\in F_{x}M$). For
$\widehat{V}$ this is just the usual translational velocity $V\in T_{x}M$:
\[
V^{i}=\frac{dx^{i}}{dt}.
\]

The quantity $\widehat{\Omega}$ and the frame $e$ give rise to the
object
\[
\Omega=\widehat{\Omega}^{A}{}_{B}e_{A}\otimes e^{B},\qquad
\Omega^{i}{}_{j}=V^{i}{}_{A}e^{A}{}_{j},
\]
i.e., explicitly in terms of derivatives:
\[
\Omega=\frac{De_{A}}{Dt}\otimes e^{A},\qquad
\Omega^{i}{}_{j}=\frac{De^{i}{}_{A}}{Dt}e^{A}{}_{j}=
e^{i}{}_{A}\widehat{\Omega}^{A}{}_{B}e^{B}{}_{j}.
\]
Geometrically, they are linear transformations in tangent spaces,
i.e.,
\[
\Omega\in{\rm L}(T_{x}M)\simeq T_{x}M\otimes T^{\ast}_{x}M={\rm GL}(T_{x}M)^{\prime},
\]
where, obviously, $x=\pi(e)$ ($e\in F_{x}M$).

Just as in mechanics of extended affine bodies, $\Omega^{i}{}_{j}$
are spatial (laboratory) components of the affine velocity, i.e.,
they represent what Eringen used to call ``gyration"
\cite{Erin62,Erin68}. Geometrically it is important that
$\widehat{\Omega}$, $\Omega$ belong respectively to Lie algebras
of the groups GL$(n,\mathbb{R})$, GL$(T_{x}M)$.

The same may be done for the covariant canonical momenta $P$ and
$P^{A}$:
\[
P=\widehat{P}_{A}e^{A},\qquad P^{A}=\widehat{\Sigma}^{A}{}_{B}e^{B},
\]
i.e.,
\[
\widehat{P}_{A}=\langle P,e_{A}\rangle=P_{i}e^{i}{}_{A},\qquad
\widehat{\Sigma}^{A}{}_{B}=\langle P^{A},e_{B}\rangle=P^{A}{}_{i}e^{i}{}_{B}.
\]
Just as previously, $\widehat{P}\in\mathbb{R}^{n}$,
$\widehat{\Sigma}\in$ L$(n,\mathbb{R})$, but strictly speaking one
means here $\mathbb{R}^{n}$ as identified in a Cartesian way with
its own dual $\mathbb{R}^{n\ast}$; similarly, L$(n,\mathbb{R})$
plays here a role of the Lie co-algebra L$(n,\mathbb{R})^{\ast}$
of GL$(n,\mathbb{R})$. One does not notice this subtle distinction
because L$(n,\mathbb{R})^{\ast}$ and L$(n,\mathbb{R})$ are
canonically identified via the pairing
\[
\langle A,B\rangle={\rm Tr}(AB).
\]

In analogy to mechanics of extended bodies in flat spaces we say that the matrix elements of
$\widehat{\Sigma}$ are co-moving components of the affine spin (hypermomentum). Just as in
flat-space theory, the laboratory (spatial) description is based on the quantity
\[
\Sigma=\widehat{\Sigma}^{A}{}_{B}e_{A}\otimes e^{B},\qquad
\Sigma^{i}{}_{j}=e^{i}{}_{A}P^{A}{}_{j}=e^{i}{}_{A}\widehat{\Sigma}^{A}{}_{B}e^{B}{}_{j}.
\]
So, $\widehat{\Sigma}^{A}{}_{B}$ are Hamiltonian generators of the
structural group GL$(n,\mathbb{R})$ (micromaterial global
transformations) (\ref{a11}), (\ref{b15}). And similarly,
$\Sigma^{i}{}_{j}$ are Hamiltonian generators of the local group
of microspatial transformations (\ref{a14a}), (\ref{a14b}),
(\ref{a15}).

If the configuration $e\in FM$ is fixed, then the velocities
$V^{i}$, $V^{i}{}_{A}$ contain exactly the same information as
$V^{i}$, $\Omega^{i}{}_{j}$ or $\widehat{V}^{A}$,
$\widehat{\Omega}^{A}{}_{B}$. Similarly, the momenta $P_{i}$,
$P^{A}{}_{i}$ are equivalent to $P_{i}$, $\Sigma^{i}{}_{j}$ and
$\widehat{P}_{A}$, $\widehat{\Sigma}^{A}{}_{B}$. Using more
sophisticated terms we would say that the state space
$\mathcal{T}FM$ is naturally diffeomorphic with the manifolds
\begin{eqnarray}
\mathcal{T}_{\Omega}FM&:=&\bigcup_{x\in M}F_{x}M\oplus T_{x}M\oplus{\rm L}(T_{x}M),
\nonumber\\
\mathcal{T}_{\widehat{\Omega}}FM&:=&\bigcup_{x\in M}F_{x}M\oplus \mathbb{R}^{n}\oplus{\rm
L}(n,\mathbb{R}).\nonumber
\end{eqnarray}
Similarly, the phase space $\mathcal{T}^{\ast}FM$ is canonically diffeomorphic with
\begin{eqnarray}
\mathcal{T}^{\ast}_{\Sigma}FM&:=&\bigcup_{x\in M}F_{x}M\oplus T^{\ast}_{x}M\oplus{\rm
L}(T_{x}M),
\nonumber\\
\mathcal{T}^{\ast}_{\widehat{\Sigma}}FM&:=&\bigcup_{x\in M}F_{x}M\oplus
\mathbb{R}^{n}\oplus{\rm L}(n,\mathbb{R}).\nonumber
\end{eqnarray}
Let us notice that there is a subtle distinction between
$\mathcal{T}_{\widehat{\Omega}}FM$ and
$\mathcal{T}^{\ast}_{\widehat{\Sigma}}FM$ indistinguishable in our
simplified notation. Namely, in the latter space the
$\mathbb{R}^{n}$-term consists of row numerical vectors, i.e.,
linear functionals on the space of numerical column vectors (so,
strictly speaking, one deals with $\mathbb{R}^{n\ast}$). And
L$(n,\mathbb{R})$ is canonically isomorphic with
L$(n,\mathbb{R})^{\ast}$. ({\bf Remark:} this isomorphism is
canonical for L$(V)$, L$(V)^{\ast}$, where $V$ is an arbitrary
linear space. This is no longer the case for $V$, $V^{\ast}$
themselves.) Obviously, the mentioned identifications between
state manifolds may be interpreted as natural ones only on the
basis of some fixed affine connection $\Gamma$ on $M$.

The cotangent bundle $P=T^{\ast}FM$ carries a natural symplectic
structure \cite{Abr-Mars78,JJS03,Ster64} and is used as the
mechanical phase space of our problem. As just mentioned, the
affine connection $\Gamma$ fixes some diffeomorphism of $P$ onto
the manifold $\mathcal{P}=\mathcal{T}^{\ast}FM$. This
diffeomorphism enables one to carry over to $\mathcal{P}$ the
intrinsic symplectic geometry of $P=T^{\ast}FM$. But there is one
delicate point which, when overlooked, may lead to serious
mistakes. Namely, unlike the previously used coordinates in
$T^{\ast}FM$, the quantities $x^{i}$, $e^{i}{}_{A}$, $P_{i}$,
$P^{A}{}_{i}$ fail to be canonical (Darboux) coordinates for the
symplectic structure $\Gamma$-transferred to
$\mathcal{T}^{\ast}FM$. After some essentially easy although
sometimes technically embarrassing calculations one obtains the
system of basic Poisson brackets. So, just like in mechanics of
extended affine bodies, we have the obvious rules
\begin{eqnarray}
&&\{x^{i},x^{j}\}=0,\quad \{e^{i}{}_{A},e^{j}{}_{B}\}=0,\quad \{x^{i},e^{j}{}_{A}\}=0,\quad
\{P^{A}{}_{i},P^{B}{}_{j}\}=0,\nonumber\\
&&\{P^{A}{}_{i},x^{j}\}=0,\qquad
\{e^{i}{}_{A},P^{B}{}_{j}\}=\delta^{i}{}_{j}\delta^{B}{}_{A},\qquad
\{x^{i},P_{j}\}=\delta^{i}{}_{j}.\nonumber
\end{eqnarray}
However, further on one obtains more complicated expressions explicitly dependent on the
$\Gamma$-geometry of $M$:
\[
\{P_{i},P_{j}\}=\Sigma^{k}{}_{l}\mathcal{R}^{l}{}_{kij},\quad
\{P_{i},P^{A}{}_{j}\}=-P^{A}{}_{k}\Gamma^{k}{}_{ji},\quad
\{P_{i},e^{j}{}_{A}\}=e^{k}{}_{A}\Gamma^{j}{}_{ki},
\]
where, obviously, $\mathcal{R}$ denotes the curvature tensor of
$\Gamma$; we use the convention
\[
\mathcal{R}^{a}{}_{bij}=\Gamma^{a}{}_{bj,i}-\Gamma^{a}{}_{bi,j}+
\Gamma^{a}{}_{ci}\Gamma^{c}{}_{bj}-\Gamma^{a}{}_{cj}\Gamma^{c}{}_{bi}
\]
(comma, as usual in the tensor calculus, denotes the partial
differentiation with respect to the indicated coordinate). The
latter three Poisson brackets mean, roughly speaking, that $P_{i}$
are Hamiltonian generators of parallel transports of our state
variables. One can also show that
\[
\{\Sigma^{i}{}_{j},\Sigma^{k}{}_{l}\}=\delta^{i}{}_{l}\Sigma^{k}{}_{j}-
\delta^{k}{}_{j}\Sigma^{i}{}_{l}.
\]
In these formulas we recognize structure constants of the linear
group. As expected, this means that $\Sigma^{i}{}_{j}$ are basic
Hamiltonian generators of (\ref{a14a}), (\ref{a14b}), i.e.,
roughly speaking, of the system of groups GL$(T_{x}M)$. It is easy
to obtain the relationship
\[
\{P_{i},\Sigma^{k}{}_{j}\}=\Sigma^{l}{}_{j}\Gamma^{k}{}_{li}-\Sigma^{k}{}_{l}\Gamma^{l}{}_{ji},
\]
that is also compatible with the mentioned interpretation of $P_{i}$ as Hamiltonian generators
of parallel transports.

If some function $F$ depends only on the configuration, i.e., on
the $FM$-variables (but not on $P_{i}$ and $P^{A}{}_{i}$), then
\[
\{\Sigma^{i}{}_{j},F\}=-E^{i}{}_{j}F=-e^{i}{}_{A}\frac{\partial F}{\partial e^{j}{}_{A}},
\]
where
\[
E^{i}{}_{j}=e^{i}{}_{K}e^{L}{}_{j}E^{K}{}_{L}=e^{i}{}_{K}\frac{\partial}{\partial
e^{j}{}_{K}}.
\]
The Poisson brackets involving co-moving components are as
follows:
\begin{eqnarray}
\{\widehat{P}_{A},\widehat{P}_{B}\}=
\widehat{\Sigma}^{K}{}_{L}\mathcal{R}^{L}{}_{KAB}-2\widehat{P}_{K}S^{K}{}_{AB},&\quad&
\{\widehat{\Sigma}^{A}{}_{B},\widehat{P}_{C}\}=-\widehat{P}_{B}\delta^{A}{}_{C},\nonumber\\
\{\widehat{\Sigma}^{A}{}_{B},\widehat{\Sigma}^{C}{}_{D}\}=
\delta^{C}{}_{B}\widehat{\Sigma}^{A}{}_{D}-\delta^{A}{}_{D}\widehat{\Sigma}^{C}{}_{B},&\quad&
\{\Sigma^{i}{}_{j},\widehat{\Sigma}^{A}{}_{B}\}=0,\nonumber
\end{eqnarray}
where $\mathcal{R}^{L}{}_{KAB}$, $S^{K}{}_{AB}$ are respectively
co-moving components of the curvature and torsion tensors of
$\Gamma$ with respect to the instantaneous internal configuration
$e$, i.e.,
\begin{eqnarray}
\mathcal{R}^{L}{}_{KAB}&:=&e^{L}{}_{i}\mathcal{R}^{i}{}_{jmn}e^{j}{}_{K}e^{m}{}_{A}e^{n}{}_{B},\nonumber\\
S^{K}{}_{AB}&:=&e^{K}{}_{i}S^{i}{}_{jm}e^{j}{}_{A}e^{m}{}_{B},\qquad
S^{i}{}_{jm}=\frac{1}{2}\left(\Gamma^{i}{}_{jm}-\Gamma^{i}{}_{mj}\right).\nonumber
\end{eqnarray}

Let us observe a characteristic difference between the spatial and
co-moving representations of Poisson brackets. Namely, the latter
ones are ``almost" identical with the basic commutation relations
(structure constants) for the affine group GAf$(n,\mathbb{R})$.
The ``almost" concerns the brackets
$\{\widehat{P}_{A},\widehat{P}_{B}\}$ which do not vanish if the
connection $\Gamma$ is not completely flat (in the sense that both
the curvature and torsion tensors vanish). The co-moving brackets
enable one to interpret $\widehat{P}_{A}$ as a Hamiltonian
generator of parallel transports along the $A$-th legs of the
frames $e$.

Let us also note certain additional and convenient Poisson brackets. If $F$ depends only on
the configuration, then
\begin{eqnarray}
\{\widehat{\Sigma}^{A}{}_{B},F\}&=&-E^{A}{}_{B}F=
-e^{i}{}_{B}\frac{\partial F}{\partial e^{i}{}_{A}},\nonumber\\
\{\widehat{P}_{A},F\}&=&-H_{A}F,\qquad
\{P_{i},F\}=-H_{i}F,\nonumber
\end{eqnarray}
where
\[
H_{i}=e^{A}{}_{i}H_{A}=\frac{\partial}{\partial x^{i}}-
\Gamma^{k}{}_{ji}e^{j}{}_{B}\frac{\partial}{\partial e^{k}{}_{B}}.
\]
Again we conclude that $P_{i}$ are Hamiltonian generators of
parallel transports along the $i$-th coordinate axes, and
$\widehat{P}_{A}$ are generators of parallel transports along the
$A$-th legs of the frames $e$.

In the theory of an extended affinely-rigid body in Euclidean or
affine space the quantities $P_{i}$, $P^{A}{}_{i}$,
$\Sigma^{i}{}_{j}$, $\widehat{\Sigma}^{A}{}_{B}$ have a very
natural geometric interpretation based on some transformation
groups acting in the configuration space. Let us remind that in
the flat affine case the configuration space $FM$ trivializes to
the Cartesian product $M\times F(V)$, where $F(V)$ is the manifold
of frames in $V$. The most important transformation groups acting
in $M\times F(V)$ are the following ones:
\begin{itemize}
\item[($i$)] {\bf spatial translations}
$(x,e)\mapsto(t_{v}(x),e)$; they shift the centre of mass $x\in M$
along the vector $v\in V$ without affecting the internal
configuration $e\in F(V)$. Analytically:
\[
\left(x^{i},e^{i}{}_{A}\right)\mapsto\left(x^{i}+v^{i},e^{i}{}_{A}\right).
\]

\item[($ii$)] {\bf additive translations of internal degrees of
freedom:} any $\xi\in V^{n}$ gives rise to the mapping
$(x,e)\mapsto(x,e+\xi)$, i.e.,
\[
(x^{i},e^{i}{}_{A})\mapsto(x^{i},e^{i}{}_{A}+\xi^{i}{}_{A}).
\]
Of course, such transformations act only locally in $M\times
F(V)$, because they may produce linearly dependent $n$-tuples of
vectors from independent ones. The Hamiltonian generators are
given by $P^{A}{}_{i}$.

\item[($iii$)] {\bf spatial affine transformations of internal
degrees of freedom:} any $L\in$ GL$(V)$ acts on the configuration
space as follows:
\[
(x;\ldots,e_{A},\ldots)\mapsto(x;\ldots,Le_{A},\ldots)
\]
or shortly $(x,e)\mapsto(x,Le)$. Analytically,
\[
(x^{i},e^{i}{}_{A})\mapsto(x^{i},L^{i}{}_{j}e^{j}{}_{A}).
\]
The centre-of-mass position is not effected. The Hamiltonian
generators of this group are given by $\Sigma^{i}{}_{j}$. Because
of this the object $\left[\Sigma^{i}{}_{j}\right]$ is referred to
as affine spin. If $g\in V^{\ast}\otimes V^{\ast}$ is the metric
tensor of $V$, then the quantity
\[
S^{i}{}_{j}=\Sigma^{i}{}_{j}-g_{jk}\Sigma^{k}{}_{l}g^{li},
\]
i.e., the doubled $g$-skew-symmetric part of $\Sigma$, is the
usual canonical spin generating rigid spatial rotations of
internal (relative) degrees of freedom.

\item[($iv$)] {\bf material affine transformations of internal
degrees of freedom:} any not singular matrix $L\in$
GL$(n,\mathbb{R})$ acts on the configuration space as follows:
\[
(x;\ldots,e_{A},\ldots)\mapsto(x;\ldots,e_{B}L^{B}{}_{A},\ldots).
\]
As in ($iii$), we use the shorthand $(x,e)\mapsto(x,eL)$. {\it
\textbf{Important in this context:} do not confuse conceptually
linear mappings in $V$ with matrices!} This group is generated in
the Hamiltonian sense by $\widehat{\Sigma}^{A}{}_{B}$, i.e., by
co-moving components of the hypermomentum. Let us observe that the
above right-acting transformations may be interpreted as ones
induced by linear mappings $L\in$ GL$(n,\mathbb{R})$ acting in the
"material space" $\mathbb{R}^{n}$. As usual, the frames themselves
may be interpreted as linear mappings $e:\mathbb{R}^{n}\rightarrow
V$. The label space $\mathbb{R}^{n}$ is endowed with the standard
metric $\delta$, sometimes denoted also by $\eta$, to stress the
link with the usual formulation of the mechanics of extended
affine systems. The $\delta$-skew-symmetric part of
$\widehat{\Sigma}^{A}{}_{B}$,
\[
V^{A}{}_{B}:=\widehat{\Sigma}^{A}{}_{B}-\delta_{BC}\widehat{\Sigma}^{C}{}_{D}\delta^{DA},
\]
i.e., vorticity, generates rigid material rotations of the body.

\item[($v$)] {\bf total affine transformations in space.} They act
both on the translational and internal degrees of freedom. Let
$\varphi:M\rightarrow M$ be an arbitrary affine transformation
acting in $M$, and $L[\varphi]:V\rightarrow V$ be its linear
(homogeneous) part. Explicitly, for any pair of points $a,b\in M$,
the radius vector $\overrightarrow{\varphi(a)\varphi(b)}$ is
obtained from the vector $\overrightarrow{ab}$ through the action
of $L[\varphi]$:
\[
\overrightarrow{\varphi(a)\varphi(b)}=L[\varphi]\overrightarrow{ab}.
\]
The action of $\varphi$ on the configuration space $M\times F(V)$ is given by
$(x,e)\mapsto\left(\varphi(x),L[\varphi]e\right)$. Analytically, in Cartesian coordinates
$x^{i}$,
\[
\left(x^{i},e^{i}{}_{A}\right)\mapsto
\left(L^{i}{}_{j}x^{j}+a^{i},L^{i}{}_{j}e^{j}{}_{A}\right).
\]
Infinitesimal Hamiltonian generators are given by
$\left(P_{i},\mathcal{J}^{i}{}_{j}\right)$, where the total
canonical affine momentum $\mathcal{J}^{i}{}_{j}$ with respect to
the origin of coordinates is defined as follows:
\[
\mathcal{J}^{i}{}_{j}=x^{i}P_{j}+\Sigma^{i}{}_{j},
\]
i.e., it consists of the ``orbital" and ``internal" parts,
respectively, $x^{i}P_{j}$ and $\Sigma^{i}{}_{j}$. Taking the
skew-symmetric part of the above expression we obtain the usual
splitting of the angular momentum onto its ``orbital" and
"internal" (spin) parts:
\[
\Im^{i}{}_{j}=L^{i}{}_{j}+S^{i}{}_{j},
\]
where, obviously,
\begin{eqnarray}
\Im^{i}{}_{j}&=&\mathcal{J}^{i}{}_{j}-g^{ik}g_{jl}\mathcal{J}^{l}{}_{k}=
\mathcal{J}^{i}{}_{j}-\mathcal{J}_{j}{}^{i},\nonumber\\
L^{i}{}_{j}&=&x^{i}P_{j}-g^{ik}g_{jl}x^{l}P_{k}=x^{i}P_{j}-x_{j}P^{i},\nonumber\\
S^{i}{}_{j}&=&\Sigma^{i}{}_{j}-g^{ik}g_{jl}\Sigma^{l}{}_{k}=
\Sigma^{i}{}_{j}-\Sigma_{j}{}^{i}.\nonumber
\end{eqnarray}

\item[($vi$)] {\bf total affine transformations in the material
space.} If the configuration space of extended affinely-rigid body
is identified with $M\times F(V)$, then, as mentioned, the
material space may be identified simply with $\mathbb{R}^{n}$. The
current position of the constituent labelled by $a\in
\mathbb{R}^{n}$ is displaced with respect to the instantaneous
position of the centre of mass by the spatial vector $u$ with
coordinates $u^{i}=e^{i}{}_{K}a^{K}$. The affine group
GAf$(n,\mathbb{R})$ in $\mathbb{R}^{n}$ is canonically identical
with the semidirect product
GL$(n,\mathbb{R})\times_{s}\mathbb{R}^{n}$, i.e., with the set of
pairs $(B,b)$ multiplied (composed) according to the group rule
\[
(B_{1},b_{1})(B_{2},b_{2})=(B_{1}B_{2},b_{1}+B_{1}b_{2}).
\]
The action of $(B,b)$ on the configuration $(x,e)$ is given by
\[
(x,e)\mapsto\left(t_{eb}(x),eB\right),
\]
where in the expression $eb$, $e\in F(V)$ is identified with a linear isomorphism of
$\mathbb{R}^{n}$ onto $V$. Analytically,
\[
\left(x^{i},e^{i}{}_{A}\right)\mapsto\left(x^{i}+e^{i}{}_{K}b^{K},e^{i}{}_{K}B^{K}{}_{A}\right).
\]
The corresponding system of infinitesimal Hamiltonian generators is given by
$\left(\widehat{P}_{A},\widehat{\Sigma}^{K}{}_{L}\right)$.
\end{itemize}

If $(M,\Gamma)$ is not flat, then the above picture of the
transformation groups changes in an essential way. There is no
counterpart of the $n(n+1)$-dimensional group of affine
transformations acting in $M$. In particular, there is no concept
of spatial translations and radius vectors. Only the
infinite-dimensional group Diff$(M)$ of all diffeomorphisms of $M$
onto $M$ is well defined. As a rule, its elements do not preserve
the parallel transport and covariant differentiation. There are no
extended affinely-rigid bodies and affine degrees of freedom may
be considered only as internal ones. Translations in the
micromaterial space $\mathbb{R}^{n}$ are well defined, however,
they lose the physical meaning that they had in the theory of
extended affine bodies. The only transformations which survive are
those dealing with internal degrees of freedom only, without any
affecting of translational motion in $M$. Let us describe them in
some details:
\begin{itemize}
\item[($i$)] {\bf additive translations of internal degrees of
freedom.} They act in any fibre $F_{x}M=\pi^{-1}(x)\subset FM$ of
the bundle of frames exactly as their flat-space counterparts do
in $F(V)$. There is, however, an essential novelty when these
transformations are considered globally in $FM$, not in separate
fibres. Namely, in a curved manifold there is no canonical
identification between different tangent spaces. Therefore, the
concept of ``the same" translation in different fibres is missing,
and the group becomes infinite-dimensional. Every ordered
$n$-tuple of vector fields in $M$,
\[
M\ni x\mapsto
\xi(x)=\left(\ldots,\xi_{A}(x),\ldots\right)\in\left(T_{x}M\right)^{n},
\]
gives rise to the local transformation of $FM$
\begin{equation}\label{a36}
F_{x}M\ni e\mapsto e+\xi(x)\in F_{x}M
\end{equation}
(local, because some frames are mapped into linearly dependent
$n$-tuples of vectors). Obviously, the sum in the last formula is
meant pointwisely, i.e., for any $A=\overline{1,n}$ the element
$e_{A}$ is replaced by $e_{A}+\xi_{A}(x)$. Canonical momenta
$P^{A}{}_{i}$ are Hamiltonian generators of such transformations.
More precisely, for any ordered $n$-tuple of covector fields
$X_{A}$ on $M$, the quantity $X^{i}{}_{A}P^{A}{}_{i}$ generates
some one-parameter group of transformations (\ref{a36}).

\item[($ii$)] {\bf spatial affine transformations of internal
degrees of freedom.} Here we are dealing with the same phenomenon
as previously, namely, the $n^{2}$-dimensional Lie groups
GL$(T_{x}M)$ acting separately in tangent spaces $T_{x}M$ give
rise to some infinite-dimensional group of transformations acting
globally in $FM$. This group is ``parameterized" by mixed
second-order tensor fields $T$ on $M$, cf. (\ref{a14a}),
(\ref{a14b}), (\ref{a15}). The quantities $\Sigma^{i}{}_{j}$ are
Hamiltonian generators of this group. More precisely, for any
mixed second-order tensor field $\alpha$ on $M$, the quantity
$\alpha^{i}{}_{j}\Sigma^{j}{}_{i}$ generates some one-parameter
group of transformations (\ref{a14a}), (\ref{a14b}), (\ref{a15}).
There is no escape from the infinite dimension because in a curved
space $(M,\Gamma)$ there is no canonical identification of tangent
spaces at different points.

\item[($iii$)] {\bf micromaterial affine transformations of
internal degrees of freedom.} Now, in a complete analogy to the
extended model in a flat space, the $n^{2}$-dimensional Lie group
GL$(n,\mathbb{R})$ acts on the configuration space $FM$ according
to the usual rule (\ref{a11}). The quantities
$\widehat{\Sigma}^{A}{}_{B}$ are Hamiltonian generators of the
corresponding Lie groups of extended point transformations.

One should stress that the quantities $L^{A}{}_{B}$ describing micromaterial transformations
may be defined as constants, but they need not be so. Namely, following the pattern developed
in gauge theories we can replace the matrices $L$ by GL$(n,\mathbb{R})$-valued functions on
$M$. They act on the configuration space $FM$ according to the rule (\ref{a13}), (\ref{b15}).
Any matrix-valued function $\alpha:M\rightarrow$ L$(n,\mathbb{R})$ gives rise to the
one-parameter group of transformations with the Hamiltonian generator
$\alpha^{A}{}_{B}\widehat{\Sigma}^{B}{}_{A}$. In this way we obtain again, just as in previous
examples, some infinite-dimensional functionally-"parameterized" transformation group.
\end{itemize}

We shall almost not deal with the diffeomorphism group Diff$(M)$ or its special
(volume-preserving) subgroup SDiff$(M)$ as a symmetry of our model. It does not play any
essential role in mechanics of material points with internal degrees of freedom. On the
contrary, it is rather relevant for the theory of micromorphic continua.

Here let us only mention that any diffeomorphism $f:M\rightarrow
M$ acts on the configuration space $FM$ according to the rule
\[
F_{x}M\ni e\mapsto Df_{x}\cdot e\in F_{f(x)}M,
\]
i.e., analytically,
\[
\left(x^{i},e^{i}{}_{A}\right)\mapsto\left(f^{i}(x),\frac{\partial
f^{i}}{\partial x^{j}}(x)e^{j}{}_{A}\right),
\]
where the functions $f^{i}(x)$ provide analytical description of
$f$ (they express coordinates of $f(x)$ in terms of those of $x$).

Due to the essentially local (``$x$-dependent") character of
transformations (\ref{a14a}), (\ref{a14b}), (\ref{b15}),
(\ref{a36}), and admissibly local character of (\ref{a13}), their
action on velocities and canonical momenta is (respectively, may
be) different from that could be naively expected on the basis of
analogy with the flat-space theory.

Additive translations in internal degrees of freedom (\ref{a36}) affect internal velocities as
follows:
\[
V^{i}{}_{A}\mapsto{}^{\prime}V^{i}{}_{A}=V^{i}{}_{A}+V^{j}\nabla_{j}\xi^{i}{}_{A},
\]
where, as previously, $V^{i}=dx^{i}/dt$ denotes the translational
velocity in $M$. This rule reduces to the invariance of
$\left(\ldots,V_{A},\ldots\right)$ (characteristic for the
flat-space theory) only when all vector fields $\xi_{A}$ are
parallel in the $\Gamma$-sense. Translational velocity is
invariant under transformations of internal degrees of freedom,
${}^{\prime}V^{i}=V^{i}$. On the contrary, the covariant canonical
momenta transform according to the rules
\[
{}^{\prime}P_{i}=P_{i}-P^{A}{}_{j}\nabla_{i}\xi^{j}{}_{A},\qquad
{}^{\prime}P^{A}{}_{i}=P^{A}{}_{i}.
\]
Microspatial affine transformations of internal degrees of freedom
(\ref{a14a}), (\ref{a14b}), (\ref{a15}) also do not affect
translational velocity, i.e.,
\[
{}^{\prime}V^{i}=V^{i},
\]
but the internal velocities are transformed in the following way:
\begin{equation}\label{a39}
{}^{\prime}V^{i}{}_{A}=
T^{i}{}_{j}V^{j}{}_{A}+V^{k}\left(\nabla_{k}T^{i}{}_{j}\right)e^{j}{}_{A}.
\end{equation}
This implies that
\begin{eqnarray}
{}^{\prime}\Omega^{i}{}_{j}&=&T^{i}{}_{l}\Omega^{l}{}_{m}T^{-1m}{}_{j}+
V^{k}\left(\nabla_{k}T^{i}{}_{m}\right)T^{-1m}{}_{j}\nonumber\\
&=&T^{i}{}_{l}\Omega^{l}{}_{m}T^{-1m}{}_{j}+
\left(\nabla_{V}T^{i}{}_{m}\right)T^{-1m}{}_{j},\label{b39}
\end{eqnarray}
and similarly,
\begin{equation}\label{c39}
{}^{\prime}\widehat{\Omega}^{A}{}_{B}=\widehat{\Omega}^{A}{}_{B}+
e^{A}{}_{l}T^{-1l}{}_{i}\left(\nabla_{V}T^{i}{}_{j}\right)e^{j}{}_{B}.
\end{equation}
On the contrary, translational canonical momenta $P_{i}$ suffer
the transformation
\begin{equation}\label{d39}
{}^{\prime}P_{i}=P_{i}-\Sigma^{k}{}_{l}T^{-1l}{}_{j}\nabla_{i}T^{j}{}_{k},
\end{equation}
whereas the internal ones obey the well-known global rule:
\[
{}^{\prime}P^{A}{}_{i}=P^{A}{}_{j}T^{-1j}{}_{i},
\]
therefore,
\[
{}^{\prime}\Sigma^{i}{}_{j}=T^{i}{}_{k}\Sigma^{k}{}_{m}T^{-1m}{}_{j},\qquad
{}^{\prime}\widehat{\Sigma}^{A}{}_{B}=\widehat{\Sigma}^{A}{}_{B}.
\]
The formulas (\ref{a39}), (\ref{b39}), (\ref{c39}), (\ref{d39})
reduce to the corresponding global rules from the mechanics of
extended affine bodies if and only if the field $T$ is
$\Gamma$-parallel, i.e.,
\[
\nabla_{k}T^{i}{}_{j}=0.
\]
The global micromaterial transformations (\ref{a11}) act just like in the flat-space theory of
extended affine bodies:
\begin{eqnarray}
{}^{\prime}V^{i}=V^{i},&\quad& {}^{\prime}V^{i}{}_{A}=V^{i}{}_{B}L^{B}{}_{A},\nonumber\\
{}^{\prime}P_{i}=P_{i},&\quad& {}^{\prime}P^{A}{}_{i}=L^{-1A}{}_{B}P^{B}{}_{i},\nonumber
\end{eqnarray}
therefore,
\begin{eqnarray}
{}^{\prime}\Omega^{i}{}_{j}=\Omega^{i}{}_{j},&\quad&
{}^{\prime}\widehat{\Omega}^{A}{}_{B}=L^{-1A}{}_{C}\widehat{\Omega}^{C}{}_{D}L^{D}{}_{B},
\nonumber\\
{}^{\prime}\Sigma^{i}{}_{j}=\Sigma^{i}{}_{j},&\quad&
{}^{\prime}\widehat{\Sigma}^{A}{}_{B}=L^{-1A}{}_{C}\widehat{\Sigma}^{C}{}_{D}L^{D}{}_{B}.
\nonumber
\end{eqnarray}
For local, i.e., $x$-dependent micromaterial transformations we
have that
\begin{eqnarray}
{}^{\prime}V^{i}&=&V^{i},\nonumber\\
{}^{\prime}V^{i}{}_{A}&=&V^{i}{}_{B}L^{B}{}_{A}+
e^{i}{}_{B}L^{B}{}_{A,k}V^{k},\nonumber\\
{}^{\prime}\Omega^{i}{}_{j}&=&\Omega^{i}{}_{j}+
e^{i}{}_{B}\left(L^{B}{}_{A,k}L^{-1A}{}_{C}\right)e^{C}{}_{j}V^{k},\nonumber\\
{}^{\prime}\widehat{\Omega}^{A}{}_{B}&=&L^{-1A}{}_{C}\widehat{\Omega}^{C}{}_{D}L^{D}{}_{B}+
L^{-1A}{}_{C}L^{C}{}_{B,k}V^{k},\nonumber
\end{eqnarray}
where comma denotes the partial differentiation. Again these
formulas reduce to the preceding ones when the field $L$ is
constant, i.e., micromaterial transformations are global. By
duality rule,
\[
{}^{\prime}P_{i}{}^{\prime}V^{i}+
{}^{\prime}\widehat{\Sigma}^{B}{}_{A}{}^{\prime}\widehat{\Omega}^{A}{}_{B}=P_{i}V^{i}+
\widehat{\Sigma}^{B}{}_{A}\widehat{\Omega}^{A}{}_{B},
\]
one can easily show that
\begin{eqnarray}
{}^{\prime}P_{i}&=&P_{i}-\widehat{\Sigma}^{A}{}_{K}L^{K}{}_{B,i}L^{-1B}{}_{A},\nonumber\\
{}^{\prime}\widehat{\Sigma}^{A}{}_{B}&=&
L^{-1A}{}_{C}\widehat{\Sigma}^{C}{}_{D}L^{D}{}_{B},\nonumber\\
{}^{\prime}P^{A}{}_{i}&=&L^{-1A}{}_{B}P^{B}{}_{i}.\nonumber
\end{eqnarray}

\section{Metrical concepts}

We have described above kinematics and canonical formalism for affine model of internal
degrees of freedom in a manifold $M$ endowed with a general affine connection $\Gamma$. The
metrical concepts were yet not used at all. They become necessary when we aim at constructing
dynamical models. And of course, in realistic physical problems one usually deals with some
metric structure. For example, in the standard General Relativity, i.e., in the Einstein
theory of gravitation, the dynamical metric tensor of the four-dimensional space-time is used
for describing gravitational field. Even in more sophisticated alternative theories, like
gauge models of gravitation, there exists always some physically relevant metrical aspect of
the gravitational field.

So, from now on, we shall usually (nevertheless, not always)
assume that the manifold $M$ is endowed both with the metrical
tensor $g$ and affine connection $\Gamma$. In principle, one can
consider the most general structure, where $\Gamma$ and $g$ are
unrelated, independent on each other. Nevertheless, both from the
geometrical and physical point of view of particular interest are
situations when some kind of compatibility between affine and
metrical structures is assumed.

Riemann-Cartan space is one in which all metrical relations like length of vectors, angles (in
particular, orthogonality) are preserved by parallel transports, i.e.,
\[
\nabla_{k}g_{ij}=0;
\]
the covariant derivative of the metric tensor vanishes. It is well
known from differential geometry that then
\begin{equation}\label{a42}
\Gamma^{i}{}_{jk}=\left\{\begin{array}{c} i \\ jk
\end{array}\right\}+K^{i}{}_{jk}=\left\{\begin{array}{c} i \\ jk
\end{array}\right\}+S^{i}{}_{jk}+S_{jk}{}^{i}+S_{kj}{}^{i},
\end{equation}
where
\[
\left\{\begin{array}{c} i \\ jk
\end{array}\right\}=\frac{1}{2}g^{im}\left(g_{mj,k}+g_{mk,j}-g_{jk,m}\right)
\]
is the Levi-Civita (Christoffel) symbol,
\[
S^{i}{}_{jk}=\Gamma^{i}{}_{[jk]}=\frac{1}{2}\left(\Gamma^{i}{}_{jk}-\Gamma^{i}{}_{kj}\right)
\]
is the torsion tensor, and all indices are moved from their
natural positions with the help of the metric tensor $g$. The
quantity $K^{i}{}_{jk}$ is referred to as the contortion tensor;
it satisfies
\[
K^{i}{}_{jk}+K_{j}{}^{i}{}_{k}=0.
\]

The structure $(M,\Gamma,g)$ reduces to the Riemann space when it is torsion-free,
$S^{i}{}_{jk}=0$, i.e., the object $\Gamma$ is symmetric and automatically coincides with
Levi-Civita symbol,
\[
\Gamma^{i}{}_{jk}=\left\{\begin{array}{c} i \\ jk
\end{array}\right\}.
\]
In this model parallelism is derived from the metric concepts.

Riemann-Cartan-Weyl space is one in which angles between vectors,
but not necessarily their lengths are preserved, i.e.,
\[
\nabla_{k}g_{ij}=-Q_{k}g_{ij},
\]
where $Q_{k}$ is referred to as the Weyl covector, and its contravariant counterpart
$Q^{k}=g^{kj}Q_{j}$ as the Weyl vector. It is known from differential geometry that in
Riemann-Cartan-Weyl spaces
\[
\Gamma^{i}{}_{jk}=\left\{\begin{array}{c} i \\ jk
\end{array}\right\}+\mathcal{K}^{i}{}_{jk},
\]
where
\[
\mathcal{K}^{i}{}_{jk}=S^{i}{}_{jk}+S_{jk}{}^{i}+S_{kj}{}^{i}+
\frac{1}{2}\left(\delta^{i}{}_{j}Q_{k}+\delta^{i}{}_{k}Q_{j}-g_{jk}Q^{i}\right)
\]
with the same as previously convention concerning the raising and lowering of indices.

If $\Gamma$ is symmetric, we are dealing with the Weyl space,
\[
\Gamma^{i}{}_{jk}=\left\{\begin{array}{c} i \\ jk
\end{array}\right\}+
\frac{1}{2}\left(\delta^{i}{}_{j}Q_{k}+\delta^{i}{}_{k}Q_{j}-g_{jk}Q^{i}\right).
\]

In the micromaterial space $\mathbb{R}^{n}$ the standard Kronecker metric $\delta_{AB}$ is
used. If for any reason we admit non-orthogonal rectilinear coordinates in $\mathbb{R}^{n}$,
we shall use the symbol $\eta_{AB}$ to denote the micromaterial metric.

For any $e\in F_{x}M$ the corresponding Green deformation tensor in $\mathbb{R}^{n}$ is given
by
\[
G[e]_{AB}=g_{ij}(x)e^{i}{}_{A}e^{j}{}_{B}
\]
or, using the obvious abbreviations,
\[
G[e]=e^{\ast}\cdot g\in\mathbb{R}^{n\ast}\otimes\mathbb{R}^{n\ast}.
\]
Similarly, the Cauchy deformation tensor $C[e]\in T_{x}^{\ast}M\otimes T_{x}^{\ast}M$ is given
by
\[
C[e]_{ij}=\eta_{AB}e^{A}{}_{i}e^{B}{}_{j}=\delta_{AB}e^{A}{}_{i}e^{B}{}_{j},
\]
or symbolically,
\[
C[e]=\widetilde{e}^{\ast}\cdot\eta.
\]
The corresponding contravariant inverses will be denoted by $\widetilde{G}[e]$,
$\widetilde{C}[e]$,
\[
\widetilde{G}^{AC}G_{CB}=\delta^{A}{}_{B},\qquad \widetilde{C}^{ik}C_{kj}=\delta^{i}{}_{j}.
\]
Obviously,
\[
\widetilde{G}[e]^{AB}=e^{A}{}_{i}e^{B}{}_{j}g(x)^{ij},\qquad
\widetilde{C}[e]=e^{i}{}_{A}e^{j}{}_{B}\eta^{AB}=e^{i}{}_{A}e^{j}{}_{B}\delta^{AB}.
\]
Lagrange and Euler deformation tensors are respectively given by
\[
\mathcal{E}[e]=\frac{1}{2}\left(G-\eta\right),\qquad
\varepsilon[e]=\frac{1}{2}\left(g-C\right).
\]
Obviously, the quantities $G[e]_{AB}$ are scalar products of vectors $e_{A}$, $e_{B}$:
\[
G[e]_{AB}=g(x)\left(e_{A},e_{B}\right)=\langle e_{A}|e_{B}\rangle=G[e]_{BA}.
\]
We say that motion is metrically rigid if $\mathcal{E}$ (equivalently $\varepsilon$) vanishes
along all admissible trajectories. In other words, only such configurations are admissible
that
\[
G[e]_{AB}=\eta_{AB}=\delta_{AB}.
\]
Equivalently, this means that
\[
C[e]=g(x),\qquad e\in F_{x}M.
\]
Infinitesimal affinely-rigid body becomes then the infinitesimal gyroscope, i.e., the
configuration space $FM$ becomes restricted to the constraints submanifold $F(M,g)\subset FM$
consisting of $g$-orthonormal frames. More precisely, one admits the connected submanifold
$F^{+}(M,g)$ consisting of $g$-orthonormal frames positively oriented with respect to some
fixed orientation in $M$.

As mentioned, constraints of gyroscopic motion may be described analytically by any of the
following systems of equations:
\begin{eqnarray}
&&g_{ij}e^{i}{}_{A}e^{j}{}_{B}=\eta_{AB}=\delta_{AB},\label{c45}\\
&&\eta_{AB}e^{A}{}_{i}e^{B}{}_{j}=\delta_{AB}e^{A}{}_{i}e^{B}{}_{j}=g_{ij},\label{b45}
\end{eqnarray}
obviously, these systems are equivalent.

Subjecting the first system to the operation $D/Dt$, i.e.,
covariant differentiation along the orbital trajectory, and
contracting the resulting equation with $e^{A}{}_{i}e^{B}{}_{j}$,
we obtain that
\begin{equation}\label{a45}
g_{ki}\Omega^{i}{}_{l}+g_{li}\Omega^{i}{}_{k}=-V^{m}\nabla_{m}g_{kl}=-\nabla_{V}g_{kl}.
\end{equation}
Equivalently, differentiating the second system in the
$D/Dt$-sense, we obtain the equivalent form
\begin{equation}\label{a46}
\eta_{AC}\widehat{\Omega}^{C}{}_{B}+\eta_{BC}\widehat{\Omega}^{C}{}_{A}=
-\left(V^{m}\nabla_{m}g_{ij}\right)e^{i}{}_{A}e^{j}{}_{B}=
-\left(\nabla_{V}g_{ij}\right)e^{i}{}_{A}e^{j}{}_{B}.
\end{equation}
Let us notice that the $D/Dt$-differentiation of the first system is identical with the usual
$d/dt$-differentiation, because from the point of view of geometry of $M$ the left- and
right-hand sides are scalar quantities (they are tensors in the micromaterial sense of
$\mathbb{R}^{n}$).

If $\Gamma$ is metrical, i.e., if $(M,\Gamma,g)$ is a Riemann-Cartan space, in particular,
just a Riemann space $(M,\{\},g)$, then the right-hand sides of equations (\ref{a45}),
(\ref{a46}) vanish and $\Omega$, $\widehat{\Omega}$ become respectively $g$- and
$\eta$-skew-symmetric,
\[
g_{km}\Omega^{m}{}_{l}+g_{lm}\Omega^{m}{}_{k}=0,\qquad
\eta_{KM}\widehat{\Omega}^{M}{}_{L}+\eta_{LM}\widehat{\Omega}^{M}{}_{K}=0.
\]
If the standard coordinates in $\mathbb{R}^{n}$ are used, $\eta_{AB}=\delta_{AB}$, then the
last condition means simply that $\widehat{\Omega}$ is literally skew-symmetric.

Therefore, $\Omega$, $\widehat{\Omega}$ are respectively elements
of Lie algebras SO$(T_{x}M,g_{x})^{\prime}$,
SO$(n,\mathbb{R})^{\prime}$ of the corresponding orthogonal
groups. They are gyroscopic angular velocities respectively in the
spatial and co-moving (material) representations. For a
unconstrained affine motion, when $\Omega$, $\widehat{\Omega}$ are
general linear mappings, angular velocity may be defined as the
corresponding $g$- or $\eta$-skew-symmetric part of the affine
velocity, i.e.,
\begin{eqnarray}
\omega^{i}{}_{j}&:=&\frac{1}{2}\left(\Omega^{i}{}_{j}-g^{ik}g_{jl}\Omega^{l}{}_{k}\right),\nonumber\\
\widehat{\omega}^{A}{}_{B}&:=&\frac{1}{2}\left(\widehat{\Omega}^{A}{}_{B}-
\eta^{AC}\eta_{BD}\widehat{\Omega}^{D}{}_{C}\right).\nonumber
\end{eqnarray}
In the case of rigid motion they are identical respectively with
$\Omega^{i}{}_{j}$ and $\widehat{\Omega}^{A}{}_{B}$. {\bf But
attention!} The general rule
\[
\Omega^{i}{}_{j}=e^{i}{}_{A}\widehat{\Omega}^{A}{}_{B}e^{B}{}_{j}
\]
is not valid any longer for $\omega^{i}{}_{j}$, $\widehat{\omega}^{A}{}_{B}$, i.e.,
\[
\omega^{i}{}_{j}\neq e^{i}{}_{A}\widehat{\omega}^{A}{}_{B}e^{B}{}_{j},
\]
unless the motion is metrically rigid (gyroscopic), i.e.,
(\ref{c45}), (\ref{b45}) hold. Therefore, in non-gyroscopic motion
$\widehat{\omega}^{A}{}_{B}$ fail to be co-moving components of
the angular velocity. The same concerns kinematical distortions,
\begin{eqnarray}
d^{i}{}_{j}&:=&\frac{1}{2}\left(\Omega^{i}{}_{j}+g^{ik}g_{jl}\Omega^{l}{}_{k}\right),\nonumber\\
\widehat{d}^{A}{}_{B}&:=&\frac{1}{2}\left(\widehat{\Omega}^{A}{}_{B}+
\eta^{AC}\eta_{BD}\widehat{\Omega}^{D}{}_{C}\right).\nonumber
\end{eqnarray}

As seen from the formulas (\ref{a45}), (\ref{a46}), if affine
connection $\Gamma$ used in the definitions of $\Omega$ and
$\widehat{\Omega}$ is not metrical (i.e., $(M,\Gamma,g)$ is not
the Riemann-Cartan space), then $\Omega$ and $\widehat{\Omega}$
are not skew-symmetric in the $g$- and $\eta$-sense (i.e., are not
elements of the Lie algebras of SO$(T_{x}M,g_{x})$,
SO$(n,\mathbb{R})$) even if motion is purely gyroscopic.
Therefore, it is confusing to interpret them as angular
velocities. Taking their $g$- and $\eta$-skew-symmetric parts also
does not seem convincing. If $\Gamma$ is metrical, i.e., $\nabla
g=0$, then, as mentioned, $\Omega$ and $\widehat{\Omega}$ are
respectively $g$- and $\eta$-skew-symmetric. Nevertheless, their
particular form depends explicitly on the torsion tensor $S$. So,
no doubt, the use of gyroscopic concepts is clean only when the
Levi-Civita affine connection $\{\}$ is used.

One can wonder whether gyroscopic motion could not be defined
without any use of the fixed metric tensor $g$ on $M$, thus,
without any problems like those mentioned above (definition of
angular velocity, $\Gamma$-$g$ compatibility, and so on).
Apparently, the idea might look both not physical and
mathematically inconsistent. However, although the physical
usefulness question is still open, the mathematical correctness
may be easily shown. In a sense, gyroscopic constraints may be
defined on the basis of affine connection structure $(M,\Gamma)$,
quite independently of the metric tensor concept. However, the
micromaterial metric $\eta_{AB}$ ($\delta_{AB}$) may be used.
Namely, it gives rise to the Cauchy deformation tensor
\[
C_{ij}=\eta_{AB}e^{A}{}_{i}e^{B}{}_{j}
\]
and its inverse
\[
\widetilde{C}^{ij}=e^{i}{}_{A}e^{j}{}_{B}\eta^{AB}.
\]
Obviously, $C[e]\in T^{\ast}_{\pi(e)}M\otimes T^{\ast}_{\pi(e)}M$
has all formal properties of the metric tensor in the
instantaneous tangent space $T_{\pi(e)}M$ (symmetry and positive
definiteness). But it is defined only at the point $x=\pi(e)\in M$
and is not induced by any metric field $g$ living globally all
over in $M$. And without such a field even the very term
"deformation tensor" is not very adequate, because we do not have
any metrical standard which might be compared with $C[e]$; thus,
we cannot decide to which extent $e\in FM$ ``deforms" $\eta$.
Nevertheless, it is meaningful to say that some motion in $FM$ is
free of deformations when $C[e]$ is covariantly constant along the
curve describing translational motion in $M$, i.e.,
\[
\frac{DC_{ij}}{Dt}=0.
\]
After simple calculations we obtain the mutually equivalent
conditions
\begin{eqnarray}
\widehat{\Omega}^{A}{}_{B}&=&-\eta^{AK}\eta_{BL}\widehat{\Omega}^{L}{}_{K},\label{a49}\\
\Omega^{i}{}_{j}&=&-\widetilde{C}^{ik}C_{jl}\Omega^{l}{}_{k},\label{b49}
\end{eqnarray}
i.e., $\widehat{\Omega}$ and $\Omega$ are skew-symmetric
respectively with respect to $\eta$ and $C$ used as metrics. There
is however an important point, namely, if $(M,\Gamma)$ is
non-Euclidean (in the local sense), then the above constraints are
aholonomic. If $(M,\Gamma)$ is locally Euclidean, these
constraints become quasi-holonomic, i.e., the manifold $FM$ is
foliated by a family of $n(n+1)/2$-dimensional integral manifolds
of the above Pfaff systems. Any of these mutually disjoint
manifolds is a possible configuration space of rigid motions. And
any particular choice is equivalent to fixing some metric tensor
field $g$ on $M$. And indeed, in an $n$-dimensional linear space
the manifold of possible metric tensors is $n(n+1)/2$-dimensional,
thus, leaving $n(n-1)/2$ independent parameters in the manifold of
linear frames reducing the metric to its standard Kronecker-delta
form. Together with $n$ translational degrees of freedom we obtain
exactly $n(n+1)/2$ degrees of freedom of a rigid body moving in
$n$-dimensional Euclidean space.

It is quite a different question whether the Cauchy tensor $C[e]$
may be meaningfully used as a metric-like tensor of the
instantaneous tangent space $T_{\pi(e)}M$. We return later on to
the question of possible physical applications. In a moment we
stress interesting geometrical aspects of aholonomic constraints
(\ref{a49}), (\ref{b49}).

\section{Dynamical affine models}

Let us now turn to describing dynamical models. We begin from the
models of kinetic energy, i.e., roughly speaking, metrics
(Riemannian structures) on the configuration space $FM$. When we
deal with internal degrees of freedom, the problem becomes
delicate because the d'Alembert method of deriving the kinetic
energy from the model of extended affinely constrained system
becomes unjustified, not reliable, perhaps just misleading. It is
so even in models of essentially internal degrees of freedom in a
flat space, but in curved manifolds the problem becomes very
essential, fundamentally embarrassing, and the only reliable
method is one based on appropriate symmetry principle. As we saw
in \cite{all-book04}, it is the case also in some non-standard
applications in the usual continuum mechanics and the dynamics of
structured bodies and defects.

Nevertheless, to begin with, we discuss as first the models (metrics on $FM$) following
formally the expressions known from the d'Alembert mechanics of extended affine bodies in flat
spaces, and in any case based on simple analogies. Later on we discuss mathematical structure
of more general models following the non-d'Alembert ideas in a flat space \cite{all-book04}.
And finally, some perspectives of physical applications will be suggested and preliminarily
discussed.

Repeating formally the d'Alembert expression from mechanics of
affine bodies in flat spaces, we obtain that
\begin{equation}\label{a52}
T=T_{\rm tr}+T_{\rm int}=\frac{m}{2}g_{ij}\frac{dx^{i}}{dt}\frac{dx^{j}}{dt}+
\frac{1}{2}g_{ij}\frac{De^{i}{}_{A}}{Dt}\frac{De^{j}{}_{B}}{Dt}J^{AB}.
\end{equation}
In this formula the descriptors ``tr" and ``int" refer obviously
to the translational and internal parts, $m>0$ is the mass, and
the symmetric and positively definite micromaterial tensor $J\in
\mathbb{R}^{n}\otimes\mathbb{R}^{n}$ describes the internal
inertia. The main difference in comparison with the flat-space
formula for extended bodies is that now the covariant along-curve
derivatives of ``directors" $e_{A}$ are used. And there is of
course one subtle point; namely, now $J$ is not derived from the
model of constrained extended system, but just postulated as
something primary. In certain calculations the following
equivalent expressions are convenient:
\begin{eqnarray}
T_{\rm tr}&=&\frac{m}{2}G[e]_{AB}\widehat{V}^{A}\widehat{V}^{B},\label{b52}\\
T_{\rm int}&=&\frac{1}{2}G[e]_{KL}\widehat{\Omega}^{K}{}_{A}\widehat{\Omega}^{L}{}_{B}
J^{AB}.\label{c52}
\end{eqnarray}
Obviously, now the coefficients in quadratic forms of co-moving velocities depend on the
internal configuration $e$. Alternatively, $T_{\rm int}$ may be expressed in spatial terms:
\begin{equation}\label{d52}
T_{\rm int}=\frac{1}{2}g_{ij}\Omega^{i}{}_{k}\Omega^{j}{}_{l}J[e]^{kl},
\end{equation}
where $J[e]$ is the configuration-dependent spatial representation of the internal inertia:
\[
J[e]^{kl}=e^{k}{}_{A}e^{l}{}_{B}J^{AB}.
\]
This second-order moment (quadrupole of the internal inertia) in
extended body dynamics is often used in nuclear physics.

\noindent{\bf Remark:} $T_{\rm int}$ in (\ref{a52}), (\ref{c52})
is invariant under translations (\ref{a36}) if and only if the
field $\xi$ of additive translations is parallel under $\Gamma$,
i.e., $\nabla_{[\Gamma]}\xi=0$.

Denoting the system of generalized coordinates of affine bodies by
\[
(\ldots,q^{\mu},\ldots)=(\ldots,x^{i},\ldots;\ldots,e^{i}{}_{A},\ldots),
\]
and writing symbolically (\ref{a52}) in the following form:
\[
T=\frac{1}{2}\mathcal{G}_{\mu\nu}\frac{dq^{\mu}}{dt}\frac{dq^{\nu}}{dt},
\]
we see that the underlying Riemannian metric $\mathcal{G}$ on $F(M)$ is flat if $(M,g)$ is
locally Euclidean. It is no longer the case in a curved manifold.

In non-dissipative models with the velocity-independent
Lagrangians $L=T-V(x,e)$ the resulting Euler-Lagrange equations
\begin{equation}\label{b53}
\frac{d}{dt}\frac{\partial L}{\partial\dot{q}^{\mu}}-\frac{\partial L}{\partial q^{\mu}}=0
\end{equation}
may be written down in the following form:
\begin{eqnarray}
m\frac{DV^{a}}{Dt}&=&\Sigma^{k}{}_{l}R^{l}{}_{k}{}^{a}{}_{j}V^{j}-
mV^{b}\mathcal{K}_{b}{}^{a}{}_{c}V^{c}\nonumber\\
&&-\mathcal{K}_{mk}{}^{a}
\frac{De^{m}{}_{A}}{Dt}\frac{De^{k}{}_{B}}{Dt}J^{AB}+F^{a},\label{a53}\\
e^{a}{}_{K}\frac{D^{2}e^{b}{}_{L}}{Dt^{2}}J^{KL}&=&-e^{a}{}_{K}g^{bm}
\frac{Dg_{mc}}{Dt}\frac{De^{c}{}_{L}}{Dt}J^{KL}+N^{ab},\label{c53}
\end{eqnarray}
where the meaning of symbols is as follows:
\begin{eqnarray}
\mathcal{K}^{a}{}_{bc}&=&\Gamma^{a}{}_{bc}-\left\{\begin{array}{c}
a \\ bc \end{array}\right\},\nonumber\\
F^{a}&=&g^{ab}F_{b}=-g^{ab}H_{b}V=-g^{ab}\left(\frac{\partial
V}{\partial x^{b}}-
\Gamma^{i}{}_{jb}e^{j}{}_{B}\frac{\partial V}{\partial e^{i}{}_{B}}\right),\nonumber\\
N^{ab}&=&N^{a}{}_{c}g^{cb}=-g^{bc}E^{a}{}_{c}V=-g^{bc}e^{a}{}_{K}\frac{\partial V}{\partial
e^{c}{}_{K}},\nonumber
\end{eqnarray}
and obviously, all shifts of tensor indices from their natural positions are meant in the
sense of $g$.

Let us stress an important fact that in general the translational force $F^{a}$ does not equal
$-g^{ab}\partial V/\partial x^{b}$. The equality holds only when $V$ does not depend on
internal degrees of freedom, i.e., when $V$ is a $\pi$-pull-back of some function on $M$, or
if connection $\Gamma$ is flat and local Cartesian coordinates on $M$ are used. If $V$ is not
a $\pi$-vertical function on $M$, then $\partial V/\partial x^{b}$ is not a covariant vector
in $M$ at all, and $g^{ab}\partial V/\partial x^{b}$ fails to be a contravariant $M$-vector.
Let us observe that $F$ may be written as follows:
\[
F^{a}=-g^{ab}\left(\frac{\partial V}{\partial
x^{b}}+N^{j}{}_{i}\Gamma^{i}{}_{jb}\right).
\]
The co-moving representations of $F$, $N$ are given as follows:
\begin{eqnarray}
\widehat{F}_{A}&=&F_{i}e^{i}{}_{A}=-H_{A}V,\nonumber\\
\widehat{N}^{A}{}_{B}&=&e^{A}{}_{i}N^{i}{}_{j}e^{j}{}_{B}=-E^{A}{}_{B}V=-e^{i}{}_{B}
\frac{\partial V}{\partial e^{i}{}_{A}}.\nonumber
\end{eqnarray}

\noindent{\bf Remark:} the contravariant objects
$\widehat{F}^{A}=e^{A}{}_{i}F^{i}$,
$\widehat{N}^{AB}=e^{A}{}_{i}e^{B}{}_{j}N^{ij}$ have the forms
\[
\widehat{F}^{A}=\widetilde{G}^{AB}\widehat{F}_{B},\qquad
\widehat{N}^{AB}=\widehat{N}^{A}{}_{C}\widetilde{G}^{CB},
\]
i.e., the lower-case indices are raised with the help of the Green deformation tensor.

When non-potential interactions, e.g., dissipative ones, are admitted, the above general form
of equations of motion in principle remains valid, however the dynamical terms $F^{a}$,
$N^{ab}$ must be replaced by more general expressions postulated on independent basis (e.g.,
additional friction forces linear in generalized velocities with symmetric negatively-definite
coefficients matrices).

It would be technically very difficult to derive equations (\ref{a53}) directly from the
second-kind Lagrange equations (\ref{b53}). It is much more convenient to use the canonical
formalism and Poisson-bracket techniques. For the potential systems with Lagrangians
$L=T-V\left(x^{i},e^{i}{}_{A}\right)$, the Legendre transformation
\[
p_{\mu}=\frac{\partial L}{\partial \dot{q}^{\mu}}=\frac{\partial T}{\partial \dot{q}^{\mu}}=
\mathcal{G}_{\mu\nu}(q)\dot{q}^{\nu}
\]
leads to the Hamiltonian
\[
H=\mathcal{T}+V\left(x^{i},e^{i}{}_{A}\right),
\]
where the kinetic term
\[
\mathcal{T}=\frac{1}{2}\mathcal{G}^{\mu\nu}(q)p_{\mu}p_{\nu}
\]
has the form
\begin{equation}\label{a56}
\mathcal{T}=\frac{1}{2m}g^{ij}P_{i}P_{j}
+\frac{1}{2}\widetilde{J}_{AB}P^{A}{}_{i}P^{B}{}_{j}g^{ij}.
\end{equation}
Let us remind that $\widetilde{J}$ is the inverse of $J$,
\[
J^{AC}\widetilde{J}_{CB}=\delta^{A}{}_{B},
\]
and the explicit expression for Legendre transformation reads that
\[
P_{i}=\frac{\partial T}{\partial V^{i}}=mg_{ij}V^{j},\qquad P^{A}{}_{i}=\frac{\partial
T}{\partial V^{i}{}_{A}}=g_{ij}V^{j}{}_{B}J^{BA}.
\]
The formerly quoted basic Poisson brackets enable one to write down explicitly Hamiltonian
equations of motion,
\begin{eqnarray}
\frac{dP_{i}}{dt}=\{P_{i},H\},&\quad&\frac{dP^{A}{}_{i}}{dt}=\{P^{A}{}_{i},H\},\nonumber\\
\frac{dx^{i}}{dt}=\{x^{i},H\},&\quad&\frac{de^{i}{}_{A}}{dt}=\{e^{i}{}_{A},H\},\nonumber
\end{eqnarray}
which after some manipulations may be reduced to the form (\ref{a53}). Let us stress that the
covariant derivative in (\ref{a53}) is meant in the sense of $\Gamma$, not in the
$g$-Levi-Civita sense; an important fact to be kept in mind when they do not coincide (i.e.,
when $(M,\Gamma,g)$ is not a Riemann space).

As expected, equations of motion (\ref{a53}) simplify in a remarkable way when $(M,\Gamma,g)$
is a Riemann-Cartan space, $\nabla g=0$. In this case $\mathcal{K}$ becomes so-called
contortion,
\[
\mathcal{K}^{a}{}_{bc}=S^{a}{}_{bc}+S_{bc}{}^{a}+S_{cb}{}^{a},
\]
and equations (\ref{a53}) reduce to
\begin{eqnarray}
m\frac{DV^{a}}{Dt}&=&\frac{1}{2}S^{k}{}_{l}\mathcal{R}^{l}{}_{k}{}^{a}{}_{j}V^{j}
+2mV^{b}V^{c}S_{bc}{}^{a}+F^{a},\label{a57}\\
e^{a}{}_{K}\frac{D^{2}e^{b}{}_{L}}{Dt^{2}}J^{KL}&=&N^{ab}.\label{b57}
\end{eqnarray}
In the first (translational) equation only the $g$-skew-symmetric part of $\Sigma^{i}{}_{j}$,
i.e., $S^{i}{}_{j}$, survives, because in Riemann-Cartan spaces the curvature tensor is
$g$-skew-symmetric in the first pair of indices. Besides of the usual external force $F^{a}$,
the right-hand side of (\ref{a57}) involves two geometric forces describing the coupling
between spatial geometry and kinematical quantities of the particle motion. Namely,
translational velocity is quadratically coupled to the torsion, and spin is coupled to the
curvature. This is in a nice way compatible with the geometric interpretation of curvature and
torsion respectively in terms of rotations and translations. The geometric force
\[
F^{a}_{\rm
geom}:=\frac{1}{2}S^{k}{}_{l}\mathcal{R}^{l}{}_{k}{}^{a}{}_{j}V^{j}
+2mV^{b}V^{c}S_{bc}{}^{a}
\]
is so-to-speak magnetic-like in the sense that due to the
corresponding anti-symmetries of $R$ and $S$ they are
$g$-orthogonal to velocities and do not do any work, i.e.,
\[
g_{ab}V^{a}F^{b}_{\rm geom}=0,
\]
therefore, they do not contribute to the energy balance. In
particular, they do not accelerate the absolute value of $V^{a}$,
$\|V\|=\sqrt{g_{ab}V^{a}V^{b}}$, but only change the direction of
$V^{a}$, i.e., give rise to the bending of translational
trajectory. Even in the force-free case the motion is not
geodetic; there is some link between this phenomenon and geodetic
deviation. Let us observe that (\ref{a57}) may be further
simplified to the form
\[
m\frac{D_{[g]}V^{a}}{Dt}=\frac{1}{2}S^{k}{}_{l}\mathcal{R}^{l}{}_{k}{}^{a}{}_{j}V^{j}+F^{a},
\]
where $D_{[g]}$ denotes the along-curve covariant differentiation
in the Levi-Civita sense. However, if $(M,\Gamma,g)$ is
non-Riemannian, i.e., the torsion does not vanish, this
simplification is a rather seeming one, because the curvature
tensor $\mathcal{R}$ of $\Gamma$ contains both the term
corresponding to the Riemannian curvature of $\{\}$ and the terms
involving torsion. And in the internal equation (\ref{b57}) there
is no simplification at all. Obviously, in the purely Riemannian
case, when the torsion does vanish, we obtain the maximally simple
and clear system of equations of motion:
\begin{eqnarray}
m\frac{DV^{a}}{Dt}&=&\frac{1}{2}S^{k}{}_{l}\mathcal{R}^{l}{}_{k}{}^{a}{}_{j}V^{j}+F^{a},\label{a59}\\
e^{a}{}_{K}\frac{D^{2}e^{b}{}_{L}}{Dt^{2}}J^{KL}&=&N^{ab},\label{d59}
\end{eqnarray}
and now both the covariant derivative and the curvature tensor are
meant in the $g$-Levi-Civita sense.

It is instructive and convenient to write down these equations in terms of some balance laws.
Let
\begin{eqnarray}
K^{a}&=&mV^{a}=m\frac{dx^{a}}{dt},\nonumber\\
K^{ab}&=&e^{a}{}_{K}V^{b}{}_{L}J^{KL}=e^{a}{}_{K}\frac{De^{b}{}_{L}}{Dt}J^{KL}\nonumber
\end{eqnarray}
denote respectively the kinematical linear momentum and kinematical affine spin (kinematical
hypermomentum), just as in \cite{all-book04}. Then (\ref{a59}), (\ref{d59}) may be written as
follows:
\begin{eqnarray}
\frac{DK^{a}}{Dt}&=&\frac{1}{2m}S^{k}{}_{l}\mathcal{R}^{l}{}_{k}{}^{a}{}_{j}K^{j}+F^{a}
=\frac{1}{m}K^{k}{}_{l}\mathcal{R}^{l}{}_{k}{}^{a}{}_{j}K^{j}+F^{a},\label{b59}\\
\frac{DK^{ab}}{Dt}&=&\frac{De^{a}{}_{K}}{Dt}\frac{De^{b}{}_{L}}{Dt}J^{KL}+N^{ab}.\label{c59}
\end{eqnarray}
We by purpose use the symbols $K^{a}$, $K^{ab}$, not $P^{a}$,
$\Sigma^{ab}$, because the latter symbols are reserved for
canonical linear momentum and affine spin in contravariant
representation. In the very special case of Lagrangian
$L=T-V(x^{i},e^{i}{}_{A})$, where $T$ is given as in (\ref{a52}),
it is really so that these concepts are mutually identified via
Legendre transformation, i.e.,
\[
P^{a}=g^{ab}P_{b}=K^{a}=mV^{a},\qquad \Sigma^{ab}=\Sigma^{a}{}_{c}g^{cb}=K^{ab}.
\]
However, for more general Lagrangians and for non-d'Alembertian models of the kinetic energy
$T$, the above relationships become false, whereas (\ref{b59}), (\ref{c59}) remain true.

The power, i.e., the time rate of work, is given by the following formula obtained by analogy
with the mechanics of extended bodies \cite{JJS82_2,all-book04}:
\[
\mathcal{P}=g_{ab}F^{a}V^{b}+g_{bc}N^{ac}\Omega^{b}{}_{a}=\mathcal{P}_{\rm
tr}+\mathcal{P}_{\rm int}.
\]
We can consider internal affine dynamics subject to additional constraints, just as in the
case of extended affine bodies. First of all, let us consider gyroscopic constraints, i.e.,
assume the moving frame $e$ to be permanently $g$-orthonormal. Then, as mentioned above,
$\Omega$ is permanently $g$-skew-symmetric ($\widehat{\Omega}$ is permanently
$\eta$-($\delta$-)skew-symmetric).

We assume here the validity of the d'Alembert model of constrained
dynamics. Therefore, our equations of motion remain valid when on
their right-hand sides we introduce some extra reaction forces
responsible for keeping the constraints, but, roughly speaking,
not influencing the along-constraints motion. Let us denote the
corresponding expression by $F^{a}_{R}$, $N^{ab}_{R}$. They are to
be added to the ``true" applied dynamical quantities $F^{a}$,
$N^{ab}$. According to the d'Alembert principle they are
completely passive controls, i.e., they do not do work along any
virtual motion compatible with constraints, i.e.,
\[
\mathcal{P}_{R}=g_{ab}F^{a}_{R}V^{b}+g_{bc}N^{ac}_{R}\Omega^{b}{}_{a}=0
\]
for any $V$ and for any $g$-skew-symmetric $\Omega$. This means
that $F^{a}_{R}=0$ and $N^{ab}_{R}=N^{ba}_{R}$, i.e., $F_{R}$
vanishes and $N_{R}$ is symmetric. Therefore, the effective,
reaction-free system of equations of motion consists of
(\ref{b59}) and the skew-symmetric part of (\ref{c59}) with
algebraically substituted gyroscopic constraints
$g_{ij}e^{i}{}_{A}e^{j}{}_{B}=\eta_{AB}=\delta_{AB}$, i.e.,
\begin{eqnarray}
\frac{DK^{a}}{Dt}&=&\frac{1}{2m}S^{k}{}_{l}\mathcal{R}^{l}{}_{k}{}^{a}{}_{j}K^{j}+F^{a},\nonumber\\
\frac{DS^{ab}}{Dt}&=&\mathcal{N}^{ab},\label{b61}
\end{eqnarray}
where, obviously,
\[
S^{ab}=K^{ab}-K^{ba}
\]
is the internal angular momentum (spin) and
\[
\mathcal{N}^{ab}=N^{ab}-N^{ba}
\]
is the skew-symmetric moment, torque, i.e., generalized force coupled to rotational degrees of
freedom. In this way one obtains the system of $n(n+1)/2$ equations of motion imposed on the
time-dependence of $n(n+1)/2$ degrees of freedom of the infinitesimal gyroscope moving in $M$
($n$ translational degrees of freedom and $n(n-1)/2$ gyroscopic ones).

Returning to the explicit description in terms of the
configuration variables we obtain that
\begin{eqnarray}
m\frac{DV^{a}}{Dt}=m\frac{D^{2}x^{a}}{Dt^{2}}&=&
\frac{1}{2}S^{k}{}_{l}\mathcal{R}^{l}{}_{k}{}^{a}{}_{j}\frac{dx^{j}}{dt}+F^{a},\nonumber\\
e^{a}{}_{K}\frac{D^{2}e^{b}{}_{L}}{Dt^{2}}J^{KL}-e^{b}{}_{K}\frac{D^{2}e^{a}{}_{L}}{Dt^{2}}J^{KL}&=&
\mathcal{N}^{ab}=N^{ab}-N^{ba}.\label{b62}
\end{eqnarray}
In these equations only given (reaction-free) forces and moments
are present and obviously
\[
S^{ij}=K^{ij}-K^{ji}=e^{i}{}_{A}\frac{De^{j}{}_{B}}{Dt}J^{AB}-
e^{j}{}_{A}\frac{De^{i}{}_{B}}{Dt}J^{AB}.
\]
In analogy to extended affine bodies one can also consider
internal dynamics with other kinds of constraints. The most
natural ones are those with the lucid group-theoretical structure.
For example, for incompressible bodies, (\ref{d59}), (\ref{c59})
are to be replaced by their $g$-traceless parts:
\[
e^{a}{}_{K}\frac{D^{2}e^{b}{}_{L}}{Dt^{2}}J^{KL}-
\frac{1}{n}g_{cd}e^{c}{}_{K}\frac{D^{2}e^{d}{}_{L}}{Dt^{2}}g^{ab}J^{KL}=
N^{ab}-\frac{1}{n}g_{cd}N^{cd}g^{ab},
\]
i.e.,
\begin{eqnarray}
&&\frac{D}{Dt}\left(K^{ab}-\frac{1}{n}g_{cd}K^{cd}g^{ab}\right)=
N^{ab}-\frac{1}{n}g_{cd}N^{cd}g^{ab}\nonumber\\
&&+\frac{De^{a}{}_{K}}{Dt}\frac{De^{b}{}_{L}}{Dt}J^{KL}
-\frac{1}{n}g_{cd}\frac{De^{c}{}_{K}}{Dt}\frac{De^{d}{}_{L}}{Dt}J^{KL}g^{ab}.\nonumber
\end{eqnarray}
Incompressibility constraints may be described analytically by
equations
\[
\det\left[e^{A}{}_{i}\right]=\sqrt{\det\left[g_{ij}\right]},
\]
or infinitesimally in any of two equivalent forms
\[
\Omega^{i}{}_{i}=0,\qquad \widehat{\Omega}^{A}{}_{A}=0.
\]
Let us recall that gyroscopic constraints in Riemannian space may
be also described in terms of equivalent infinitesimal conditions:
\[
\Omega^{i}{}_{j}=-\Omega_{j}{}^{i}=-g^{ik}g_{jl}\Omega^{l}{}_{k},\qquad
\widehat{\Omega}^{A}{}_{B}=-\widehat{\Omega}_{B}{}^{A}=-\eta^{AC}\eta_{BD}\widehat{\Omega}^{D}{}_{C}.
\]
If only purely dilatational internal motion is admitted, i.e.,
\[
\Omega^{i}{}_{j}=\lambda\delta^{i}{}_{j},\qquad
\widehat{\Omega}^{A}{}_{B}=\lambda\delta^{A}{}_{B},
\]
or, equivalently,
\[
\Omega^{i}{}_{j}-\frac{1}{n}\Omega^{k}{}_{k}\delta^{i}{}_{j}=0,\qquad
\widehat{\Omega}^{A}{}_{B}-\frac{1}{n}\widehat{\Omega}^{C}{}_{C}\delta^{A}{}_{B}=0,
\]
then the internal motion is (on the basis of d'Alembert principle)
completely described by the $g$-trace of (\ref{d59}) or
(\ref{c59}):
\begin{eqnarray}
g_{ab}e^{a}{}_{K}\frac{D^{2}e^{b}{}_{L}}{Dt^{2}}J^{KL}&=&g_{ab}N^{ab},\label{a63}\\
g_{ab}\frac{DK^{ab}}{Dt}&=&g_{ab}\frac{De^{a}{}_{K}}{Dt}\frac{De^{b}{}_{L}}{Dt}J^{KL}+g_{ab}N^{ab}.\label{b63}
\end{eqnarray}
If the body is subject to Weyl constraints, i.e., its internal
degrees of freedom undergo only rigid rotations and dilatations,
then equations of internal motion are given by (\ref{b61}) and
(\ref{b63}), or equivalently (\ref{b62}) and (\ref{a63}).

A very interesting example of constraints is that of rotation-less
motion in $M$,
\begin{equation}\label{c63}
\Omega^{i}{}_{j}=\Omega_{j}{}^{i}=g_{jk}g^{il}\Omega^{k}{}_{l}.
\end{equation}
Then the d'Alembert principle implies that equations of internal
motion (effective ones, free of reaction forces) are given by the
symmetric part of (\ref{d59}) or (\ref{c59}), i.e.,
\[
e^{a}{}_{K}\frac{D^{2}e^{b}{}_{L}}{Dt^{2}}J^{KL}+
e^{b}{}_{K}\frac{D^{2}e^{a}{}_{L}}{Dt^{2}}J^{KL}=N^{ab}+N^{ba},
\]
i.e.,
\[
\frac{D}{Dt}\left(K^{ab}+K^{ba}\right)=
\frac{De^{a}{}_{K}}{Dt}\frac{De^{b}{}_{L}}{Dt}J^{KL}
+\frac{De^{b}{}_{K}}{Dt}\frac{De^{a}{}_{L}}{Dt}J^{KL}+
N^{ab}+N^{ba}.\nonumber
\]
We have mentioned that when the geometry of $M$ is non-Euclidean
(curved), then in a natural way aholonomic velocities and other
aholonomic concepts appear. In the last example situation is even
much more complicated, because, as mentioned in our papers about
extended affine bodies \cite{JJS82_1,JJS82_2}, constraints
(\ref{c63}) are very essentially aholonomic even in mechanics of
extended affine bodies in flat spaces; the more so in a
non-Euclidean manifold.

From now on we concentrate on the mechanics of infinitesimal
gyroscopes and infinitesimal affine bodies without additional
constraints, but very often with the special and strong stress on
the mutual coupling between rotational and deformative motion.
This induces us to develop certain analytical procedures. They are
even more than analytical procedures, because the underlying
geometric techniques are interesting in themselves and
simultaneously they reveal certain very interesting mechanical
facts.

So, let us begin with the special case of infinitesimal gyroscope
in a Riemann space $(M,\{\},g)$, where equations of motion are
given by (\ref{b61}) or, more explicitly, by (\ref{b62}).

Let us stress an important fact. In mechanics of unconstrained
infinitesimal affine bodies, equations of motion (\ref{a59}),
(\ref{d59}) or (\ref{b59}), (\ref{c59}) are directly applicable
even in the purely technical sense, because
$\left(x^{i},e^{i}{}_{A}\right)$ are ``good" independent
(unconstrained) generalized coordinates. It is also so in
expressions (\ref{a52}), (\ref{a56}) for the kinetic energy.
However, after the gyroscopic constraints are imposed,
\[
g_{ij}e^{i}{}_{A}e^{j}{}_{B}=\eta_{AB}\left(=\delta_{AB}\right),
\]
the quantities $e^{i}{}_{A}$ are no longer independent and cannot
be used as generalized coordinates. Even when we study the mutual
coupling of rotational and deformative degrees of freedom, the
quantities $e^{i}{}_{A}$ are inconvenient, although well defined
as generalized coordinates. In flat spaces the procedure is
obvious: the system $[e^{i}{}_{A}]$ as a matrix is subject to
various polar, two-polar \cite{all-book04}, and similar
decompositions, and then the mutual interplay between rotations
and deformations is easily treatable. Something similar must be
done here, but the direct methods based on the flat-space geometry
are not applicable any longer. Instead, some geometric techniques
based on orthonormal aholonomic reference frames may be developed.
In the special case of gyroscopic constraints one can use then
various well-known coordinates on the special orthogonal group
SO$(n,\mathbb{R})$ as a subsystem of well-defined generalized
coordinates on $(FM,g)$.
\begin{center}
\includegraphics[scale=0.35]{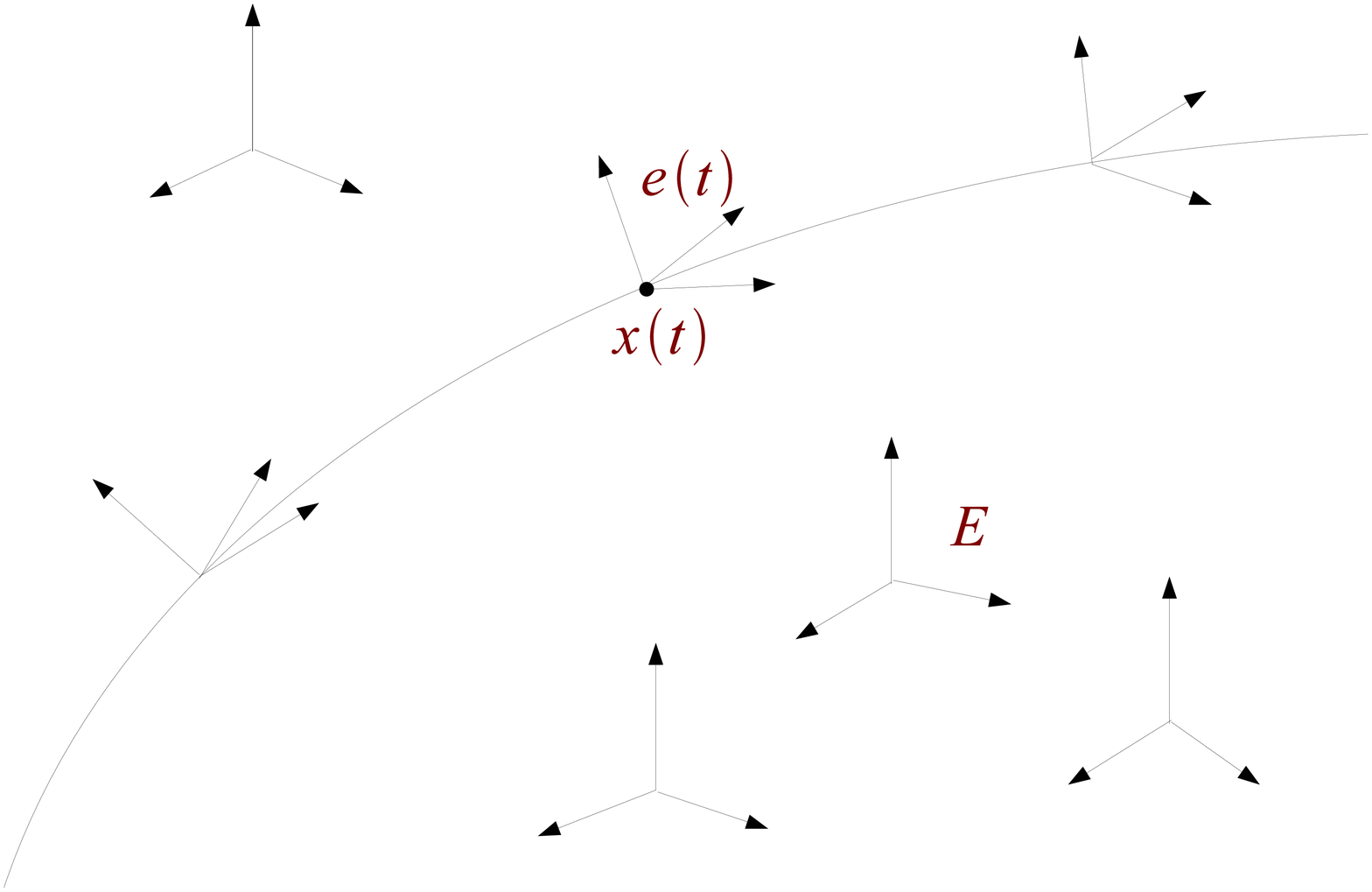}

{\rm Fig. 3}
\end{center}
\bigskip

So, let $E$ be some preestablished, fixed once for all fields of
linear orthonormal frames on $M$, i.e., a cross-section of the
subbundle $(FM,g)\subset FM$ over $M$. By the way, the
orthonormality demand is a technically simplifying one, but in
certain problems some general frames would do. But let us assume
$E$ to be orthonormal in the $g$-sense, i.e.,
\[
g(E_{A},E_{B})=g_{ij}E^{i}{}_{A}E^{j}{}_{B}=\eta_{AB}=\delta_{AB}.
\]
The dual co-frame $\widetilde{E}=(\ldots,E^{A},\ldots)$ will be
also used,
\[
\langle E^{A},E_{B}\rangle
=E^{A}{}_{i}E^{i}{}_{B}=\delta^{A}{}_{B}.
\]
It is obviously orthonormal with respect to the reciprocal
(contravariant) metric $\widetilde{g}$ on $M$,
\[
E^{A}{}_{i}E^{B}{}_{j}g^{ij}=\eta^{AB}\left(=\delta^{AB}\right).
\]

\noindent {\bf Remark:} this is a field of frames, and it is
defined all over in $M$ and kept fixed. The natural question
arises as to the choice of this aholonomic frame in $M$. In
general the particular structure of the Riemann space $(M,g)$
suggests some choices, both technically convenient and
geometrically lucid, which are well suited to the problem.

So, $M$ is ``inhabited" by continuum of orthonormal frames. The
metric field on $M$ may be expressed as follows:
\[
g=\eta^{AB}E_{A}\otimes E^{B}=\delta^{AB}E_{A}\otimes E^{B}.
\]

In the course of time the body moves in $M$. At the time instant
$t\in \mathbb{R}$ it is instantaneously placed at the geometric
point $x(t)\in M$ and its internal configuration is then given by
$e(t)\in F_{x(t)}M$. Obviously, the vectors $e_{A}(t)\in
T_{x(t)}M$, $A=\overline{1,n}$, may be expanded with respect to
the just passed frame $E_{x(t)}\in F_{x(t)}M$, i.e.,
\[
e_{A}(t)=E_{x(t)B}R^{B}{}_{A}(t),
\]
where, in the gyroscopic case, the matrix $R$ is orthogonal,
\begin{equation}\label{a67}
\delta_{AB}=\delta_{CD}R^{C}{}_{A}R^{D}{}_{B},
\end{equation}
or if for any reason we use a more general (non-standard) basis in
$\mathbb{R}^{n}$, then
\[
\eta_{AB}=\eta_{CD}R^{C}{}_{A}R^{D}{}_{B}.
\]
Roughly, (\ref{a67}) may be written in the usual matrix terms:
\[
R^{T}R=I_{n},
\]
where the $n\times n$ identity matrix is on the right-hand side.
In this way the configuration space $F(M,g)$, or more precisely
$F^{+}(M,g)$ (orthonormal frames positively oriented with respect
to some fixed standard of orientation) is identified by the field
$E$ with the product manifold
\[
M\times {\rm SO}(n,\mathbb{R}).
\]
Obviously, this diffeomorphism depends explicitly on the choice of
$E$. And of course everything is based on some topological
assumption about $M$, namely, that it admits a globally defined
smooth field of frames (and then, obviously, infinity of them).
Instead of describing motion in terms of time-dependent quantities
$x^{i}(t)$, $e^{i}{}_{A}(t)$, we describe it in terms of
quantities $x^{i}(t)$, $R^{A}{}_{B}(t)$. In other words, the
motion is represented by a pair of curves: the one in $M$, i.e.,
the translational motion, and the other in
SO$(n,\mathbb{R})$-internal rotational motion. And let us repeat
that this trivialization (splitting) is possible only when the
reference field $E$ is fixed and depends explicitly on it. One of
the main advantages is a possibility of introducing good
generalized coordinates instead of redundant quantities
$(x^{i},e^{i}{}_{A})$. Namely, SO$(n,\mathbb{R})$ admits many
well-investigated and geometrically convenient coordinatizations
like, first of all, canonical coordinates of the first kind
(components of the rotation vector if $n=3$), canonical
coordinates of the second kind, and also some other possibilities.
For example, in the physically interesting special case one uses
also Euler angles or spherical variables in the space of rotation
vectors. So, in any case we have at disposal various natural
choices of ``good", i.e., not redundant generalized coordinates
$(\ldots,x^{i},\ldots;\ldots,\xi^{\mu},\ldots)$, where
$\xi^{\mu}$, $\mu=\overline{1,n(n-1)/2}$, are the mentioned
coordinates on SO$(n,\mathbb{R})$.

The same procedure may be applied in the general case of the
infinitesimal affinely-rigid body. The prescribed field of
reference frames $E=(\ldots,E_{A},\ldots)$ may be in principle
quite general, but usually it is chosen as $g$-orthonormal, even
if deformations are admitted. The reason is that such a
description is physically convenient when dealing with mutual
interactions of rigid rotations and finite or infinitesimal
homogeneous deformations. Just as previously we expand
\[
e_{A}(t)=E_{x(t)B}\varphi^{B}{}_{A}(t),\qquad
\varphi^{A}{}_{B}(t)=\left\langle E^{A},e_{B}(t)\right\rangle,
\]
but now $\varphi(t)\in$ GL$(n,\mathbb{R})$ is a general
non-singular matrix, or usually a general positive-determinant
matrix, $\varphi(t)\in$ GL$^{+}(n,\mathbb{R})$. In analytical
procedures describing realistic and simple isotropic problems we
often represent $\varphi$ in terms of the polar or two-polar
decomposition \cite{all-book04}.

In this way the bundle of linear frames $FM$ ($F^{+}M$) is
represented by a trivialization $M\times {\rm GL}(n,\mathbb{R})$
($M\times {\rm GL}^{+}(n,\mathbb{R})$), also explicitly dependent
on the choice of the frame $E$. Analytically, configurations are
parameterized by generalized coordinates
$(\ldots,x^{i},\ldots;\ldots,\varphi^{A}{}_{B},\ldots)$. Let us
stress once again that now it is not logically necessary, because
the quantities $(\ldots,x^{i},\ldots;\ldots,e^{i}{}_{A},\ldots)$
themselves are good, independent generalized coordinates. The
point is only that that, as mentioned, in many realistic problems
generalized coordinates
$(\ldots,x^{i},\ldots;\ldots,\varphi^{A}{}_{B},\ldots)$ are
geometrically and technically more convenient and computationally
effective. The more so it is when $\varphi^{A}{}_{B}$ are replaced
by coordinates appearing in the polar and two-polar decompositions
of $\varphi$.

Let us begin with the general description of affinely-rigid body
in terms of prescribed reference frame $E$ and later on consider
the special case of infinitesimal gyroscopes. This will be also
convenient for making the explicit use of the polar and two-polar
decompositions, in particular, for discussing integrable and
superintegrable models in a two-dimensional space. The latter
problem, i.e., two-dimensional infinitesimal gyroscopes or
homogeneously deformable ones sliding over two-dimensional curved
manifolds may be useful in the theory of microstructured
(micropolar or micromorphic) plates.

Obviously, the primary dynamical variables are the moving-frame
vectors $e_{A}(t)$, or equivalently the moving-frame covectors,
$e^{A}(t)$, whereas the reference frames $E$ are only the
auxiliary quantities, does not matter how important for
computational purposes. Let us express
\begin{equation}\label{a70}
e_{A}(t)=E_{B}(x(t))\varphi^{B}{}_{A}(t),\qquad
e^{A}(t)=\varphi^{-1A}{}_{B}(t)E^{B}(x(t)),
\end{equation}
where, obviously, $e(t)\in F_{x(t)}M$, i.e., $x(t)=\pi(e(t))$.

We shall also need inverse formulas, i.e.,
\begin{equation}\label{b70}
E_{A}(x(t))=e_{B}(t)\varphi^{-1B}{}_{A}(t),\qquad
E^{A}(x(t))=\varphi^{A}{}_{B}(t)e^{B}(t).
\end{equation}
To avoid unnecessary crowd of symbols, when it is not confusing,
we shall omit the arguments $t,x(t)$ in the above quantities.

To express explicitly affine velocities we must find formulas for
the covariant derivatives $De_{A}/Dt$ in terms of the quantities
$E,\varphi$.

The above expressions (\ref{a70}) imply that the covariant
differentiation of $\mathbb{R}\ni t\mapsto e_{A}\in FM$ along the
curve of translational motion $\mathbb{R}\ni t\mapsto x(t)\in M$
is given by
\begin{equation}\label{b71}
\frac{De_{A}}{Dt}=\frac{D}{Dt}\left(E_{B}\varphi^{B}{}_{A}\right)=
\frac{DE_{B}}{Dt}\varphi^{B}{}_{A}+E_{B}\frac{d\varphi^{B}{}_{A}}{dt}.
\end{equation}
Let us express the along-curve differentiation of $E_{B}$ through
its field-differen\-tiation (well defined because $E$ is defined
as a field all over in $M$),
\begin{equation}\label{a71}
\frac{DE_{B}}{Dt}=\left(\nabla_{i}E_{B}\right) \frac{dx^{i}}{dt}=
\left(\nabla_{C}E_{B}\right)E^{C}{}_{i}\frac{dx^{i}}{dt},
\end{equation}
where $\nabla_{C}E_{B}$ is an abbreviation for
$\nabla_{E_{C}}E_{B}$, and let us remind that for any vector field
$Y$ and tensor field $T$, $\nabla_{Y}T$ denotes the covariant
derivative of $T$ along $Y$. Analytically,
\[
\nabla_{Y}T=Y^{i}\nabla_{i}T.
\]
It is convenient to use aholonomic coefficients of our affine
connection with respect to the field $E$,
\begin{equation}\label{c71}
\nabla_{C}E_{B}=\Gamma^{A}{}_{BC}E_{A}.
\end{equation}
The usual holonomic coefficients of $\Gamma$ with respect to
coordinates $x^{i}$ are given by the expression
\[
\Gamma^{i}{}_{jk}=E^{i}{}_{A}\Gamma^{A}{}_{BC}E^{B}{}_{j}E^{C}{}_{k}+E^{i}{}_{A}E^{A}{}_{j,k},
\]
where the comma as usual denotes the partial differentiation. The
second term
\begin{equation}\label{a72}
\Gamma[E]^{i}{}_{jk}:=E^{i}{}_{A}E^{A}{}_{j,k}
\end{equation}
denotes the teleparallelism connection induced by $E$. It is
defined as follows:
\begin{equation}\label{b72}
\nabla_{[\Gamma[E]]}E_{A}=0,
\end{equation}
i.e., it is the only affine connection with respect to which all
the fields $E_{A}$ (of course, also their dual co-fields $E^{A}$)
are parallel. Its curvature tensor vanishes and the torsion equals
\begin{equation}\label{c72}
S[E]^{i}{}_{jk}=E^{i}{}_{A}E^{A}{}_{[j,k]}=
\frac{1}{2}E^{i}{}_{A}\left(E^{A}{}_{j,k}-E^{A}{}_{k,j}\right).
\end{equation}
The parallel transport with respect to $\Gamma[E]$ is
path-independent. A tensor field $T$ is $\Gamma[E]$-parallel,
$\nabla_{[\Gamma[E]]}T=0$, if its aholonomic components with
respect to $E$ are constant in $M$,
\begin{eqnarray}
T^{i_{1}\ldots i_{k}}{}_{j_{1}\ldots j_{l}}&=&T^{I_{1}\ldots
I_{k}}{}_{J_{1}\ldots J_{l}}E_{I_{1}}\otimes\cdots\otimes
E_{I_{k}}\otimes E^{J_{1}}\otimes\cdots\otimes
E^{J_{l}},\nonumber\\
T^{I_{1}\ldots I_{k}}{}_{J_{1}\ldots J_{l}}&=&{\rm const}.
\end{eqnarray}

If the field of frames $E$ is holonomic, in particular, if it is
given by the system of tangent vectors $\partial/\partial x^{i}$
of $x^{i}$, then
\[
\nabla_{\partial/\partial x^{k}}\frac{\partial}{\partial x^{j}}=
\Gamma^{i}{}_{jk}\frac{\partial}{\partial x^{i}},
\]
which means that we obtain the usual components of $\Gamma$ with
respect to local coordinates $x^{i}$.

Obviously, the quantity
\[
\Gamma^{i}{}_{jk}-\Gamma[E]^{i}{}_{jk}=E^{i}{}_{A}\Gamma^{A}{}_{BC}E^{B}{}_{j}E^{C}{}_{k},
\]
i.e., in the coordinate-free form
\[
\Gamma-\Gamma[E]=\Gamma^{A}{}_{BC}E_{A}\otimes\widetilde{E}^{B}\otimes\widetilde{E}^{C}
\]
is a tensor field, once contravariant and twice covariant, as it
is always the case with the difference of affine connections.

The teleparallelism torsion is directly related to what is well
known in differential geometry as the aholonomic object of $E$,
i.e.,
\[
S[E]=\frac{1}{2}\Omega^{A}{}_{BC}E_{A}\otimes\widetilde{E}^{B}\otimes\widetilde{E}^{C}.
\]
In other words the doubled $E$-co-moving components of $S[E]$
coincide with the Schouten aholonomic complex $\Omega$; the latter
may be defined in terms of Lie brackets of basic vector fields,
\begin{equation}\label{a73}
\Omega^{A}{}_{BC}=2\widehat{S}^{A}{}_{BC}=\left\langle
E^{A},\left[E_{B},E_{C}\right]\right\rangle,
\end{equation}
i.e.,
\begin{equation}\label{b73}
[E_{A},E_{B}]=\widehat{\Omega}^{C}{}_{AB}E_{C},\qquad
dE^{A}=\frac{1}{2}\widehat{\Omega}^{A}{}_{BC}E^{C}\wedge E^{B}.
\end{equation}
After some calculations performed on (\ref{b71}) with the use of
(\ref{a70}), (\ref{b70}), (\ref{a71}), (\ref{c71}), we finally
obtain the formula
\[
\frac{De_{A}}{Dt}=e_{B}\widehat{\Omega}^{B}{}_{A},
\]
where the co-moving affine velocity $\widehat{\Omega}$ may be
expressed as follows:
\begin{equation}\label{a74}
\widehat{\Omega}^{B}{}_{A}=\varphi^{-1B}{}_{F}\Gamma^{F}{}_{DC}
\varphi^{D}{}_{A}\varphi^{C}{}_{E}\widehat{V}^{E}+
\varphi^{-1B}{}_{C}\frac{d\varphi^{C}{}_{A}}{dt},
\end{equation}
where, as previously,
\[
\widehat{V}^{K}=e^{K}{}_{i}\frac{dx^{i}}{dt}=e^{K}{}_{i}V^{i}
\]
denotes the co-moving components of translational velocity. In
this way the quantity
\[
\widehat{\Omega}^{B}{}_{A}=\left\langle
e^{B},\frac{De_{A}}{Dt}\right\rangle=e^{B}{}_{i}\frac{De^{i}{}_{A}}{Dt}
\]
is expressed through the matrix $\varphi$ and its time
derivatives. This representation becomes very convenient when
gyroscopic constraints are imposed. Let us observe that in
addition to the second term $\varphi^{-1}d\varphi/dt$ familiar
from the flat-space theory, the above expression (\ref{a74})
contains the first term explicitly depending on the geometry of
$M$ (more precisely, on $(M,\Gamma)$) and on the choice of the
auxiliary reference frame $E$. Roughly speaking, the quantity
$\varphi^{-1}d\varphi/dt$ is the co-moving affine velocity of
internal degrees of freedom, as seen from the point of view of the
just passed frame $E_{x(t)}$. And the first term represents the
contribution to $\widehat{\Omega}$ coming from the
rotation-deformation of $E$ itself.

It is interesting to rewrite the equations (\ref{a57}),
(\ref{b57}), (\ref{a59}), (\ref{d59}), and (\ref{b59}),
(\ref{c59}) in purely co-moving terms. After some calculations we
obtain that
\begin{eqnarray}
m\frac{d\widehat{V}^{A}}{dt}&=&-m\widehat{\Omega}^{A}{}_{B}\widehat{V}^{B}+
2m\widehat{V}^{B}\widehat{V}^{C}S_{BC}{}^{A}\nonumber\\
&&+\frac{1}{2}\widehat{\Omega}^{C}{}_{D}
R^{D}{}_{C}{}^{A}{}_{B}\widehat{V}^{B}+\widehat{F}^{A},\\
\frac{d\widehat{\Omega}^{D}{}_{C}}{dt}J^{CA}&=&-\widehat{\Omega}^{B}{}_{D}
\widehat{\Omega}^{D}{}_{C}J^{CA}+\widehat{N}^{AB}.
\end{eqnarray}
Obviously, the torsion here is taken into account, so strictly
speaking they are co-moving counterparts of (\ref{a57}),
(\ref{b57}).

\noindent {\bf Remark:} do not confuse the three-index torsion
tensor with the two-index spin one. The two-index quantity in the
curvature term is vorticity.

Using explicitly the balance form of the equations of motion for
the linear momentum and affine spin we obtain that
\begin{eqnarray}
\frac{d\widehat{P}^{A}}{dt}&=&-\widehat{P}^{B}\widetilde{J}_{BC}\widehat{K}^{CA}+
\frac{2}{m}\widehat{P}^{B}\widehat{P}^{C}S_{BC}{}^{A}\nonumber\\
&&+\frac{1}{2m}\widehat{K}^{C}{}_{D}
R^{D}{}_{C}{}^{A}{}_{B}\widehat{P}^{B}+\widehat{F}^{A},\\
\frac{d\widehat{K}^{AB}}{dt}&=&-\widehat{K}^{AC}
\widetilde{J}_{CD}\widehat{K}^{DB}+\widehat{N}^{AB},\label{a75}
\end{eqnarray}
where, as previously, $\widetilde{J}_{BC}J^{CA}=\delta^{A}{}_{B}$.

Obviously, we are allowed to use the usual time derivatives at the
left-hand side of equations because $\widehat{P}^{A}$,
$\widehat{K}^{AB}$, $\widehat{V}^{A}$,
$\widehat{\Omega}^{A}{}_{B}$ are scalars from the point of view of
geometry of $M$ (although they are tensors in the micromaterial
space $\mathbb{R}^{n}$).

These are, so to speak, affine Euler equations. By the way, when
we impose metrical constraints, i.e., assume that permanently
\begin{equation}\label{a76}
g(e_{A},e_{B})=g_{ij}e^{i}{}_{A}e^{j}{}_{B}=\delta_{AB},\qquad
\varphi\in{\rm SO}(n,\mathbb{R}),
\end{equation}
and according to the d'Alembert principle the reaction-free
equations of internal motion, i.e., the skew-symmetric part of
(\ref{a75}) is only left, then we exactly obtain Euler equations
for infinitesimal gyroscope in a non-Euclidean space.

Let us observe that all the above equations of motion, in
particular (\ref{a57}), (\ref{b57}) and the expression for the
kinetic energy (\ref{a52}), or its equivalent form (\ref{b52}),
(\ref{c52}) may be expressed in terms of (\ref{a74}), and in the
case of gyroscopic motion this becomes technically unavoidable.
The point is that $e^{i}{}_{A}$ are not then independent
generalized quantities, just due to the constraints (\ref{a76}).
And although equations of motion in their general balance form
(\ref{b61}) are correct and free of reaction forces, in particular
dynamical problems they are not directly useful in analytical
calculations and analysis of the phase portraits. The quantities
$\varphi^{K}{}_{L}$ are not generalized coordinates either,
because the matrix $\varphi$ is orthogonal. However, as mentioned,
$\varphi\in$ SO$(n,\mathbb{R})$ may be easily parameterized on the
basis of group-theoretical considerations, and the corresponding
parameters are just proper generalized coordinates to be
effectively used in equations of motion. Let $q^{\alpha}$,
$\alpha=\overline{1,n(n-1)/2}$, be such group parameters. Then
taking some independent subsystem of (\ref{b61}), e.g., one given
by $a<b$ (or conversely), we obtain a system of $n(n-1)/2$
second-order differential equations imposed on the time-dependence
of $n(n-1)/2$ generalized coordinates. In variational problems it
is more convenient to substitute the expressions $\varphi(q)$
directly to the formula for the kinetic energy (\ref{a52}) or its
Hamiltonian version (\ref{a53}), and then obtain directly the
corresponding Euler-Lagrange equations or (more convenient)
Hamilton ones formulated in terms of Poisson brackets. Such an
approach is better suited to the study for integrability problems
and action-angle variables. Geometrically the most suggestive
choice of generalized coordinates $q^{\alpha}$ is that based on
the exponential representation of $\varphi$. The quantities
$q^{\alpha}$ are then canonical coordinates of the first kind on
the group manifold of SO$(n,\mathbb{R})$. More precisely, let
$M_{KL}\in$ SO$(n,\mathbb{R})^{\prime}$ be basic skew-symmetric
matrices; they have only two not vanishing entries, namely $\pm 1$
in the rows and columns labelled by $K,L$ \cite{all-book04}. Then
$R\in$ SO$(n,\mathbb{R})$ may be represented as follows:
\[
R(\varepsilon)=\exp\left(\frac{1}{2}\varepsilon^{KL}M_{LK}\right),
\]
where $\varepsilon$ is an arbitrary real skew-symmetric matrix,
$\varepsilon^{KL}=-\varepsilon^{LK}$, and, e.g.,
$\varepsilon^{KL}$ with $K<L$ may be chosen as independent
coordinates. The corresponding $M_{KL}$, $K<L$, form a basis of
the Lie algebra SO$(n,\mathbb{R})^{\prime}$.

In the special case of $n=3$, due to the exceptional isomorphism
between axial vectors and skew-symmetric second-order tensors in
$\mathbb{R}^{3}$, it is customary to write \cite{all-book04}
\[
R\left(\overline{k}\right)=\exp\left(k^{A}M_{A}\right),
\]
where $M_{A}$, $A=1,2,3$, are basic skew-symmetric matrices,
\[
\left(M_{A}\right)_{BC}=-\varepsilon_{ABC}
\]
($\varepsilon$ is the totally skew-symmetric Levi-Civita-Ricci
symbol normalized by $\varepsilon_{123}=1$), and $k^{A}$ are
components of the so-called rotation vector in $\mathbb{R}^{3}$.
Its length
\[
k=\sqrt{\left(k^{1}\right)^{2}+\left(k^{2}\right)^{2}+\left(k^{3}\right)^{2}}
\]
satisfies $k\leq \pi$ if all possible direction versors
$\overline{n}=\overline{k}/k$ are admitted. Obviously, for any
versor $\overline{n}$ we have
\[
R\left(\pi\overline{n}\right)=R\left(-\pi\overline{n}\right).
\]
Therefore, the group manifold is represented in the space of
$\mathbb{R}^{3}$-vectors $\overline{k}$ by the ball of radius
$\pi$, with the proviso that antipodal points on the limiting
sphere $k=\pi$ are identified. This provides a very nice form of
understanding that the group SO$(3,\mathbb{R})$ is doubly
connected.

In the three-dimensional case one uses also other
parameterizations, depending on dynamical details of the
considered problem. For example, in certain dynamical models of
the spherical rigid body it is convenient to use the spherical
variables $(k,\phi,\theta)$ in the space of the rotation vectors
$\overline{k}$ \cite{all-book04}:
\[
k^{1}=k\sin\theta\cos\phi,\qquad k^{2}=k\sin\theta\sin\phi,\qquad
k^{3}=k\cos\theta.
\]

In the theory of a free or heavy symmetric top ($I_{1}=I_{2}$, but
in general $I_{1}\neq I_{3}$), it is convenient to use Euler
angles $(\varphi,\vartheta,\psi)$,
\[
R[\varphi,\vartheta,\psi]=\exp(\varphi M_{3})\exp(\vartheta
M_{1})\exp(\psi M_{3}).
\]
One can also use canonical coordinates of the second kind
$(\alpha,\beta,\gamma)$, i.e.,
\[
R[\alpha,\beta,\gamma]=\exp(\alpha M_{1})\exp(\beta
M_{2})\exp(\gamma M_{3}).
\]
Surprisingly enough, for the spherical rigid body
$(I_{1}=I_{2}=I_{3})$ the kinetic energy expressions have almost
the same form in coordinates $(\varphi,\vartheta,\psi)$ and
$(\alpha,\beta,\gamma)$.

If $n=2$, the situation is much more simple, because then
SO$(2,\mathbb{R})$ is one-dimensional (thus, obviously,
commutative) and $R(\varphi)=\exp(\varphi\epsilon)$, where
$\epsilon_{11}=\epsilon_{22}=0$, $\epsilon_{12}=-\epsilon_{21}=1$,
i.e., explicitly
\[
R(\varphi)=\left[
\begin{array}{cc}
  \cos\varphi & -\sin\varphi \\
  \sin\varphi & \cos\varphi
\end{array}\right].
\]
In the case of rigid motion the second term in (\ref{a74}),
$\varphi^{-1}d\varphi/dt$ is always skew-symmetric, i.e., an
element of the Lie algebra SO$(n,\mathbb{R})^{\prime}$. More
precisely, if we use artificial not orthogonal coordinates in
$\mathbb{R}^{3}$, it is $\eta$-skew-symmetric, i.e.,
\[
\varphi^{-1B}{}_{C}\frac{d\varphi^{C}{}_{A}}{dt}=-\eta_{AK}\eta^{BL}
\varphi^{-1K}{}_{C}\frac{d\varphi^{C}{}_{L}}{dt}.
\]
If the connection $\Gamma$ is metrical, $\nabla_{[\Gamma]}g=0$
(Riemann-Cartan space), then, as we saw, the total
$\widehat{\Omega}$ is also skew-symmetric ($\eta$-skew-symmetric)
and therefore, so is the first term of (\ref{a74}). By the way,
the aholonomic connection coefficients $\Gamma^{F}{}_{DC}$ are
then also skew-symmetric (more precisely, $\eta$-skew-symmetric).
Just as in the general affine motion, we can say that
$\widehat{\Omega}$ is then the co-moving angular velocity,
$\varphi^{-1}d\varphi/dt$ is the angular velocity with respect to
the pre-fixed aholonomic frame $E$, and the first term is, in a
sense, the angular velocity with which $E$ itself rotates along
the trajectory of the structured material point.

In both the general affine and constrained gyroscopic motion the
expression of kinetic energy through the above quantities becomes
relatively lucid and computationally effective when the manifold
$(M,\Gamma,g)$ has some special structure, e.g., if it is a
constant-curvature space and if the auxiliary field of frames $E$
is appropriately chosen. The proper choice depends on the
particular geometry of $M$, and it is also a matter of some
inventive intuition. In such special cases the problem may be
effectively studied on the rigorous analytical level and, in
particular, interesting results concerning integrability and
degeneracy (superintegrability) may be obtained
\cite{Gol01,Gol02,Gol03,Gol04,Gol04S,Gol06,Mart02,Mart03,Mart04_1,Mart04_2}.

Unless otherwise stated, from now on we concentrate on Riemannian spaces, when $\Gamma$ is the
Levi-Civita connection built of $g$. The general case, when $\Gamma,g$ are unrelated may be
also interesting in itself, but certainly there are some difficulties to be overcome, because
even in the rigid motion the affine velocity $\Omega$ is not $g$-skew-symmetric and its
co-moving representation $\widehat{\Omega}$ is not $\eta$-($\delta$-)skew-symmetric. The fact
that they do not belong to the corresponding Lie algebras of the rotational groups
SO$(T_{x}M,g_{x})$, SO$(n,\mathbb{R})$ obscures their interpretation. If $(M,\Gamma,g)$ is a
Riemann-Cartan space but $\Gamma$ is not symmetric (i.e., it is not Levi-Civita connection),
then affine velocities are skew-symmetric (respectively in a $g$- or $\eta$-sense), but they
are explicitly dependent on the torsion tensor. This also leads to certain interpretation
problems. Namely, one obtains different expressions for the kinetic energy and different
equations of motion when one uses respectively the affine connections $\left\{\begin{array}{c}
i \\ jk \end{array}\right\}$ ($g$-Levi-Civita) and
\[
\Gamma^{i}{}_{jk}=\left\{\begin{array}{c}
  i \\
  jk
\end{array}\right\}+S^{i}{}_{jk}+S_{jk}{}^{i}+S_{kj}{}^{i}
\]
in the definition of angular velocity.

In the general case of affine motion the formula (\ref{a74}) will
be written in any of the abbreviated forms
\[
\widehat{\Omega}=\widehat{\Omega}_{\rm dr}+\widehat{\Omega}_{\rm
rl}=\widehat{\Omega}(\rm dr)+\widehat{\Omega}(\rm rl),
\]
where
\begin{eqnarray}
\widehat{\Omega}_{\rm dr}{}^{A}{}_{B}=\widehat{\Omega}(\rm dr)^{A}{}_{B}&=&
\varphi^{-1}{}^{A}{}_{F}\Gamma^{F}{}_{DC}\varphi^{D}{}_{B}\varphi^{C}{}_{E}\widehat{V}^{E},\label{a81} \\
\widehat{\Omega}_{\rm rl}{}^{A}{}_{B}=\widehat{\Omega}(\rm
rl)^{A}{}_{B}&=&\varphi^{-1}{}^{A}{}_{C}\frac{d\varphi^{C}{}_{B}}{dt}.
\end{eqnarray}
The labels ``dr" and ``rl" refer respectively to ``drift" (or
"drive") and ``relative". The reason is that, as mentioned,
$\varphi$ refers to affine rotations with respect to the just
passed prescribed reference frame $E$; the first term describes
the time rate of affine rotations contained in the field $E$
itself. When gyroscopic constraints are imposed, all these
$\widehat{\Omega}$-objects become skew-symmetric angular
velocities. To stress this, sometimes, but not always, we shall
use then the symbols $\widehat{\omega}$, $\widehat{\omega}_{\rm
dr}=\widehat{\omega}(\rm dr)$ and $\widehat{\omega}_{\rm
rl}=\widehat{\omega}(\rm rl)$.

One of analytical advantages following from the prescribed reference frame $E$ is the possibility of using the polar
and two-polar decompositions \cite{all-book04},
\[
\varphi=UA=BU=LDR^{-1},
\]
where $U$, $L$, $R$ are orthogonal (more precisely, $\eta$-orthogonal when artificial non-orthonormal coordinates are
used in $\mathbb{R}^{n}$), $A$, $B$ are symmetric ($\eta$-symmetric), $D$ is diagonal, and obviously,
\[
B=UAU^{-1}.
\]

As usual, $U$, $L$, $R$ denote fictitious gyroscopic degrees of
freedom extracted from $\varphi\in$ GL$(n,\mathbb{R})$
\cite{all-book04}. The corresponding ``co-moving" angular
velocities are given by the expressions
\begin{equation}\label{a82}
\widehat{\omega}_{\rm rl}=U^{-1}\frac{dU}{dt},\qquad \widehat{\chi}_{\rm rl}=L^{-1}\frac{dL}{dt},\qquad
\widehat{\vartheta}_{\rm rl}=R^{-1}\frac{dR}{dt}.
\end{equation}
Obviously, as usual, the ``spatial" representation may be used:
\[
\omega_{\rm rl}=\frac{dU}{dt}U^{-1},\qquad \chi_{\rm rl}=\frac{dL}{dt}L^{-1},\qquad \vartheta_{\rm
rl}=\frac{dR}{dt}R^{-1}.
\]
However, in calculations appearing in practical problems the
``co-moving" objects are more convenient. Obviously, in the
two-dimensional world, when $n=2$, these representations coincide.

After some calculations one can show that the kinetic energy of internal motion $T_{\rm int}$
(\ref{a52}) may be expressed in the following way in terms of the polar decomposition:
\begin{equation}\label{a83}
T_{\rm int}=-\frac{1}{2}{\rm
Tr}\left(AJA\widehat{\omega}^{2}\right)+{\rm
Tr}\left(AJ\frac{dA}{dt}\widehat{\omega}\right) +\frac{1}{2}{\rm
Tr}\left(J\left(\frac{dA}{dt}\right)^{2}\right),
\end{equation}
where
\begin{equation}\label{c83}
\widehat{\omega}=\widehat{\omega}_{\rm dr}+\widehat{\omega}_{\rm rl}=\widehat{\omega}_{\rm dr}+U^{-1}\frac{dU}{dt},
\end{equation}
and obviously $\widehat{\omega}_{\rm dr}$ is the restriction of $\widehat{\Omega}_{\rm dr}$ (\ref{a81}) to the
$U$-rigid motion:
\begin{equation}\label{b83}
\widehat{\omega}_{\rm dr}{}^{A}{}_{B}=U^{-1A}{}_{F}\Gamma^{F}{}_{DC}U^{D}{}_{B}U^{C}{}_{E}\widehat{V}^{E}.
\end{equation}
The formula (\ref{a83}) is written in the standard orthonormal coordinates in $\mathbb{R}^{n}$. Otherwise, when
$\eta_{AB}$ is admitted to be different than $\delta_{AB}$, we have to replace the symbol $J$ in (\ref{a83}) by
$J_{\eta}$, where
\[
J_{\eta}{}^{A}{}_{B}:=J^{AC}\eta_{CB}.
\]
Geometrically, $J_{\eta}$ is a mixed tensor in $\mathbb{R}^{n}$, i.e., linear endomorphism of $\mathbb{R}^{n}$,
$J_{\eta}\in$ L$(n,\mathbb{R})$, whereas $J$ itself is a twice contravariant tensor.

The two-polar decomposition becomes analytically useful in
doubly-iso\-tropic dynamical problems, i.e., ones isotropic both
in the physical space $M$ and the micromaterial space. This double
isotropy imposes certain restrictions both on the kinetic and
potential energies. What concerns the very kinetic energy, the
inertial tensor must be proportional to $\widetilde{\eta}$, i.e.,
\[
J^{AB}=I\eta^{AB},\qquad J_{\eta}{}^{A}{}_{B}=I\delta^{A}{}_{B}.
\]
Then one can show that (\ref{a83}) becomes
\begin{equation}\label{a84}
T_{\rm int}=-\frac{I}{2}{\rm
Tr}\left(D^{2}\widehat{\chi}^{2}\right)-\frac{I}{2}{\rm
Tr}\left(D^{2}\widehat{\vartheta}^{2}\right)+I{\rm
Tr}\left(D\widehat{\chi}D\widehat{\vartheta}\right),
\end{equation}
where now
\begin{eqnarray}
\widehat{\vartheta}&=&R^{-1}\frac{dR}{dt},\label{c84}\\
\widehat{\chi}&=&\widehat{\chi}_{\rm dr}+\widehat{\chi}_{\rm
rl}=\widehat{\chi}_{\rm dr}+L^{-1}\frac{dL}{dt},
\label{b84}\\
\widehat{\chi}_{\rm
dr}{}^{A}{}_{B}&=&L^{-1A}{}_{F}\Gamma^{F}{}_{DC}L^{D}{}_{B}L^{C}{}_{E}\widehat{V}^{E}.
\label{d84}
\end{eqnarray}
The last formula is quite analogous to (\ref{b83}). Just like
there, $\widehat{\chi}$ contains the ``drive" term built of the
connection coefficients. It is only the $L$-rotation that is
coupled in this way to spatial geometry; the $R$-rotation is
geometry-independent.

Again we conclude that (\ref{a84}) is structurally identical with
the corresponding formula for extended affine bodies
\cite{JJS82_2} with the proviso however that $\widehat{\chi}$
contains the drive-term. The expression for $\widehat{\vartheta}$
is free of such a correction. Everything that has to do with
$(M,\Gamma,g)$-geometry is absorbed by the $\widehat{\chi}$-term.

\section{Special two-dimensional problems}

We shall consider now some special two-dimensional cases, i.e.,
when $n=2$. Therefore, for the infinitesimal rigid body
(infinitesimal gyroscope) we are dealing with three degrees of
freedom: two translational and one internal, rotational. If no
gyroscopic constraints are imposed and the internal motion is
affine, then of course there are four internal degrees of freedom;
together with translational motion one obtains six degrees of
freedom. The resulting models are interesting in themselves from
the point of view of pure analytical mechanics, in particular,
some integrability and hyperintegrability (degeneracy) problems
may be effectively studied. Obviously, the explicit analytical
results exist only in Riemann manifolds $(M,g)$ with some peculiar
structure, first of all (but not only) in constant-curvature
spaces. Some practical applications of two-dimensional models also
seem to be possible, e.g., in geophysical problems, in mechanics
of structured micropolar and micromorphic shells, etc. What
concerns geophysics, we mean, e.g., motion of continental plates.
Motion of pollutions like oil spots on the oceanic surface is
another suggestive example.

Below we consider in some details three kinds of two-dimensional
problems, namely motion of structured material points on the
sphere, pseudo-sphere (Lobatchevski space), and torus manifolds
with geometry induced by injections in the three-dimensional
Euclidean space $\mathbb{R}^{3}$. Many interesting dynamical
models, including some quite realistic ones, may be effectively
investigated in analytical terms. It is not very surprising in
spherical and pseudo-spherical geometries because of exceptional
properties of constant-curvature spaces. But there exist also nice
completely integrable models on the $\mathbb{R}^{3}$-injected
torus. Perhaps this may have something to do with that that all
these manifolds are algebraic ones (of the second degree in the
spherical and pseudo-spherical cases, and of the fourth degree in
toroidal geometry).

\subsection{Spherical case}

Let us begin with the spherical geometry. The two-dimensional
"world" will be realized as the two-dimensional sphere in
$\mathbb{R}^{3}$ with the radius $R$ and the centre at the
beginning of coordinates. Of course, the radius has the intrinsic
sense, namely the scalar curvature equals $2/R^{2}$.

We introduce the ``polar" coordinates $(r,\varphi)$. Obviously,
$r$ is the geodetic distance measured from the ``North Pole",
i.e., $\vartheta=r/R$ is a modified ``geographic latitude", and
$\varphi$ is ``geographic longitude", i.e., $R\varphi$ is the
distance measured along ``equator". So, $\varphi$, $\vartheta$ are
usual polar angles in $\mathbb{R}^{3}$, and $R$ is the fixed
radius-distance from the origin. Euclidean coordinates of points
on $S^{2}(0,R)\in\mathbb{R}^{3}$ are given by the usual
expressions:
\[
x=R\sin\vartheta\cos\varphi,\quad y=R\sin\vartheta\sin\varphi,\quad z=R\cos\vartheta,\quad x^{2}+y^{2}+z^{2}=R^{2}.
\]
Obviously, $r$ runs over the range $[0,\pi R]$ from the ``North
Pole" to the ``South Pole", and $\varphi$ has the usual range
$[0,2\pi]$ of the polar angle. Coordinates are singular at $r=0$,
$r=\pi R$, and there is the obvious $0$, $2\pi$ ambiguity of the
longitude $\varphi$.

Restricting the metric tensor of $\mathbb{R}^{3}$ (roughly
speaking, the metric element $dx^{2}+dy^{2}+dz^{2}$) to
$S^{2}(0,R)$ we obtain the obvious expression for the arc element
on the sphere
\begin{equation}\label{a87}
ds^{2}=dr^{2}+R^{2}\sin^{2}\frac{r}{R}\ d\varphi^{2}.
\end{equation}
So, the kinetic energy of translational motion is given by
\begin{equation}\label{a88}
T_{\rm tr}=\frac{m}{2}\left(\left(\frac{dr}{dt}\right)^{2}
+R^{2}\sin^{2}\frac{r}{R}\left(\frac{d\varphi}{dt}\right)^{2}\right),
\end{equation}
where $m$ denotes the mass of the material point.

Even for the purely translational motion some interesting questions arise, e.g., what are spherically symmetric
potentials $V(r)$ for which all orbits are closed? Obviously we mean problems based on Lagrangians
\[
L_{\rm tr}=T_{\rm tr}-V(r).
\]
This is a counterpart of the famous Bertrand problem in
$\mathbb{R}^{2}$. And it may be shown that the answer is similar
\cite{Koz-Har92,JJS80,JJS00_2}, i.e., the possible potentials are
as follows:
\begin{itemize}
\item[$(i)$] {\bf oscillatory potentials:}
\begin{equation}\label{b88}
V(r)=\frac{\varkappa}{2}R^{2}{\rm tg}^{2}\frac{r}{R},
\end{equation}

\item[$(ii)$] {\bf Kepler-Coulomb potentials:}
\begin{equation}\label{c88}
V(r)=-\frac{\alpha}{R}{\rm ctg}\frac{r}{R}.
\end{equation}
\end{itemize}
Obviously, with the spherical topology also the geodetic problem belongs here:
\begin{itemize}
\item[$(iii)$] $V(r)=0$, i.e., (in a sense) the special case of
$(i)$ or $(ii)$ when $\varkappa=0$, $\alpha=0$.
\end{itemize}
There is an obvious correspondence with the flat-space Bertrand
problem; it is suggested by the very asymptotics for $r\approx 0$,
i.e.,
\[
V(r)\approx\frac{\varkappa}{2}r^{2}, \qquad V(r)\approx -\frac{\alpha}{r}.
\]
Obviously, this is a rough argument, but it may be shown
\cite{JJS80,JJS00_2} that there exists a rigorous isomorphism
based on the projective geometry.

The mentioned Bertrand models lead to completely integrable and
maximally degenerate (hyperintegrable) problems. But even for the
simplest, i.e., geodetic, models with the internal degrees of
freedom the situation drastically changes. There exist interesting
and practically applicable integrable models, but as a rule
interaction with internal degrees of freedom reduces or completely
removes degeneracy.

Let us begin with the gyroscopic model of internal motion. Unlike the general case, in
two-dimensional problems with the constant-curvature spaces, gyroscopic problem is simpler
than affine one. Moreover, it very simplifies study of the affine case.

The first step is to introduce an appropriate field $E$, i.e., the
auxiliary and fixed once for all orthonormal aholonomic reference
frame. This is often the matter of inventive guessing. In this
case it is natural to expect that the natural base (holonomic one)
tangent to coordinate lines is a good starting point. It consists
of the vector fields
\[
\mathcal{E}_{r}=\frac{\partial}{\partial r},\qquad
\mathcal{E}_{\varphi}=\frac{\partial}{\partial \varphi}.
\]
Obviously, we mean here the well-known identification between vector fields and first-order
differential operators, i.e.,
\[
X=X^{i}\frac{\partial}{\partial x^{i}}
\]
($X^{i}$ are components with respect to the manifold coordinates $x^{i}$).

Using more traditional language: if coordinates are ordered as
$(r,\varphi)$, then $\mathcal{E}_{r}$ and $\mathcal{E}_{\varphi}$
have respectively components $[1,0]^{T}$ and $[0,1]^{T}$. This
holonomic system is not appropriate however, because it is not
orthonormal (no holonomic system may be so in a curved Riemann
space). However, it is evidently orthogonal ($r$, $\varphi$ are
orthogonal coordinates):
\[
g_{r\varphi}=g_{\varphi r}=0,\qquad g_{rr}=1,\qquad
g_{\varphi\varphi}=R^{2}\sin^{2}\frac{r}{R}.
\]
The lengths of $\mathcal{E}_{r}$, $\mathcal{E}_{\varphi}$ are
obviously given by $\|\mathcal{E}_{r}\|=1$,
$\|\mathcal{E}_{\varphi}\|=R\sin\left(r/R\right)$. So, it is
natural to expect that the normalized fields
\begin{equation}\label{a90}
E_{r}=\frac{\partial}{\partial r}=\mathcal{E}_{r},\qquad
E_{\varphi}=\frac{1}{R\sin\frac{r}{R}}\frac{\partial}{\partial
\varphi}=\frac{1}{R\sin\frac{r}{R}}\mathcal{E}_{\varphi}
\end{equation}
will form a convenient orthonormal frame $E$. Let us stress that
$E$ is aholonomic, although directions of $E_{r}$, $E_{\varphi}$
are tangent respectively to coordinate lines of $r$, $\varphi$.
The reason is just the ``crossed-variables" normalization, i.e.,
$\mathcal{E}_{\varphi}$ is multiplied by a function dependent on
$r$. There are no coordinates $r^{\prime}$, $\varphi^{\prime}$ for
which we would have $E_{r}=\partial/\partial r^{\prime}$,
$E_{\varphi}=\partial/\partial \varphi^{\prime}$. The orthonormal
frame $e=(e_{1},e_{2})$ describing the internal configuration is
obtained from the fixed aholonomic frame $E=(E_{r},E_{\varphi})$
with the help of some time-dependent orthogonal matrix
$\varphi=U$, i.e.,
\begin{eqnarray}
e_{1}&=&E_{1}U^{1}{}_{1}+E_{2}U^{2}{}_{1}=
E_{r}U^{r}{}_{1}+E_{\varphi}U^{\varphi}{}_{1},\nonumber\\
e_{2}&=&E_{1}U^{1}{}_{2}+E_{2}U^{2}{}_{2}=
E_{r}U^{r}{}_{2}+E_{\varphi}U^{\varphi}{}_{2}.\nonumber
\end{eqnarray}
Obviously, $U$ may be parameterized in the usual way, i.e.,
\[
U=\left[\begin{array}{cc}
  \cos \psi & -\sin \psi \\
  \sin \psi & \cos \psi
\end{array}\right]
\]
and
\[
\widehat{\omega}_{\rm rl}=\omega_{\rm rl}\left[\begin{array}{cc}
  0 & -1 \\
  1 & 0
\end{array}\right]=\frac{d\psi}{dt}\left[\begin{array}{cc}
  0 & -1 \\
  1 & 0
\end{array}\right].
\]
{\bf Remark:} do not confuse the scalar quantity $\omega_{\rm rl}$
with the second order skew-symmetric tensor $\widehat{\omega}_{\rm
rl}$ parameterized by it.

Similarly we write that
\[
\widehat{\omega}_{\rm dr}=\omega_{\rm dr}\left[\begin{array}{cc}
  0 & -1 \\
  1 & 0
\end{array}\right],\qquad \widehat{\omega}=\omega\left[\begin{array}{cc}
  0 & -1 \\
  1 & 0
\end{array}\right],
\]
where
\begin{eqnarray}
\omega_{\rm dr}&=&\cos\frac{r}{R}\frac{d\varphi}{dt}=
\cos\vartheta\frac{d\varphi}{dt},\nonumber\\
\omega&=&\omega_{\rm rl}+\omega_{\rm
dr}=\frac{d\psi}{dt}+\cos\frac{r}{R}\frac{d\varphi}{dt}=
\frac{d\psi}{dt}+\cos\vartheta\frac{d\varphi}{dt}. \label{a91}
\end{eqnarray}
Obviously, the above expressions are the special cases of
(\ref{a82}), (\ref{c83}), (\ref{b83}) and follow easily from the
formulas for Christoffel symbols (Levi-Civita connection) on the
sphere, i.e.,
\[
\Gamma^{r}{}_{\varphi\varphi}=-\frac{R}{2}\sin\frac{r}{R},\qquad
\Gamma^{\varphi}{}_{r\varphi}=\Gamma^{\varphi}{}_{\varphi
r}=\frac{1}{R}{\rm ctg}\frac{r}{R}
\]
(the remaining coefficients vanish). In the aholonomic
representation:
\[
\Gamma^{r}{}_{\varphi\varphi}=-\frac{1}{R}{\rm
ctg}\frac{r}{R},\qquad
\Gamma^{\varphi}{}_{r\varphi}=\frac{1}{R}{\rm ctg}\frac{r}{R},
\]
but let us notice that
\[
\Gamma^{\varphi}{}_{\varphi r}=0\neq\Gamma^{\varphi}{}_{r\varphi},
\]
i.e., the aholonomic coefficients are in general not symmetric.
The remaining aholonomic coefficients also vanish. Obviously, the
above $\Gamma^{a}{}_{bc}$ are exactly the formerly used
$\Gamma^{A}{}_{BC}$; simply in some particular concrete cases this
way of writing is more suggestive and intuitive.

After simple calculations based on the formula (\ref{a52}) or rather on its restriction to the
gyroscopic motion, one obtains the expected expression:
\begin{eqnarray}
T&=&T_{\rm tr}+T_{\rm int}\label{a92}\\
&=&\frac{m}{2}\left(\left(\frac{dr}{dt}\right)^{2}+
R^{2}\sin^{2}\frac{r}{R}\left(\frac{d\varphi}{dt}\right)^{2}\right)
+\frac{I}{2}\left(\frac{d\psi}{dt}+
\cos\frac{r}{R}\frac{d\varphi}{dt}\right)^{2},\nonumber
\end{eqnarray}
i.e., briefly,
\[
T=T_{\rm tr}+T_{\rm int}=\frac{mv^{2}}{2}+\frac{I\omega^{2}}{2},
\]
where $\omega$ is given by (\ref{a91}). The scalar quantity $I$,
i.e., the inertial moments of the plane rotator is related to the
tensor $J$ by the formula
\[
I=\eta_{AB}J^{AB}={\rm Tr}J_{\eta},
\]
i.e., in Euclidean material coordinates ($\eta_{AB}=\delta_{AB}$)
simply $I={\rm Tr}J$. In the absence of deformations the internal
inertia is controlled only by this single scalar. This is the
peculiarity of the ``two-dimensional world".

For certain reasons it will be convenient to rewrite the formula (\ref{a92}) in terms of the
variable $\vartheta=r/R$, i.e.,
\begin{eqnarray}
T&=&T_{\rm tr}+T_{\rm int}\label{a93}\\
&=&\frac{mR^{2}}{2}\left(\left(\frac{d\vartheta}{dt}\right)^{2}+
\sin^{2}\vartheta\left(\frac{d\varphi}{dt}\right)^{2}\right)
+\frac{I}{2}\left(\frac{d\psi}{dt}+
\cos\vartheta\frac{d\varphi}{dt}\right)^{2}.\nonumber
\end{eqnarray}

It is seen that if formally $(\varphi,\vartheta,\psi)$ are interpreted as Euler angles
(respectively the precession, nutation, and rotation), the above expression is formally
identical with the kinetic energy of the three-dimensional symmetric rigid body (without
translations) with the main moments of inertia given respectively by
\[
I_{1}=I_{2}=mR^{2},\qquad I_{3}=I.
\]
If $I=mR^{2}$ one obtain the expression for the spherical top.

There is nothing surprising in the mentioned isomorphism because
the quotient manifold SO$(3,\mathbb{R})/$SO$(2,\mathbb{R})$ may be 
identified with S$^{2}(0,1)$ (or with any
S$^{2}(0,R)$) in a natural way. Projecting the motion of the three-dimensional
symmetric top onto the quotient sphere-manifold we obtain
two-dimensional translational motion; the one-dimensional subgroup
of rotations about the $z$-axis refers to the internal motion of
the two-dimensional rotator.

The projection procedure is exactly compatible with the mentioned
correspondence between Euler angles in SO$(3,\mathbb{R})$ and our
generalized coordinates $\varphi$, $\vartheta=r/R$, $\psi$ of the
infinitesimal rotator in S$^{2}(0,R)$.

Let $U(\varphi,\vartheta,\psi)\in$ SO$(3,R)$ be just the element
labelled by the Euler angles $\varphi$, $\vartheta$, $\psi$, thus
\begin{equation}\label{b93}
U(\varphi,\vartheta,\psi)=U_{z}(\varphi)U_{x}(\vartheta)U_{z}(\psi),
\end{equation}
where $U_{z}$, $U_{x}$ are rotations respectively around the $z$-
and $x$-axes; angles of rotations are indicated as arguments.
Calculating the ``co-moving angular velocity"
\begin{equation}\label{b94}
\widehat{\varkappa}=U^{-1}\frac{dU}{dt}
\end{equation}
of this fictitious three-dimensional top one obtains that
\begin{equation}\label{c94}
\widehat{\varkappa}=\widehat{\varkappa}_{1}\left[\begin{array}{ccc}
  0 & 0 & 0 \\
  0 & 0 & -1 \\
  0 & 1 & 0
\end{array}\right]+\widehat{\varkappa}_{2}\left[\begin{array}{ccc}
  0 & 0 & 1 \\
  0 & 0 & 0 \\
  -1 & 0 & 0
\end{array}\right]+\widehat{\varkappa}_{3}\left[\begin{array}{ccc}
  0 & -1 & 0 \\
  1 & 0 & 0 \\
  0 & 0 & 0
\end{array}\right],
\end{equation}
where
\begin{eqnarray}
\widehat{\varkappa}_{1}&=&\sin\vartheta\sin\psi\frac{d\varphi}{dt}
+\cos\psi\frac{d\vartheta}{dt},\label{e94}\\
\widehat{\varkappa}_{2}&=&\sin\vartheta\cos\psi\frac{d\varphi}{dt}
-\sin\psi\frac{d\vartheta}{dt},\label{f94}\\
\widehat{\varkappa}_{3}&=&\cos\vartheta\frac{d\varphi}{dt}+\frac{d\psi}{dt}.\label{a94}
\end{eqnarray}
In expression (\ref{a94}) we easily recognize (\ref{a91}), i.e.,
the expression for the one-component angular velocity of the
two-dimensional rotator. Calculating formally the kinetic energy
of the three-dimensional symmetric SO$(3,\mathbb{R})$-top, i.e.,
\begin{equation}\label{d94}
T=\frac{K}{2}\left(\widehat{\varkappa}_{1}\right)^{2}
+\frac{K}{2}\left(\widehat{\varkappa}_{2}\right)^{2}
+\frac{I}{2}\left(\widehat{\varkappa}_{3}\right)^{2},
\end{equation}
and substituting $K=mR^{2}$, $\vartheta=r/R$, we obtain exactly (\ref{a92}), i.e.,
(\ref{a93}).

The principal fibre bundle of orthonormal frames F$({\rm S}^{2},g)$ over the two-dimensional
sphere (with its induced metric $g$) has SO$(2,\mathbb{R})$ as the structural group and it may
be itself identified with SO$(3,\mathbb{R})$. This is just the essence of the above
identification between SO$(3,\mathbb{R})$ and its quotient manifold S$^{2}\simeq$
SO$(3,\mathbb{R})/$SO$(2,\mathbb{R})$. Some important invariance problems appear there.
Namely, as usual in analytical mechanics, the kinetic energy (\ref{a92}), (\ref{a93}) may be
identified with some Riemannian structure on the configuration space. Let us write down our
kinetic energy in the following form with the explicitly separated mass factor:
\[
T=\frac{m}{2}G_{ij}(q)\frac{dq^{i}}{dt}\frac{dq^{j}}{dt}.
\]
Just as above, our generalized coordinates $q^{i}$, $i=1,2,3$, are
the variables $(r,\varphi,\psi)$ written just in this direction.
On the level of SO$(3,\mathbb{R})$ as identified with F$({\rm
S}^{2},g)$, they are equivalent to the Euler angles
$(\vartheta,\varphi,\psi)$, where $\vartheta=r/R$.

After some calculations one obtains that
\[
\left[G_{ij}\right]=\left[\begin{array}{ccc}
  1 & 0 & 0 \\
  &&\\
  0 & R^{2}\sin^{2}\frac{r}{R}
  +\frac{I}{m}\cos^{2}\frac{r}{R} &
  \frac{I}{m}\cos\frac{r}{R} \\
  &&\\
  0 & \frac{I}{m}\cos\frac{r}{R} & \frac{I}{m}
\end{array}\right].
\]
In the special case $I=mR^{2}$ one obtains that $G$ simplifies to
$\breve{G}$, where
\begin{equation}\label{a95}
\left[\breve{G}_{ij}\right]=\left[\begin{array}{ccc}
  1 & 0 & 0 \\
  0 & R^{2} & R^{2}\cos\frac{r}{R} \\
  0 & R^{2}\cos\frac{r}{R} & R^{2}
\end{array}\right].
\end{equation}
The corresponding expressions for the weight-one volume densities are as follows:
\[
\sqrt{G}=R\sqrt{\frac{I}{m}}\sin\frac{r}{R},\qquad
\sqrt{\breve{G}}=R^{2}\sin\frac{r}{R}.
\]

The contravariant inverse metric $G^{ij}$
$(G^{ik}G_{kj}=\delta^{i}{}_{j})$ is given as follows:
\[
\left[G^{ij}\right]= \left[
\begin{array}{ccc}
  1 & 0 & 0 \\
  &&\\
  0 & \frac{1}{R^{2} \sin^{2}\frac{r}{R}} & -\frac{\cos\frac{r}{R}}{R^{2} \sin^{2}\frac{r}{R}} \\
  &&\\
  0 & -\frac{\cos\frac{r}{R}}{R^{2} \sin^{2}\frac{r}{R}} & \frac{m}{I}+ \frac{1}{R^{2}}{\rm ctg}^{2}\frac{r}{R} \\
\end{array}
\right],
\]
and, obviously,
\[
\left[\breve{G}^{ij}\right]= \left[
\begin{array}{ccc}
  1 & 0 & 0 \\
  & & \\
  0 & \frac{1}{R^{2} \sin^{2}\frac{r}{R}} & -\frac{\cos\frac{r}{R}}{R^{2} \sin^{2}\frac{r}{R}} \\
  & & \\
  0 & -\frac{\cos\frac{r}{R}}{R^{2} \sin^{2}\frac{r}{R}} & \frac{1}{R^{2}\sin^{2}\frac{r}{R}} \\
\end{array}
\right].
\]
If we once identify the bundle manifold $F(S^{2},g)$ with the
group SO$(3, \mathbb{R})$, then we can consider the action of two
transformation groups. They are, obviously, represented by the
left and right regular actions of SO$(3, \mathbb{R})$ on itself,
i.e.,
\[
X \mapsto VX, \qquad X \mapsto XV,
\]
where $X,V \in {\rm SO}(3,\mathbb{R})$. From the point of view of
the original configuration space $F(S^{2},g)$, these groups act
both on the positions of material point in $S^{2}$ and
orientations of the attached bases, i.e., internal configurations.

The above metrics $G$ are invariant under the total left-acting
group SO$(3, \mathbb{R})$. What concerns the right actions, in
general they are invariant only under the subgroup ${\rm SO}(2,
{\mathbb{R}}) \subset {\rm SO}(3,\mathbb{R})$ interpreted as the
group of rotations about the $z$-axis in ${\mathbb{R}}^{3}$.
Therefore, $G$ has the four-dimensional isometry group ${\rm
SO}(3, {\mathbb{R}}) \times {\rm SO}(2,\mathbb{R})$. Only in the
special case $I=m R^{2}$ the metric tensor $G$ is invariant under
${\rm SO}(3, {\mathbb{R}}) \times {\rm SO}(3,\mathbb{R})$, i.e.,
under all left and right regular translations. Therefore, on the
level of ${\rm SO}(3,\mathbb{R})$-description we are dealing then
with the spherical top and up to a constant multiplier $G$ becomes
the Killing metric on ${\rm SO}(3,\mathbb{R})$. Of course, this
special case is not interesting from the point of view of our
primary model for an infinitesimal rigid body moving in $S^{2}(0,
\mathbb{R})$ because it is there physically too exotic and too
exceptional.

For the potential systems with Lagrangians $L= T - V(q)$ the
Legendre transformation has the usual form:
\[
p_{i}=\frac{\partial L}{\partial\dot{q}^{i}}
=mG_{ij}(q)\dot{q}^{j}.
\]
Inverting it,
\begin{equation}\label{a97}
\dot{q}^{i}=\frac{1}{m}G^{ij}(q)p_{j},
\end{equation}
and substituting (\ref{a97}) to the expression for energy,
\[
E=\dot{q}^{i}\frac{\partial L}{\partial\dot{q}^{i}}-L=T+V(q),
\]
one obtains the Hamiltonian $H$, i.e.,
\[
H(q,p)=\mathcal{T}(q,p)+V(q)=\frac{1}{2m}G^{ij}(q)p_{i}p_{j}+
V(q),
\]
in particular, the geodetic Hamiltonian (when $V(q)=0$):
\[
\mathcal{T}=\frac{1}{2m}G^{ij}(q)p_{i}p_{j}.
\]
As a rule, the presence of forces, i.e., not constant $V(q)$,
reduces strongly the aforementioned symmetry of geodetic models.

We are interested here in integrability and hyperintegrability
(degeneracy) problems, thus, we concentrate our attention on the
Hamilton-Jacobi equation, i.e.,
\[
\frac{\partial S}{\partial t}+H\left(q^{i},\frac{\partial
S}{\partial q^{i}}\right)=0,
\]
its separability, and the action-angle variables. We are dealing
only with time-independent problems, thus, $S(t,q)$ is always
sought in the following form:
\[
S(t,q)=-Et+S_{0}(q),
\]
where, obviously, the reduced action $S_{0}$ satisfies the
time-independent Ha\-milton-Jacobi equation, i.e.,
\begin{equation}\label{a98}
H\left(q^{i},\frac{\partial S_{0}}{\partial q^{i}}\right)=E.
\end{equation}

Let us write down explicitly a few formulas describing Legendre
transformation and kinetic energy for the rigid body moving in
two-dimensional spherical world.

Denoting canonical momenta conjugate to coordinates
$(r,\varphi,\psi)$ respectively by $(p_{r},p_{\varphi},p_{\psi})$,
we have that
\begin{eqnarray}
p_{r} &=& m\dot{r},\nonumber\\
p_{\varphi} &=& mR^{2}\left(\sin^{2}\frac{r}{R}+\frac{I}{mR^{2}}
\cos^{2}\frac{r}{R}\right)\dot{\varphi}+I\cos\frac{r}{R}\dot{\psi},\nonumber\\
p_{\psi} &=& I\dot{\psi}+I\cos\frac{r}{R}\dot{\varphi}.\nonumber
\end{eqnarray}
The resulting geodetic Hamiltonian has the following form:
\begin{eqnarray}
{\mathcal{T}} &=& \frac{p_{r}^{2}}{2m}+\frac{p_{\varphi}^{2}-2
p_{\varphi}p_{\psi} \cos \frac{r}{R}+ \left(\frac{mR^{2}}{I}
\sin^{2} \frac{r}{R}+ \cos^{2}
\frac{r}{R}\right)p_{\psi}^{2}}{2mR^{2}
\sin^{2}\frac{r}{R}} \nonumber \\
&=& \frac{p_{r}^{2}}{2m}+ \frac{p_{\varphi}^{2}-
2p_{\varphi}p_{\psi} \cos \frac{r}{R}+ p_{\psi}^{2}}{2mR^{2}
\sin^{2}\frac{r}{R}} + \frac{mR^{2}-I}{2mR^{2}I}p^{2}_{\psi}.
\nonumber
\end{eqnarray}
It is seen again how this expression simplifies in the special
case when $I=mR^{2}$.

Without gyroscopic degree of freedom, when $T$ is reduced to
$T_{\rm tr}$ given by the formula (\ref{a88}), $\mathcal{T}$ is
reduced to
\[
{\mathcal{T}}_{\rm tr} = \frac{p_{r}^{2}}{2m}+
\frac{p_{\varphi}^{2}}{2mR^{2}\sin^{2}\frac{r}{R}}.
\]
Then there exists the class of separable potentials
\[
V(r,\varphi)=V_{r}(r)+\frac{V_{\varphi}(\varphi)}{R^{2}\sin^{2}
\frac{r}{R}}.
\]
With rotational degree of freedom the metric tensor $G$ is not
diagonal in natural coordinates $(r,\varphi, \psi)$ and there is
no direct analogue of above integrable models. Of course, in
three-dimensional Riemann manifolds there are always orthogonal
coordinates, but it is fairly not obvious whether there exist
diagonalizing coordinates admitting a physically reasonable class
of potentials treatable in terms of the separation-of-variables
method.

It is seen however that $\varphi, \psi$ are cyclic variables in
the kinetic energy term; this focuses our attention on the models
where the potential energy also does not depend on $\varphi,
\psi$. These angles are then cyclic variables for the total
Lagrangian $L=T-V$ and the corresponding conjugate momenta
$p_{\varphi}, p_{\psi}$ are constants of motion. The resulting
models, including geodetic ones $(V=0)$, are completely integrable
and one can analyze them by means of the separation-of-variables
method.

As usual when dealing with cyclic variables, we seek the reduced
action $S_{0}$ in the following form:
\begin{equation}\label{a100}
S_{0}(r,\varphi,\psi;E,\ell,s)=S_{r}(r;E)+\ell\varphi+s\psi,
\end{equation}
where $\ell$, $s$ are integration constants, i.e., the dependence
of $S_{\varphi}(\varphi)$ and $S_{\psi}(\psi)$ on their arguments
is postulated as linear. Together we have three integration
constants $E, \ell, s$, just as it should be in a complete
integral for the system with three degrees of freedom. As a matter
of fact, due to the assumed symmetry, the problem reduces to the
one-dimensional one for $S_{r}$ and substituting (\ref{a100}) into
(\ref{a98}), we obtain the ordinary differential equation
\begin{eqnarray}
\left(\frac{dS_{r}}{dr}\right)^{2} &=& 2m \left(E-V(r)\right)\nonumber\\
&&- \frac{\ell^{2}-2\ell
s\cos\frac{r}{R}+s^{2}\left(\frac{mR^{2}}{I}\sin^{2}\frac{r}{R}+
\cos^{2}\frac{r}{R}\right)}{R^{2}\sin^{2}\frac{r}{R}}.\label{a101}
\end{eqnarray}
Therefore, $p_{r}=d S_{r}/dr$ equals plus-minus (depending on the
phase of motion) square root of the rigid-hand side of the
expression above. Obviously, this is well defined only in the
classically admissible region between the turning points. It is
clear that
\[
p_{\varphi}= \frac{\partial S_{0}}{\partial \varphi}= \frac{d
S_{\varphi}}{d \varphi}= \ell, \qquad p_{\psi}= \frac{\partial
S_{0}}{\partial \psi}= \frac{d S_{\psi}}{d \psi}= s.
\]
The corresponding action variables are given as follows:
\begin{eqnarray}
J_{\varphi} &=& \oint p_{\varphi} d\varphi = \int\limits_{0}^{2
\pi} \ell d \varphi = 2 \pi \ell, \nonumber \\
J_{\psi} &=& \oint p_{\psi} d\psi = \int\limits_{0}^{2 \pi} s d
\psi = 2 \pi s. \nonumber
\end{eqnarray}
The radial action variable
\[
J_{r}= \oint p_{r} dr
\]
equals the doubled integral of the square-rooted right-hand side
of (\ref{a101}) between the turning points, i.e., between nulls of
(\ref{a101}). Substituting there
\begin{equation}\label{b102}
\ell = \frac{J_{\varphi}}{2\pi},\qquad s=\frac{J_{\psi}}{2\pi},
\end{equation}
we obtain the expression
\begin{equation}\label{a102}
J_{r}=\oint\sqrt{2m\left(E-V(r)\right)-\frac{\left(J_{\varphi}-
J_{\psi}\cos\frac{r}{R}\right)^{2}}{4\pi^{2}R^{2}\sin^{2}\frac{r}{R}}
+\frac{mJ^{2}_{\psi}}{4\pi^{2}I}}dr.
\end{equation}
Let us remind that the above expression, i.e., the contour
integral of the differential one-form $p_{r}dr$ along the
corresponding orbit in the two-dimensional phase space of the $(r,
p_{r})$-variables, equals
\[
J_{r}=2\int\limits_{r_{\rm min}}^{r_{\rm max}}p_{r}(r)dr,
\]
where $r_{\rm min},r_{\rm max}$ denote respectively the left and
right turning points of the $r$-motion.

When some explicit form of $V(r)$ is assumed and substituted to
(\ref{a102}), then in principle the integral may be calculated and
one obtains the functional dependence of the quantity $J_{r}$ on
the integration constants $E, J_{\varphi}, J_{\psi}$, i.e.,
\[
J_{r}=J_{r}\left(E,J_{\varphi},J_{\psi}\right).
\]
Once calculated, this expression may be in principle solved with
respect to the energy $E$, i.e.,
\[
E= {\mathcal{H}}\left(J_{r},J_{\varphi},J_{\psi}\right).
\]
In this way the energy is expressed in terms of action variables.
Substituting this expression and (\ref{b102}) into (\ref{a100}),
we obtain the generating function of canonical transformation from
the original phase-space coordinates
$(r,\varphi,\psi;p_{r},p_{\varphi},p_{\psi})$ to the action-angle
variables
$(\Theta_{r},\Theta_{\varphi},\Theta_{\psi};J_{r},J_{\varphi},J_{\psi})$,
i.e.,
\[
S_{0}\left(r,\varphi,\psi;J_{r},J_{\varphi},J_{\psi}\right)=
S_{r}\left(r,{\mathcal{H}}\left(J_{r}, J_{\varphi},
J_{\psi}\right)\right)+ \frac{\varphi}{2 \pi} J_{\varphi}+
\frac{\psi}{2 \pi} J_{\psi}.
\]
The resulting angle quantities are given as follows:
\[
\Theta_{r}=\frac{\partial S_{0}}{\partial J_{r}}, \qquad
\Theta_{\varphi}= \frac{\partial S_{0}}{\partial J_{\varphi}},
\qquad \Theta_{\psi}=\frac{\partial S_{0}}{\partial J_{\psi}}.
\]
This ``angles" are meant modulo $1$ but perhaps more intuitive are
angles taken modulo $2 \pi$, i.e.,
\[
\widetilde{\Theta}_{r}= 2 \pi \Theta_{r},\qquad
\widetilde{\Theta}_{\varphi}= 2 \pi \Theta_{\varphi},\qquad
\widetilde{\Theta}_{\psi}= 2 \pi \Theta_{\psi}.
\]
Let us observe that usually
\[
\widetilde{\Theta}_{\varphi} \neq \varphi, \qquad
\widetilde{\Theta}_{\psi} \neq \psi.
\]

The fundamental frequencies are given as follows:
\[
\nu_{r}= \frac{\partial {\mathcal{H}}}{\partial J_{r}}, \qquad
\nu_{\varphi}= \frac{\partial {\mathcal{H}}}{\partial
J_{\varphi}}, \qquad \nu_{\psi}= \frac{\partial
{\mathcal{H}}}{\partial J_{\psi}},
\]
and their circular counterparts are the corresponding
$2\pi$-multipliers, i.e.,
\[
\omega_{r}= 2 \pi \nu_{r}, \qquad \omega_{\varphi}= 2 \pi
\nu_{\varphi}, \qquad \omega_{\psi}= 2 \pi \nu_{\psi}.
\]

As usual, the action variables are constants of motion, i.e.,
\[
\frac{d J_{r}}{dt}= \frac{\partial {\mathcal{H}}}{\partial
\Theta_{r}}=0, \qquad \frac{d J_{\varphi}}{dt}= \frac{\partial
{\mathcal{H}}}{\partial \Theta_{\varphi}}=0, \qquad \frac{d
J_{\psi}}{dt}= \frac{\partial {\mathcal{H}}}{\partial
\Theta_{\psi}}=0.
\]
And as usual, the angular variables perform uniform ``librations"
because
\[
\frac{d \Theta^{i}}{dt} = \frac{\partial {\mathcal{H}}}{\partial
J_{i}}= \nu^{i}(J)= {\rm const},
\]
thus,
\[
\Theta^{i}(t)= \nu^{i}(J)t + \alpha^{i},\qquad
\widetilde{\Theta}^{i}(t)= \omega^{i}(J)t + \beta^{i},
\]
and, obviously, the index $i$ runs over the labels $r, \varphi,
\psi$.

A very important point is hyperintegrability, i.e., degeneracy.
Let us remind that a multiply periodic system with $n$ degrees of
freedom (an integrable one with an open set of bounded
trajectories) is said to be $k$-fold degenerate (or $(n-k)$-fold
periodic) when fundamental frequencies satisfy a system of $k$
(but not more) independent equations, i.e.,
\[
n^{\alpha}{}_{i} \nu^{i}(J)=0, \qquad \alpha = \overline{1,k},
\]
where $n^{\alpha}{}_{i}$ are integers, i.e.,
$n^{\alpha}{}_{i}\in\mathbb{Z}$ (but nothing changes when we say
that they are rationals, i.e., $n^{\alpha}{}_{i} \in \mathbb{Q}$).
By ``independent equations" we mean obviously that ${\rm
Rank}\left[n^{\alpha}{}_{i}\right]=k$. Integers $n^{\alpha}{}_{i}$
are to be constant, independent of $J$, thus, the same for all
possible frequencies. If the above equations hold exceptionally
for certain exclusive values of $J$, one says about accidental
degeneracy.

Of course the $n$-dimensional manifolds $J_{i}={\rm const}$ are
tori. While $\left\{J_{i},J_{k}\right\}=0$, i.e., the action
variables are in Poisson-bracket involution, these tori are
Lagrangian submanifolds, i.e., the symplectic two-form vanishes
when evaluated on any pair of their tangent vectors at any point,
and they are maximal connected manifolds of this property. If the
system is $k$-fold degenerate ($(n-k)$-fold periodic), then the
closure of every trajectory is an $(n-k)$-dimensional isotropic
torus (the symplectic form vanishes when evaluated at any pair of
tangent vectors attached at any point). The system orbits are
dense in these tori. The mentioned $(n-k)$-tori form a regular
foliation (congruence) of partially disjoint submanifolds on the
Lagrangian tori $J_{a}={\rm const}$, $a=\overline{1,n}$. For any
$k$-fold degenerate system Hamiltonian depends on the action
variables $J_{1},\ldots,J_{n}$ in a very peculiar way, namely, it
is a function of some $(n-k)$ linear combination of
$J_{1},\ldots,J_{n}$ with integer coefficients (and of no less
number of such combinations). The extreme examples are as follows:
\begin{enumerate}
\item[$(i)$] if $k=0$, then the system is completely
nondegenerate, any orbit fills densely some torus $J_{a}={\rm
const}$, $a=\overline{1,n}$, and $H$ depends on all
$J_{a}$-variables in an essential way, i.e., there is no
possibility to superpose them into a smaller numbers of
combinations with integer coefficients,

\item[$(ii)$] if $k=n-1$, then all bounded trajectories are
periodic, i.e., they are topological circles (one-dimensional
tori) regularly foliating the tori $J_{a}={\rm const}$,
$a=\overline{1,n}$, so that the quotient manifolds are
$(n-1)$-dimensional tori. Then there exists a combination of
$J_{a}$-variables with integer coefficients, i.e., $J=
n^{a}J_{a}$, $n^{a} \in \mathbb{Z}$, such that $H$ depends on all
$J_{a}$-s only through $J$.
\end{enumerate}
The first special case is completely not resonant and has $n$
independent frequencies (periods). The second, quite opposite,
case is maximally resonant, i.e., there is only one fundamental
frequency (period) and trajectories are known from the elementary
mechanics as Lissajous figures in the original configuration
space.

If the system is $k$-fold degenerate, then the original
action-angle variables obtained from the space-time coordinates
$q^{i},p_{i}$, $i=\overline{1,n}$, may be replaced by some new
ones, better expressing the degeneracy (hyperintegrability)
structure. Namely, it is always possible to introduce some
quantities $\widetilde{\Theta}^{a},\widetilde{J}_{a}$ such that
\begin{enumerate}
\item[$(i)$] $J_{a}=\widetilde{J}_{b}N^{b}{}_{a}$,
$\widetilde{\Theta}^{a}=N^{a}{}_{b}\Theta^{b}$,

\item[$(ii)$] $N^{a}{}_{b} \in \mathbb{Z}$, i.e., entries of the
matrix $N$ are integers,

\item[$(iii)$] $\det\left[N^{a}{}_{b}\right]= \pm 1$,

\item[$(iv)$] $H=\mathcal{H}\left(\widetilde{J}_{1},
\ldots,\widetilde{J}_{n-k}\right)$, i.e., Hamiltonian depends only
on the first $(n-k)$-tuple of the action variables,

\item[$(v)$] the first $(n-k)$ basic frequencies
$\widetilde{\nu}^{r}$, $r=\overline{1,(n-k)}$, are independent
over integers, i.e., the equations
\[
\sum_{r=1}^{n-k} n^{s}{}_{r} \widetilde{\nu}^{r}= \sum_{r=1}^{n-k}
n^{s}{}_{r} \frac{\partial {\mathcal{H}}}{\partial
\widetilde{J}_{r}}
\]
imply that the $(n-k) \times (n-k)$ matrix $n^{s}{}_{r}$ vanishes,
i.e., $\mathbb{Z} \ni n^{s}{}_{r}=0$.
\end{enumerate}
Obviously, the remaining $k$ frequencies, i.e.,
$\widetilde{\nu}^{p}=\partial\mathcal{H}/\partial J_{p}=0$ with
$p=\overline{(n-k+1),n}$, vanish and the corresponding new action
variables $\widetilde{\Theta}^{p}$, $p=\overline{(n-k+1),n}$, are
constants of motion. Together we have $(n+k)$ constants of motion,
global and smooth ones:
\[
\widetilde{J}_{1},\ldots,\widetilde{J}_{n},
\widetilde{\Theta}^{n-k+1},\ldots,\widetilde{\Theta}^{n}.
\]
What concerns globality, according to the above conditions $(i)$
and $(ii)$, the quantities $\widetilde{\Theta}^{a}$,
$a=\overline{1,n}$, are ``good" angular variables (they would not
be so without these conditions). They are of course multivalued in
a ``harmless" way when property treated, just like the angular
variables. To avoid this multivaluedness, one can replace them by
unimodular complex numbers, i.e.,
\[
\zeta^{a}= \exp \left( i \Theta^{a}\right).
\]

Let us go back from these general digressions to our special
models, beginning from (\ref{a102}), thus, to $n=3$.

It happens often, e.g., in the case of symmetry in the material
point mechanics, that some partial features of the problem may be
seen from the formulas like (\ref{a102}) for $J_{r}$ even without
calculating the integral and without even assuming anything about
the shape of $V(r)$. For example, in central problems of the
material point motion, when the spherical coordinates $r,
\vartheta, \varphi$ are used, it is seen that the corresponding
action variables $J_{\vartheta}, J_{\varphi}$ enter the
under-square-root expression through the combination
$J_{\vartheta}+J_{\varphi}$, so the one-fold degeneracy is obvious
from the very beginning. The particular shapes of $V(r)$ are
necessary for answering the question concerning total degeneracy.
And then, according to what is known from somewhere else, really
one can prove after the explicit integration, that the total
degeneracy occurs for the Kepler-Coulumb (attractive) problem and
the isotropic harmonic oscillator, i.e.,
\[
V = -\frac{\alpha}{r}, \qquad V = \frac{\varkappa}{2}r^{2}, \qquad
\alpha > 0, \qquad \varkappa > 0.
\]
In the formula (\ref{a102}) nothing like this is seen, even for
the very special, highly-symmetric case $I=mr^{2}$, i.e.,
\begin{equation}\label{a108}
J_{r}= \oint \sqrt{ 2m \left(E-V(r)\right)-
\frac{J_{\varphi}^{2}-2 J_{\varphi}J_{\psi}\cos\frac{r}{R}+
J_{\psi}^{2}}{4 \pi^{2}R^{2}\sin^{2}\frac{r}{R}}}dr.
\end{equation}
Without performing some calculations, one does not see anything
even in the geodetic case, when $V=0$. But everybody knows that
the problem is then completely degenerate because it is
isomorphic, as mentioned, with the free spherical top, each orbits
of which are closed (one-parameter group and their cosets). But
without the explicit calculation nothing may be decided because
$J_{\varphi}$ and $J_{\psi}$ do not combine integer-wise under the
square root. Moreover, there are some unexpected strange technical
problems when calculating then the relationship between $E$ and
$J$-variables. This is probably the reason one can hardly find the
action-angle analysis in the rigid body mechanics. Passing to the
isomorphic problem for the rigid body we obtain that
\begin{equation}\label{b108}
J_{\vartheta}=\oint \sqrt{ 2I \left(E-V(\vartheta)\right)-
\frac{J_{\varphi}^{2}-2 J_{\varphi}J_{\psi}\cos \vartheta +
J_{\psi}^{2}}{4 \pi^{2}R^{2} \sin^{2}\vartheta}}d\vartheta,
\end{equation}
where $\vartheta=r/R$. Substituting here $x=-\cos \vartheta$, we
obtain for the geodetic case that
\begin{equation}\label{a109}
J_{\vartheta}=-\oint \sqrt{-2IE x^{2}-
\frac{J_{\varphi}J_{\psi}}{2\pi^{2}}x+2IE
-\frac{J_{\varphi}^{2}+J_{\psi}^{2}}{4\pi^{2}}}
\frac{dx}{(x-1)(x+1)}.
\end{equation}
This integral may be calculated according to the usual rule for
integrals of the form
\[
\mathcal{R} \left( \sqrt{ax^{2}+bx+c}, x \right),
\]
where $\mathcal{R}$ denotes a rational function of indicated
expression. Of course, the corresponding indefinite integral is
elementary one and can be calculated, then the resulting definite
integral (the doubled value between the turning points) may be in
principle obtained in this way. There is however plenty of
possibilities of making mistakes and it is much more convenient to
use here the method of complex integration, as elaborated by Max
Born in analytical mechanics and the old quantum theory. Replacing
$x$ by the complex variable $z$, we see that there are exactly two
branch points, i.e., just the classical turning points. Therefore,
the action variable $J_{\vartheta}$ $(J_{r}, J_{x})$ may be
calculated as the contour integral along the path infinitesimally
surrounding the cut joining the branch point along the real axis.
There are also three poles, i.e., $z=-1$, $z=+1$, $z= \infty$.
Surrounding these poles by additional contours and taking the
composed unconnected contour consisting of the above four ones, we
finally conclude that
\[
J_{\vartheta}= - 2 \pi i {\rm Res}_{-1} - 2 \pi i {\rm Res}_{1} -
2 \pi i {\rm Res}_{\infty},
\]
where, obviously, residua are calculated for the total integrand.
After some calculations one obtains that
\begin{eqnarray}
{\rm Res}_{-1}&=&-\frac{\left|J_{\varphi}-J_{\psi}\right|}{4\pi}i,\nonumber \\
{\rm Res}_{1}&=&-\frac{\left|J_{\varphi}+J_{\psi}\right|}{4\pi}i, \nonumber \\
{\rm Res}_{\infty}&=&\sqrt{2IE}i.\nonumber
\end{eqnarray}
Thus, finally,
\[
4 \pi \sqrt{2IE} = 2 J_{\vartheta}+
\left|J_{\varphi}-J_{\psi}\right| +
\left|J_{\varphi}+J_{\psi}\right|.
\]
Only now, after all calculations, the degeneracy is seen on the
level of action variables, although it was obvious from the very
beginning from the spherical top analogy. There are however some
delicate points, namely, the structure of this total degeneracy is
a little bit different in the four regions of the phase space:
\begin{enumerate}
\item[$(i)$] $\left(J_{\varphi}>J_{\psi}\right)\wedge\left(
J_{\varphi}>-J_{\psi}\right)$, i.e.,
$J_{\varphi}>\left|J_{\psi}\right|$, then opening the absolute
value sign we obtain that
\[
E=\frac{\left(J_{\vartheta} + J_{\varphi}\right)^{2}}{8 \pi^{2}I},
\]

\item[$(ii)$] $\left(J_{\varphi}<J_{\psi}\right)\wedge\left(
J_{\varphi}>-J_{\psi}\right)$, i.e.,
$\left|J_{\varphi}\right|<J_{\psi}$, then
\[
E = \frac{\left(J_{\vartheta} + J_{\psi}\right)^{2}}{8 \pi^{2}I},
\]

\item[$(iii)$] $\left(J_{\varphi}>J_{\psi}\right)\wedge\left(
J_{\varphi}<-J_{\psi}\right)$, i.e.,
$\left|J_{\varphi}\right|<\left|J_{\psi}\right|=-J_{\psi}$, then
\[
E = \frac{\left(J_{\vartheta} - J_{\psi}\right)^{2}}{8 \pi^{2}I},
\]

\item[$(iv)$] $\left(J_{\varphi}<J_{\psi}\right)\wedge\left(
J_{\varphi}<-J_{\psi}\right)$, i.e.,
$J_{\varphi}<-\left|J_{\psi}\right|$, then
\[
E = \frac{\left(J_{\vartheta} - J_{\varphi}\right)^{2}}{8
\pi^{2}I}.
\]
\end{enumerate}
It is seen that there are four open regions of the phase space in
which the formula looks slightly different. Nevertheless, the
difference is not very essential. In any of these region one can
introduce such action variables that only one of them is essential
for the energy, i.e.,
\[
E= \frac{J^{2}}{8 \pi^{2}I}.
\]
On the level of the old quantum theory according to the
Bohr-Sommerfeld postulates we have that
\[
J=nh, \qquad n \in \mathbb{Z},
\]
and we obtain that
\[
E_{n}= \frac{n^{2}\hbar^{2}}{2I},
\]
i.e., just as it should be.

By the way, it is seen from the formulas (\ref{a102}),
(\ref{a108}), (\ref{b108}), (\ref{a109}) that the above class of
integrable problems contains also interesting non-geodetic models.
Obviously, this is the case when the structure of $V$ is
appropriately sited to other (geodetic) terms occurring under the
square root sign. In any case, it is seen from (\ref{a109}) that
potentials of the type
\[
V(x)= \alpha x^{2} + \beta x, \qquad \alpha,\beta={\rm const}
\]
do not violate the explicit solvability in terms of elementary
integrals. If $\beta \neq 0$, then they also do not violate the
total degeneracy because the term with the constant $\alpha$ will
be simply absorbed by $-2IE$ and will regauge it. In other words,
in (\ref{b108}) we can simply use that
\[
V(\vartheta) = \widetilde{\alpha} \cos^{2} \vartheta +
\widetilde{\beta} \cos \vartheta,
\]
i.e.,
\[
V(r) = \widehat{\alpha} \cos^{2} \frac{r}{R} + \widehat{\beta}
\cos\frac{r}{R},
\]
with the same consequences concerning the explicit solvability and
degeneracy.

For the general case (\ref{a102}) the same is true concerning the
explicit solvability, however, as a rule, the degeneracy will be
removed there even for the special case $\beta=0$ (i.e.,
$\widetilde{\beta}=0$, $\widehat{\beta}=0$).

Let us now go back to infinitesimal affinely-rigid bodies, thus,
in addition to the gyroscopic degrees of freedom the deformative
ones are taken into account. As mentioned, when the polar
decomposition is used, then for the kinetic energy one obtains the
formula (\ref{a83}) with $\widehat{\omega}$ given by the
expressions (\ref{c83}) and (\ref{b83}). Just as in the gyroscopic
case we concentrate on the phenomena in the two-dimensional
spherical world. We use the same as previously parametrization of
this world, i.e., $(r,\varphi)$- or
$(\vartheta,\varphi)$-coordinates. What concerns internal degrees
of freedom we use coordinates $\psi$, $\xi$, $\eta$, $\zeta$,
where
\[
U(\psi)= \left[
\begin{array}{cc}
  \cos \psi & -\sin \psi \\
  \sin \psi & \cos \psi \\
\end{array}
\right], \qquad A(\xi, \eta, \zeta)= \left[
\begin{array}{cc}
  \xi & \eta \\
  \eta & \zeta \\
\end{array}
\right],
\]
and $U$, $A$ are factors of the polar decomposition.

After some calculations we obtain that
\begin{eqnarray}
T&=&T_{\rm tr}+T_{\rm int}=\frac{m}{2}
\left(\left(\frac{dr}{dt}\right)^{2}+\sin^{2}\frac{r}{R}
\left(\frac{d \varphi}{dt}\right)^{2}\right) \nonumber\\
&&+ \frac{1}{2} \left(J_{1}\xi^{2}+
\left(J_{1}+J_{2}\right)\eta^{2}+J_{2}\zeta^{2}\right)\omega^{2}
\nonumber\\
&&+ \left(-J_{1}\eta \frac{d\xi}{dt}+ \left(J_{1}\xi - J_{2}\eta
\right) \frac{d\eta}{dt} + J_{2}\eta \frac{d\zeta}{dt}\right)
\omega\nonumber \\
&&+ \frac{1}{2} \left(J_{1}\left(\frac{d\xi}{dt}\right)^{2} +
\left(J_{1}+J_{2}\right)\left(\frac{d\eta}{dt}\right)^{2}+J_{2}
\left(\frac{d\zeta}{dt}\right)^{2}\right),\nonumber
\end{eqnarray}
where the micromaterial coordinates are chosen (always possible)
in such a way that the inertial quadrupole is diagonal, i.e.,
\[
J= \left[
\begin{array}{cc}
  J_{1} & 0 \\
  0 & J_{2} \\
\end{array}
\right].
\]
Obviously, $\omega$ is given by (\ref{a91}). The first term of
$T_{\rm int}$ above, i.e., the first term of (\ref{a83}) when $n$
is specified to $2$, is the centrifugal contribution; the
coefficient at $\omega^{2}/2$ is the dynamical
deformation-dependent inertial momentum of the $U$-rigid body (the
material tensor $AJA$ in (\ref{a83})). The second term is the
Coriollis contribution to $T_{\rm int}$, i.e., the coupling
between angular and deformation velocities (${\rm
Tr}\left(AJ(dA/dt)\widehat{\omega}\right)$ in (\ref{a83})). And
the third term of $T_{\rm int}$ is the kinetic energy of pure
deformations (${\rm Tr}\left(J(dA/dt)^{2}\right)/2$ in
(\ref{a83})).

It is seen that even the purely kinetic term is rather complicated
and there is no hope for integrability (or any kind of rigorous
solvability) even if the potential term has a simple structure
adapted to that of $T$. Obviously, the most natural potentials are
those given by scalar invariants of the Green deformation tensor,
i.e., analytically
\[
G = \left[
\begin{array}{cc}
  \xi^{2}+\eta^{2} & \eta(\xi + \zeta) \\
  \eta(\xi + \zeta) & \xi^{2}+\eta^{2} \\
\end{array}
\right].
\]
Obviously, in the case of some practical motivation some
approximation or numerical procedures are possible, however, for
the general $J$ there is no hope for realistic potentials
admitting integrability or some kind of analytical procedures.

As usual, this is possible however for highly symmetric systems,
when the internal inertia is isotropic, i.e.,
\[
J^{AB}=I\eta^{AB}=I\delta^{AB}.
\]
This is explicitly seen when one uses the two-polar decomposition
for internal degrees of freedom.

For the general $n$ we have the expression (\ref{a84}) with
$\widehat{\chi}$, $\widehat{\vartheta}$ given by the expressions
(\ref{c84}), (\ref{b84}), (\ref{d84}). And there is also no hope
for integrability or any kind of explicit analytical solvability.
But again the special case $n=2$, i.e., two-dimensional spherical
world, is an exception.

Parametrization of the $S^{2}(0,R)$-world is again the same, i.e.,
$(r,\varphi)$- or $(\vartheta,\varphi)$-variables. When expressed
in terms of the two-polar decomposition, i.e.,
\[
\varphi = LDR^{-1} \in {\rm GL}(2, \mathbb{R}),
\]
then $\varphi$ is parameterized by generalized coordinates
$\alpha, \beta, \lambda, \mu$, where
\[
L(\alpha)= \left[
\begin{array}{cc}
  \cos \alpha & -\sin \alpha \\
  \sin \alpha & \cos \alpha \\
\end{array}
\right], \qquad R(\beta)= \left[
\begin{array}{cc}
  \cos \beta & -\sin \beta \\
  \sin \beta & \cos \beta \\
\end{array}
\right],
\]
\[
D(\lambda, \mu)= \left[
\begin{array}{cc}
  \lambda & 0 \\
  0 & \mu \\
\end{array}
\right].
\]
In a partial analogy to (\ref{a91}) we obtain that
\[
\widehat{\chi}= L^{-1}\frac{dL}{dt} = \chi \left[
\begin{array}{cc}
  0 & -1 \\
  1 & 0 \\
\end{array}
\right], \qquad \widehat{\vartheta}= R^{-1}\frac{dR}{dt}=
\vartheta \left[
\begin{array}{cc}
  0 & -1 \\
  1 & 0 \\
\end{array}
\right],
\]
where
\[
\chi = \chi_{\rm rl}+ \chi_{\rm dr}= \frac{d\alpha}{dt}+ \cos
\frac{r}{R}\frac{d\varphi}{dt}= \frac{d\alpha}{dt}+ \cos \vartheta
\frac{d\varphi}{dt},
\]
but $\vartheta$ has no ``drive" term, i.e.,
\[
\vartheta = \frac{d \beta}{dt}.
\]
{\bf Remark:} do not confuse this $\vartheta$ with the angle
$\vartheta$.

It turns out that to avoid some embarrassing cross-terms it is
convenient to introduce the ``mixed" coordinates
\[
x:=\frac{1}{\sqrt{2}}(\lambda - \mu), \qquad
y:=\frac{1}{\sqrt{2}}(\lambda + \mu), \qquad \gamma:= \alpha +
\beta, \qquad \delta:=\alpha - \beta.
\]
The inverse rules read that
\[
\lambda= \frac{1}{\sqrt{2}}(x+y), \qquad \mu=
\frac{1}{\sqrt{2}}(y-x), \qquad \alpha= \frac{1}{2}(\gamma +
\delta), \qquad \beta= \frac{1}{2}(\gamma - \delta).
\]
The canonical momenta satisfy the contragradient rules
\begin{eqnarray}
p_{x}=\frac{1}{\sqrt{2}}\left(p_{\lambda}-p_{\mu}\right),&\qquad&
p_{y}=\frac{1}{\sqrt{2}}\left(p_{\lambda}+p_{\mu}\right),\nonumber \\
p_{\gamma}= \frac{1}{2}\left(p_{\alpha}+p_{\beta}\right),&\qquad&
p_{\delta}= \frac{1}{2}\left(p_{\alpha}-p_{\beta}\right),\nonumber
\end{eqnarray}
and conversely,
\begin{eqnarray}
p_{\lambda}=\frac{1}{\sqrt{2}}\left(p_{x}+p_{y}\right),&\qquad&
p_{\mu}= \frac{1}{\sqrt{2}}\left(p_{y}-p_{x}\right),\nonumber\\
p_{\alpha}=p_{\gamma}+p_{\delta},&\qquad&
p_{\alpha}=p_{\gamma}-p_{\delta}.\nonumber
\end{eqnarray}
{\bf Remark}: it is well known that the two-polar decomposition is
not unique and has singular points \cite{all-book04}. Therefore,
for $n=2$ the $(\alpha,\beta,\lambda,\mu)$-parameteri\-zation is
not a diffeomorphism between GL$(2,\mathbb{R})$ and
$\mathbb{T}^{2} \times \mathbb{R}^{2}$, i.e., the Cartesian
product of the two-dimensional torus and the real plane. Rather
the representation manifold $\mathbb{T}^{2} \times \mathbb{R}^{2}$
is plagued by a complicated system of identifications
\cite{all-book04} and only the resulting quotient space may be
interpreted as a proper representation of our configuration space.
What concerns the quantities $\gamma, \delta$ the situation is
even worse. They are ``less proper" angular variables than
$\alpha,\beta$ themselves are. The reason is that the
integer-entries matrix transforming $(\alpha,\beta)$ into
$(\gamma,\delta)$ has the determinant $-2$, thus, different from
$\pm 1$. If we normalize it to the $\pm 1$ determinant, then its
entries will be no longer integer. Therefore, this matrix is not a
well-defined automorphism either of the torus $\mathbb{T}^{2}$ or
the lattice $\mathbb{Z}^{2}$. The strange status of variables
$\gamma,\delta$ is not misleading if it is not forgotten. And of
course, it is quite proper to say simply that to avoid some
cross-terms in the kinetic energy expression we replace the
angular velocities $\chi,\vartheta$ by their combinations, i.e.,
\[
a:= \chi + \vartheta, \qquad b:= \chi - \vartheta.
\]

Let us order our generalized coordinates $q^{i}$,
$i=\overline{1,6}$, as follows:
\[
r, \varphi, \gamma, \delta, x, y.
\]
Then the kinetic energy will be written as follows:
\[
T= \frac{m}{2} G_{ij}(q) \frac{dq^{i}}{dt} \frac{dq^{j}}{dt},
\]
where for the above ordering of variables the matrix
$\left[G_{ij}\right]$ of the metric tensor $G$ consists of three
blocks subsequently placed along the diagonal (looking from the
top to bottom):
\begin{enumerate}
\item[$(i)$] the $1 \times 1$ block $M_{1}$, i.e.,
\[
M_{1}= \left[1\right],
\]

\item[$(ii)$] the $3 \times 3$ block $M_{2}$ given as follows:
\[
M_{2}= \left[
\begin{array}{ccc}
  R^{2} \sin^{2}\frac{r}{R}+ \frac{I}{m}\left(x^{2}+y^{2}\right)
  \cos^{2}\frac{r}{R} &
  \frac{I}{m}x^{2}\cos \frac{r}{R} &
  \frac{I}{m}y^{2}\cos \frac{r}{R} \\
   & & \\
  \frac{I}{m}x^{2}\cos \frac{r}{R} & \frac{I}{m}x^{2} & 0 \\
  & & \\
  \frac{I}{m}y^{2}\cos \frac{r}{R} & 0 & \frac{I}{m}y^{2} \\
\end{array}
\right],
\]

\item[$(iii)$] the $2 \times 2$ isotropic block $M_{3}$, i.e.,
\[
M_{3}= \frac{I}{m}I_{2}= \left[
\begin{array}{cc}
  \frac{I}{m} & 0 \\
  0 & \frac{I}{m} \\
\end{array}
\right],
\]
where, obviously, $I_{2}$ denotes the $2 \times 2$ identity
matrix.
\end{enumerate}
It is seen that now, when deformative degrees of freedom are taken
into account, the previous special case $I=mR^{2}$ does not lead
to any essential simplification.

One can easily show that
\[
\det \left[G_{ij}\right]=
R^{2}\left(\frac{I}{m}\right)^{4}x^{2}y^{2}\sin^{2}\frac{r}{R},
\]
thus, the density of the Riemannian volume element equals
\[
\sqrt{\left|G\right|}=
R\left(\frac{I}{m}\right)^{2}\left|xy\right| \sin \frac{r}{R}.
\]
Explicitly, the block matrix $\left[G_{ij}\right]$ is given as
follows:
\[
\left[G_{ij}\right]= \left[
\begin{array}{ccc}
  M_{1} &  &  \\
   & M_{2} &  \\
   &  & M_{3} \\
\end{array}
\right].
\]
The inverse contravariant tensor matrix $\left[G^{ij}\right]$ is
obviously given by
\[
\left[G^{ij}\right]= \left[
\begin{array}{ccc}
  M_{1}^{-1} &  &  \\
   & M_{2}^{-1} &  \\
   &  & M_{3}^{-1} \\
\end{array}
\right],
\]
where the inverse blocks have the forms:
\begin{enumerate}
\item[$(i)$] $M_{1}^{-1}=\left[1\right],$

\item[$(ii)$] $M_{2}^{-1}= \left[
\begin{array}{ccc}
  \frac{1}{R^{2}\sin^{2}\frac{r}{R}} &
  -\frac{\cos \frac{r}{R}}{R^{2}\sin^{2}\frac{r}{R}} &
  -\frac{\cos \frac{r}{R}}{R^{2}\sin^{2}\frac{r}{R}} \\
  &&\\
  -\frac{\cos \frac{r}{R}}{R^{2}\sin^{2}\frac{r}{R}} &
  \frac{m}{I}\frac{1}{x^{2}}+\frac{1}{R^{2}}{\rm ctg}^{2}\frac{r}{R}
  & \frac{1}{R^{2}}{\rm ctg}^{2}\frac{r}{R} \\
  &&\\
  -\frac{\cos \frac{r}{R}}{R^{2}\sin^{2}\frac{r}{R}} &
  \frac{1}{R^{2}}{\rm ctg}^{2}\frac{r}{R} &
  \frac{m}{I}\frac{1}{y^{2}}+\frac{1}{R^{2}}{\rm
  ctg}^{2}\frac{r}{R}
\end{array}
\right],$

\item[$(iii)$] $M_{3}^{-1}= \frac{m}{I} I_{2} = \left[
\begin{array}{cc}
  \frac{m}{I} & 0 \\
  0 & \frac{m}{I} \\
\end{array}
\right].$
\end{enumerate}

For potential systems with Lagrangians of the form
\[
L=T-V(q),
\]
the corresponding kinetic (geodetic) Hamiltonian equals
\[
{\mathcal{T}}= \frac{1}{2m}G^{ij}(q)p_{i}p_{j}
\]
and the full Hamiltonian is as follows:
\[
H=\mathcal{T}+V(q).
\]
According to our convention of ordering coordinates $q^{i}$,
$i=\overline{1,6}$, i.e.,
\[
r, \varphi, \gamma, \delta, x, y,
\]
the corresponding conjugate momenta $p_{i}$, $i=\overline{1,6}$,
are denoted and ordered as follows:
\[
p_{r}, p_{\varphi}, p_{\gamma}, p_{\delta}, p_{x}, p_{y}.
\]
In certain expressions it is convenient to use the original
momenta $p_{\alpha}$, $p_{\beta}$, $p_{\lambda}$, $p_{\mu}$. First
of all this concerns $p_{\alpha}, p_{\beta}$ because of their
geometrical interpretation respectively as spin and vorticity
\cite{all-book04}.

For the above potential systems the reduced Hamilton-Jacobi
equation has the form
\[
\frac{1}{2m} G^{ij}(q) \frac{\partial S_{0}}{\partial q^{i}}
\frac{\partial S_{0}}{\partial q^{j}} + V(q) = E.
\]
We do not write it down explicitly because of the complicated and
rather obscure form of the resulting expression. The above block
form is much more readable and lucid.

Just as in the gyroscopic case we are dealing here with
non-orthogonal coordinates (the $3 \times 3$ block $M_{2}$) and it
is not clear for us whether in some hypothetic orthonormal
coordinates (they exist, of course) the system is separable. It is
perhaps a little surprising that our kinetic Hamiltonian
$\mathcal{T}$ has the separable structure. Of course, for the
system with deformative degrees of freedom as above, the geodetic
model is not physical because it admits unlimited expansion and
contraction. Therefore, some potential must be assumed and this is
just the problem, i.e., we could not determine a wide class of
potentials compatible with the separability in our non-orthogonal,
but nevertheless natural, coordinates. Just as in the gyroscopic
case we restrict ourselves to some special class of potentials,
assuming in particular that all angles $\varphi, \alpha, \beta$
(equivalently $\varphi, \gamma, \delta$) are cyclic variables.
Therefore, the reduced action $S_{0}$ is sought as a function
linear in angular variables and separable, i.e.,
\begin{eqnarray}
S_{0}(q)&=& S_{r}(r)+ S_{x}(x)+S_{y}(y)+ \ell \varphi
+C_{\gamma}\gamma +C_{\delta}\delta \nonumber \\
&=& S_{r}(r)+ S_{x}(x)+S_{y}(y)+ \ell \varphi +C_{\alpha}\alpha
+C_{\beta}\beta,
\end{eqnarray}
where $\ell$, $C_{\gamma}$, $C_{\delta}$, $C_{\alpha}$,
$C_{\beta}$ are constants. The relationship between $(\gamma,
\delta)$ and $(\alpha,\beta)$ implies that
\[
C_{\alpha}= C_{\gamma}+C_{\delta}, \qquad
C_{\beta}=C_{\gamma}-C_{\delta},
\]
i.e.,
\[
C_{\gamma}= \frac{1}{2}\left(C_{\alpha}+C_{\beta}\right), \qquad
C_{\delta}= \frac{1}{2}\left(C_{\alpha}-C_{\beta}\right).
\]
As $p_{\alpha}, p_{\beta}$ corresponding respectively to the spin
and vorticity in the flat-space theory \cite{all-book04}, we shall
also denote the constants $C_{\alpha}$, $C_{\beta}$ as $s$, $j$,
thus,
\[
S_{0}(q)= S_{r}(r)+ S_{x}(x)+S_{y}(y)+ \ell \varphi+ s \alpha +j
\beta.
\]

Let us now quote the expressions for the action variables, i.e.,
\begin{eqnarray}
J_{\alpha}&=& \oint p_{\alpha} d\alpha = C_{\alpha}
\int\limits_{0}^{2\pi} d\alpha = 2\pi C_{\alpha}=2\pi s,\nonumber \\
J_{\beta}&=& \oint p_{\beta} d\beta = C_{\beta} \int\limits_{0}^{2
\pi} d\beta = 2\pi C_{\beta}=2\pi j,\nonumber \\
J_{\varphi}&=& \oint p_{\varphi} d\varphi = \ell
\int\limits_{0}^{2 \pi} d\varphi = 2\pi \ell.\nonumber
\end{eqnarray}
Let us assume that the potential energy separates explicitly with
respect to acyclic variables, i.e.,
\[
V(r, x, y)= V_{r}(r)+V_{x}(x)+V_{y}(y).
\]
Then the resulting Hamilton-Jacobi equation also separates and
denoting the corresponding separation constants by $C_{r}$,
$C_{x}$, $C_{y}$ we have that
\[
C_{r}+C_{x}+C_{y}=E,
\]
and then
\begin{eqnarray}
\frac{1}{2m} \left(\frac{dS_{r}}{dr}\right)^{2}+
\frac{\left(J_{\varphi}-J_{\alpha}\cos\frac{r}{R}
\right)^{2}}{8\pi^{2}mR^{2}\sin^{2}\frac{r}{R}}+
V_{r}(r)&=&E-C_{x}-C_{y},\label{a122}\\
\frac{1}{2I} \left(\frac{dS_{x}}{dx}\right)^{2}+
\frac{\left(J_{\alpha}+J_{\beta} \right)^{2}}{32 \pi^{2}Ix^{2}}+
V_{x}(x)&=&C_{x},\label{b122}\\
\frac{1}{2I} \left(\frac{dS_{y}}{dy}\right)^{2}+
\frac{\left(J_{\alpha}-J_{\beta} \right)^{2}}{32 \pi^{2}Iy^{2}}+
V_{y}(y)&=&C_{y}.\label{a123}
\end{eqnarray}
Therefore, after expressing $C_{\alpha}=s$, $C_{\beta}=j$,
$C_{\varphi}=\ell$ with the help of $J_{\alpha}$, $J_{\beta}$,
$J_{\varphi}$, we obtain for the remaining action variables the
expressions
\begin{eqnarray}
J_{x}&=& \oint p_{x}dx=\oint
\sqrt{2I\left(C_{x}-V_{x}(x)\right)-\frac{\left(J_{\alpha}+J_{\beta}
\right)^{2}}{16 \pi^{2}x^{2}}}dx, \label{b123}\\
J_{y}&=& \oint p_{y}dy = \oint
\sqrt{2I\left(C_{y}-V_{y}(y)\right)-\frac{\left(J_{\alpha}-J_{\beta}
\right)^{2}}{16 \pi^{2}y^{2}}}dy, \label{c123}\\
J_{r}&=& \oint p_{r}dr\label{d123}\\
&=&\oint\sqrt{2m\left(E-C_{x}-C_{y}-V_{r}(r)\right)-\frac{\left(J_{\varphi}-
J_{\alpha}\cos\frac{r}{R} \right)^{2}}{4
\pi^{2}R^{2}\sin^{2}\frac{r}{R}}}dr.\nonumber
\end{eqnarray}
Due to the cyclic character of $\varphi$, $\alpha$, $\beta$, the
true dynamics is contained in (\ref{a122}), (\ref{b122}),
(\ref{a123}) or on the level of action variables in (\ref{b123}),
(\ref{c123}), (\ref{d123}). Obviously, the microdeformation
dynamics is described by (\ref{b122}), (\ref{a123}) or
equivalently by (\ref{c123}), (\ref{d123}). Unlike this,
(\ref{a122}) and (\ref{d123}) describe the dynamical influence of
the ``North Pole" ($r=0$) and ``South Pole" ($r=\pi R$) on the
translational motion. The dynamics of deformation invariants $x,
y$ is in many respects similar to the one in \cite{JJS82_2} (the
combinations $\left(D_{1}-D_{2}\right)/\sqrt{2}$ and
$\left(D_{1}+D_{2}\right)/\sqrt{2}$ in \cite{JJS82_2}). The
$r$-geodetic model with $V_{r}=0$ is obviously well formulated.
But in d'Alembert models the $(x,y)$-geodetic case ($V_{x}=0$,
$V_{y}=0$) would be quite not physical because of admitting
unlimited expansion and contraction of the body. This is not the
case in affine models where the ``elastic vibrations" may be
encoded in the very kinetic energy.

The procedure is now as follows. We assume some particular forms
of the deformative potentials $V_{x}$, $V_{y}$ basing both on some
physical arguments (taken, e.g., from elasticity theory) and on
the possibility of explicit analytical calculations (at least in
terms of some familiar special functions) \cite{JJS82_2}. And
later on, calculating (\ref{b123}), (\ref{c123}) and inverting the
resulting formulas, we express $C_{x}$ in terms of $J_{x}$,
$J_{\alpha}$, $J_{\beta}$ and $C_{y}$ in terms of $J_{y}$,
$J_{\alpha}$, $J_{\beta}$. These expressions show some kind of
"degeneracy" because $C_{x}$ depends on $J_{\alpha}$, $J_{\beta}$
through the integer combination $J_{\alpha}+ J_{\beta}$, and
similarly, $C_{y}$ depends on them through $J_{\alpha}-
J_{\beta}$. However, this does not prejudice the true degeneracy
of the Hamiltonian. Substituting $C_{x}\left(J_{x}, J_{\alpha}+
J_{\beta}\right)$ and $C_{y}\left(J_{y}, J_{\alpha}-
J_{\beta}\right)$ to the main dynamical formula (\ref{d123}) and
assuming again some functional form of $V_{r}$ (e.g., just
$V_{r}=0$ in the translationally-free case), we in principle
calculate $J_{r}$ in terms of $E, J_{\alpha}, J_{\beta},
J_{\varphi}, J_{x}, J_{y}$. And again, solving this expression
with respect to $E$, one finds in principle that
\[
E={\mathcal{H}}\left(J_{r}, J_{\varphi}, J_{\alpha}, J_{\beta},
J_{x}, J_{y}\right).
\]
Without assuming the particular, both physical and analytically
treatable, form of potentials one cannot a priori decide about
problems concerning hyperintegrability (degeneracy) of the above
integrable models.

Just as in the flat-space problems, there exist reasonably-looking
models separable in other coordinates in the space of deformation
invariants, moreover, separable simultaneously in several systems
of coordinates in this space, thus probably degenerate
(hyperintegrable) ones.

An interesting class of separable models is obtained when one uses
the polar coordinates $(\varrho, \varepsilon)$ in the space of
deformation invariants:
\begin{equation}\label{a125}
x=\varrho \sin \varepsilon, \quad y=\varrho \cos \varepsilon.
\end{equation}
Then the kinetic energy
\[
T=\frac{m}{2}  G_{ij}(q)\frac{dq^{i}}{dt}\frac{dq^{j}}{dt}
\]
with coordinates ordered like
\[
(q^{1}, q^{2}, q^{3}, q^{4}, q^{5}, q^{6})=(r, \varphi, \gamma,
\delta, \varrho, \varepsilon)
\]
has the block matrix of the metric components
\[
\left[G_{ij}\right]=\left[
\begin{array}{ccc}
  K_{1} &  &  \\
   & K_{2} &  \\
   &  & K_{3} \\
\end{array}
\right],
\]
where
\[
K_{1}=[1],
\]
\begin{equation}\label{a126}
K_{2}=\left[
\begin{array}{ccc}
  R^{2}\sin^{2}\frac{r}{R}+\frac{I}{m}\varrho^{2}\cos
  ^{2}\frac{r}{R}&
   \frac{I}{m}\varrho^{2}\sin ^{2}\varepsilon \cos \frac{r}{R} &
   \frac{I}{m}\varrho^{2}\cos ^{2}\varepsilon \cos \frac{r}{R}  \\
   & & \\
   \frac{I}{m}\varrho^{2}\sin ^{2}\varepsilon \cos \frac{r}{R} &
   \frac{I}{m}\varrho^{2}\sin ^{2}\varepsilon  & 0 \\
   & & \\
    \frac{I}{m}\varrho^{2}\cos ^{2}\varepsilon \cos \frac{r}{R} & 0
    &
     \frac{I}{m}\varrho^{2}\cos ^{2}\varepsilon \\
\end{array}
\right],
\end{equation}
\[
K_{3}=\left[
\begin{array}{cc}
  \frac{I}{m} & 0\\
  0 &  \frac{I}{m}\varrho^{2} \\
\end{array}
\right]= \frac{I}{m}\left[
\begin{array}{cc}
  1 & 0\\
  0 &  \varrho^{2} \\
\end{array}
\right].
\]
The inverse metric tensor $G^{ij}$ underlying the kinetic part of
the Hamiltonian is given as follows:
\[
\left[G^{ij}\right]=\left[
\begin{array}{ccc}
  K_{1}^{-1} &  &  \\
   & K_{2}^{-1} &  \\
   &  & K_{3}^{-1} \\
\end{array}
\right],
\]
where, obviously,
\[
K_{1}^{-1}=[1],
\]
\begin{equation}\label{a127}
K_{2}^{-1}=\left[
\begin{array}{ccc}
  \frac{1}{R^{2}\sin^{2}\frac{r}{R}}&
   -\frac{\cos\frac{r}{R}}{R^{2}\sin^{2}\frac{r}{R}} &
   -\frac{\cos\frac{r}{R}}{R^{2}\sin^{2}\frac{r}{R}}  \\
   & & \\
   -\frac{\cos\frac{r}{R}}{R^{2}\sin ^{2}\frac{r}{R}} &
   \frac{m}{I}\frac{1}{\varrho^{2}\sin ^{2} \varepsilon }+
   \frac{1}{R^{2}}{\rm ctg}^{2}\frac{r}{R} & \frac{1}{R^{2}}{\rm ctg}^{2}\frac{r}{R} \\
   & & \\
    -\frac{\cos\frac{r}{R}}{R^{2}\sin ^{2}\frac{r}{R}} & \frac{1}{R^{2}}{\rm ctg}^{2}\frac{r}{R} &
      \frac{m}{I}\frac{1}{\varrho^{2}\cos ^{2} \varepsilon }+
   \frac{1}{R^{2}}{\rm ctg}^{2}\frac{r}{R} \\
\end{array}
\right],
\end{equation}
\[
K_{3}^{-1}=\left[
\begin{array}{cc}
  \frac{m}{I} & 0\\
  0 &  \frac{m}{I}\frac{1}{\varrho^{2}} \\
\end{array}
\right]= \frac{m}{I}\left[
\begin{array}{cc}
  1 & 0\\
  0 &  \frac{1}{\varrho^{2}} \\
\end{array}
\right].
\]
Obviously, the purely kinetic (geodetic) Hamiltonian is given by
\[
{\mathcal{T}}=\frac{1}{2m} G^{ij}(q)p_{i}p_{j},
\]
where the canonical momenta are ordered as follows:
\[
(p_{1}, p_{2}, p_{3}, p_{4}, p_{5}, p_{6})=(p_{r}, p_{\varphi},
p_{\gamma}, p_{\delta}, p_{\varrho}, p_{\varepsilon}).
\]
Strictly speaking, our ``polar coordinates" $\varrho$,
$\varepsilon$ are rotated by $\pi/2$ with respect to the usual
convention,but for some reasons it is convenient for us.

Tensor densities built of the metric $G_{ij}$ are in these
coordinates given by the expressions
\begin{eqnarray}
G&=&\det
\left[G_{ij}\right]=R^{2}\left(\frac{I}{m}\right)^{4}\varrho^{4}\sin
^{2} \varepsilon \cos ^{2} \varepsilon \sin
^{2}\frac{r}{R},\nonumber\\
\sqrt{G}&=&R\left(\frac{I}{m}\right)^{2}\varrho^{2}\left|\sin
\varepsilon \cos \varepsilon \right| \sin \frac{r}{R}.\nonumber
\end{eqnarray}
As previously, the system is separable (thus, completely
integrable) for potentials independent of $\varphi, \alpha,
\beta$. Then the reduced action $S_{0}$ is given by
\[
S_{0}(q)=S_{r}(r)+\ell
\varphi+C_{\alpha}\alpha+C_{\beta}\beta+S_{\varrho}(\varrho)+S_{\varepsilon}(\varepsilon),
\]
where as previously $\ell$, $C_{\alpha}$, $C_{\beta}$ are
integration constants. Using again the terms of spin and vorticity
$s, j$ we can write that
\[
S_{0}(q)=S_{r}(r)+\ell
\varphi+s\alpha+j\beta+S_{\varrho}(\varrho)+S_{\varepsilon}(\varepsilon).
\]
It is easily seen that such problems with cyclic variables
$\varphi$, $\alpha$, $\beta$ are separable for deformation
potentials of the form
\begin{equation}\label{a128}
V_{\varrho}(\varrho)+\frac{V_{\varepsilon}(\varepsilon)}{\varrho^{2}},
\end{equation}
i.e., for the total potentials we have that
\begin{equation}\label{b128}
V(r,\varrho,\varepsilon)=V_{r}(r)+V_{\varrho}(\varrho)+
\frac{V_{\varepsilon}(\varepsilon)}{\varrho^{2}}.
\end{equation}
Just as previously the action variables $J_{\alpha}$, $J_{\beta}$,
$J_{\varphi}$ are given as follows:
\begin{eqnarray}
J_{\alpha}&=&\oint p_{\alpha}d \alpha=C_{\alpha}\int\limits
_{0}^{2\pi}d\alpha=2\pi C_{\alpha}=2\pi s,\nonumber\\
J_{\beta}&=&\oint p_{\beta}d \beta=C_{\beta}\int\limits
_{0}^{2\pi}d\beta=2\pi C_{\beta}=2\pi j,\nonumber\\
J_{\varphi}&=&\oint p_{\varphi}d \varphi=\ell \int\limits
_{0}^{2\pi}d\varphi=2\pi \ell.\nonumber
\end{eqnarray}
Then $S_{\varrho}$, $S_{\varepsilon}$, $S_{r}$ satisfy the
ordinary differential equations
\begin{eqnarray}
\varrho^{2}\left(\frac{dS_{\varrho}}{d\varrho}\right)^{2}+
2I\varrho^{2}(V_{\varrho}(\varrho)-C)&=&-2IA,\nonumber\\
\left(\frac{dS_{\varepsilon}}{d\varepsilon}\right)^{2}+\frac{J_{\alpha}^{2}+2
J_{\alpha}J_{\beta}\cos 2 \varepsilon
+J_{\beta}^{2}}{4\pi^{2}\sin^{2}2\varepsilon}+
2IV_{\varepsilon}(\varepsilon)&=&2IA,\nonumber\\
\frac{1}{2m}\left(\frac{dS_{r}}{dr}\right)^{2}+\frac{(J_{\varphi}-J_{\alpha}
\cos \frac{r}{R} )^{2}}{8\pi^{2}mR^{2}\sin^{2}\frac{r}{R}}&=&
E-C-V_{r}(r),\nonumber
\end{eqnarray}
where $C$ and $A$ are the separation constants. More precisely,
$2IA$ is the constant value of the Hamilton-Jacobi term depending
only on $\varepsilon$, and $(E-C)$ is the constant value of the
$r$-term. Therefore, the remaining action quantities are given by:
\begin{eqnarray}
J_{\varepsilon}&=&\oint\sqrt{2I(A-V_{\varepsilon}(\varepsilon))-\frac{J_{\alpha}^{2}+2
J_{\alpha}J_{\beta}\cos 2 \varepsilon
+J_{\beta}^{2}}{4\pi^{2}\sin^{2}2\varepsilon}}d\varepsilon,\label{a129}\\
J_{\varrho}&=&\oint\sqrt{2I(C-V_{\varrho}(\varrho))-
\frac{2IA}{\varrho^{2}}}d\varrho,\label{b129}\\
J_{r}&=&\oint\sqrt{2m(E-C-V_{r}(r))-\frac{\left(J_{\varphi}-J_{\alpha}\cos
\frac{r}{R}\right)^{2}}{4\pi^{2}R^{2}\sin^{2}\frac{r}{R}}}dr.\label{c129}
\end{eqnarray}
And again the calculation procedure is as follows. Some explicit
forms of controlling potential functions $V_{\varepsilon}$,
$V_{\varrho}$, $V_{r}$ must be specified on the basis of
microphysical, phenomenological, or simply geometrical arguments.
Later on one ``calculates" the integral (\ref{a129}) and expresses
$J_{\varepsilon}$ as a function of $A$, $J_{\alpha}$, $J_{\beta}$,
i.e.,
\[
J_{\varepsilon}=J_{\varepsilon}\left(A, J_{\alpha},
J_{\beta}\right).
\]
Solving this expression with respect to $A$, one obtains the
formula for $A$ as a function of three action variables:
\[
A=A\left(J_{\alpha}, J_{\beta}, J_{\varepsilon}\right).
\]
Substituting this to (\ref{b129}) and ``calculating" the integral,
one obtains $J_{\varrho}$ as a function of $C$, $J_{\alpha}$,
$J_{\beta}$, $J_{\varepsilon}$, i.e.,
\[
J_{\varrho}=J_{\varrho}\left(C, J_{\alpha}, J_{\beta},
J_{\varepsilon}\right).
\]
Solving this with respect to $C$, one can in principle express $C$
as function of four action variables:
\[
C=C\left(J_{\alpha}, J_{\beta}, J_{\varrho},
J_{\varepsilon}\right).
\]
And finally, this expression is substituted to (\ref{c129}) and
then we obtain the formula for $J_{r}$ as a function of $E$,
$J_{\alpha}$, $J_{\beta}$, $J_{\varrho}$, $J_{\varepsilon}$,
$J_{\varphi}$, i.e.,
\[
J_{r}=J_{r}(E, J_{\alpha}, J_{\beta}, J_{\varrho},
J_{\varepsilon}, J_{\varphi}).
\]
Solving this with respect to $E$, one obtains the concluding
formula expressing $E$ through the (six) action variables:
\[
E=\mathcal{H}(J_{r}, J_{\varphi}, J_{\alpha}, J_{\beta},
J_{\varrho}, J_{\varepsilon}).
\]
Unfortunately, unlike in the mechanics of material point moving in
a central field, even the partial degeneracy cannot be concluded
directly from the formulas (\ref{a129}), (\ref{b129}),
(\ref{c129}) without performing the integration process based on
some fixed forms of the potential controlling functions
$V_{\varepsilon}$, $V_{\varrho}$, $V_{r}$. And just (and even more
so) like in the mechanics of infinitesimal gyroscope, interaction
with internal degrees of freedom removes the total degeneracy of
the material-point Bertrand models (\ref{b88}), (\ref{c88}).

The potentials of the form (\ref{a128}), (\ref{b128}) are very
convenient from the point of view of nonlinear macroscopic
elasticity. Being compatible with the very nature of deformative
degrees of freedom they are also interesting in the theory of
infinitesimal objects. The following example is very instructive,
moreover in macroscopic nonlinear elasticity in two dimensions it
is almost canonical \cite{all-book04}:
\begin{equation}\label{a131}
V=\frac{2\varkappa}{\varrho^{2}\cos
2\varepsilon}+\frac{\varkappa}{2}\varrho^{2}=\varkappa\left(\frac{1}{\lambda
\mu}+\frac{\lambda^{2}+\mu^{2}}{2}\right), \qquad \varkappa > 0.
\end{equation}
The first term prevents any kind of collapse of the
two-dimensional body: to the point or to the straight line. The
second term of the ``harmonic oscillator" type prevents the
unlimited expansion. The natural state $\lambda=\mu=1$ (no
deformation) minimizes the potential energy $V$; it is a stable
equilibrium. Extension in one direction is accompanied by
contraction in the orthogonal one. One can invent plenty of
similar potentials just using the polar coordinates $\varrho,
\varepsilon$ in the space of deformation invariants. The above one
is particularly simple and suggestive; it is also well suited to
the analytical procedure.

\subsection{Pseudospherical case}

Let us now consider the same problems on the pseudosphere, i.e.,
on the two-dimensional Lobatshevski space. We assume it to have
the pseudoradius $R$, therefore, the constant negative curvature
$\mathcal{R}=-2/R^{2}$. The most natural realization of this space
is the ``upper" shell of the hyperboloid in ${\mathbb{R}}^{3}$,
i.e.,
\[
-x^{2}-y^{2}+z^{2}=R^{2}, \qquad z>0.
\]
We denote it as $H^{2,2,+}(0,R)$. Its metric tensor is obtained as
the restriction (injection-pull-back) of the Minkowski metric:
\[
dx^{2}+dy^{2}-dz^{2}.
\]
Again we introduce the ``polar" coordinates $(r, \varphi)$, where
$r$ is the geodetic distance measured from the ``North Pole"
$x=0$, $y=0$, $z=R$, and $\varphi$ is the ``geographic longitude",
i.e., the polar angle in the sense of the plane $z=0$.

\noindent{\bf Remark:} the ``South Pole" $x=0$, $y=0$, $z=-R$ is
not interesting for us as it is placed on the ``lower" shell of
the hyperboloid, i.e., on the other connected component.
$H^{2,2,+}(0,R)$ is not compact and the radial variable $r$ has
the infinite range $[0, \infty]$. We shall also use the
pseudoangle $\vartheta=r/R$. The parametric description of
$H^{2,2,+}(0,R)\subset {\mathbb{R}}^{3}$ is given by
\[
x=R {\rm sh}\vartheta \cos \varphi, \qquad y=R {\rm sh}\vartheta
\sin \varphi, \qquad z=R {\rm ch}\vartheta.
\]
Expressed in terms of coordinates $(r,\varphi)$, the metric
element of $H^{2,2,+}(0,R)$, in analogy to (\ref{a87}), has the
form
\[
ds^{2}=dr^{2}+R^{2}{\rm sh}^{2} \frac{r}{R} d\varphi^{2}.
\]
And in general, practically all pseudospherical formulas may be
obtained from the spherical ones by substituting hyperbolic
functions instead of the corresponding trigonometric ones. In
certain expressions one must be however careful with signs. So,
the translational kinetic energy, in analogy to (\ref{a88}), has
the form
\[
T_{\rm tr}=\frac{m}{2}\left(\left(\frac{dr}{dt}\right)^{2}+{\rm
sh}^{2} \frac{r}{R}\left(\frac{d\varphi}{dt}\right)^{2}\right).
\]
And again there are two Bertrand-type potentials
\cite{JJS80,JJS00_2}, i.e.,
\begin{enumerate}
\item[$(i)$] the ``harmonic oscillator"-type potential:
\begin{equation}\label{a133}
 V(r)=\frac{\varkappa}{2} R^{2}{\rm th}^{2}\frac{r}{R}, \qquad \varkappa>0,
\end{equation}

\item[$(ii)$] the ``attractive Kepler-Coulomb"-type one:
\begin{equation}\label{b133}
 V(r)=-\frac{\alpha}{R} {\rm cth}\frac{r}{R}, \qquad \alpha>0.
\end{equation}
\end{enumerate}
With these and only these potentials all bounded orbits are
closed. And now the term ``bounded" is essential because the
"physical space" is now not compact. And indeed, there exist
unbounded motions corresponding to energy values exceeding some
thresholds. It is interesting that unlike in the spherical world,
in Lobatshevski space the isotropic degenerate oscillator has an
open subset of unbounded trajectories because the potential
(\ref{a133}) has a finite upper bound, i.e.,
\[
{\rm Sup}\ V=\frac{\varkappa}{2}R^{2}.
\]
For energy values above this threshold all trajectories are
unbounded, the motion is infinite. Below this threshold all
trajectories are not only bounded but also periodic.

The existence of threshold in (\ref{b133}) is not surprising, it
is like in the usual Kepler in $\mathbb{R}^{2}$. But the threshold
for the isotropic degenerate oscillator is a very interesting
feature of the Lobatshevski ``world".

Let us now consider an infinitesimal gyroscope. The success of
(\ref{a90}) suggests us to use the orthonormal basis
\[
E_{r}=\frac{\partial}{\partial r}=\mathcal{E}_{r}, \qquad
E_{\varphi}=\frac{1}{R {\rm sh}
\frac{r}{R}}\frac{\partial}{\partial \varphi}=\frac{1}{R {\rm sh}
\frac{r}{R}} \mathcal{E_{\varphi}},
\]
or in terms of components:
\[
E_{r}=[1, 0]^{T}, \qquad E_{\varphi}=\frac{1}{R {\rm sh}
\frac{r}{R}} [0,1]^{T}.
\]
It is easy to see that this aholonomic basis is really orthonormal
because
\[
g_{r\varphi}=g_{\varphi r}=0, \qquad g_{rr}=1, \qquad g_{\varphi
\varphi}=R^{2}{\rm sh}^{2} \frac{r}{R}.
\]
We formally repeat all considerations of the spherical case, i.e.,
all formulas for Christoffel symbols are analogous, the
trigonometric functions are replaced by the hyperbolic ones, etc.
Preserving the same notation we obtain that
\[
\omega_{\rm rl}=\frac{d\psi}{dt}, \qquad \omega=\omega_{\rm
rl}+\omega_{\rm dr}=\frac{d\psi}{dt}+{\rm
ch}\frac{r}{R}\frac{d\varphi}{dt} =\frac{d\psi}{dt}+{\rm ch}
\vartheta \frac{d\varphi}{dt}.
\]
In analogy to (\ref{a92}), (\ref{a93}) we obtain that
\begin{eqnarray}
 T&=&T_{\rm tr}+T_{\rm int} \label{a135} \\
 &=&\frac{m}{2}\left(\left(\frac{dr}{dt}\right)^{2}+R^{2}{\rm sh}^{2}
  \frac{r}{R}\left(\frac{d\varphi}{dt}\right)^{2}\right)+
  \frac{I}{2}\left(\frac{d\psi}{dt}+
  {\rm ch}\frac{r}{R}\frac{d\varphi}{dt}\right)^{2},\nonumber
\end{eqnarray}
i.e.,
\begin{eqnarray}
 T&=&T_{\rm tr}+T_{\rm int} \label{b135} \\
 &=& \frac{mR^{2}}{2}\left(\left(\frac{d\vartheta}{dt}\right)^{2}+
 {\rm sh}^{2}\vartheta\left(\frac{d\varphi}{dt}\right)^{2}\right)+
 \frac{I}{2}\left(\frac{d\psi}{dt}+{\rm ch} \vartheta
 \frac{d\varphi}{dt}\right)^{2}.\nonumber
\end{eqnarray}

When the generalized coordinates $(q^{1},q^{2},q^{3})$ are ordered
as previously, i.e, $(r,\varphi,\psi)$, then the corresponding
metric on the configuration space is given as follows:
\begin{equation}\label{c135}
\left[G_{ij}\right]=\left[
\begin{array}{ccc}
  1 & 0 & 0 \\
  & & \\
  0 & R^{2}\left({\rm sh}^{2}\frac{r}{R}+\frac{I}{mR^{2}}{\rm ch}^{2}
  \frac{r}{R}\right) & \frac{I}{m} {\rm ch} \frac{r}{R}\\
  & & \\
  0 & \frac{I}{m} {\rm ch} \frac{r}{R} & \frac{I}{m} \\
\end{array}
\right].
\end{equation}
The kinetic energy is then given by the expression
\[
T=\frac{m}{2} G_{ij}(q)\frac{dq^{i}}{dt}\frac{dq^{j}}{dt}.
\]
The contravariant metric $\left[G^{ij}\right]$ has the form
\[
\left[G^{ij}\right]=\left[
\begin{array}{ccc}
  1 & 0 & 0 \\
  & & \\
  0 & \frac{1}{R^{2} {\rm sh}^{2}\frac{r}{R}} &
  -\frac{{\rm ch}\frac{r}{R}}{R^{2}{\rm sh}^{2}\frac{r}{R}}\\
  & & \\
  0 & -\frac{{\rm ch}\frac{r}{R}}{R^{2}{\rm sh}^{2}\frac{r}{R}} &
  \frac{m}{I}+\frac{1}{R^{2}}{\rm cth}^{2}\frac{r}{R}  \\
\end{array}
\right].
\]
For potential systems the corresponding geodetic Hamiltonian is
written as follows:
\[
\mathcal{T}=\frac{1}{2m} G^{ij}(q)p_{i}p_{j},
\]
where, obviously, the generalized momenta $(p_{1},p_{2},p_{3})$
are ordered as follows: $(p_{r},p_{\varphi},p_{\psi})$.

The weight-two density built of $G$ has the form
\begin{equation}\label{a136}
G=\det\left[G_{ij}\right]=R^{2}\frac{I}{m}{\rm sh}^{2}\frac{r}{R},
\end{equation}
and the weight-one density defining the volume element is given by
\begin{equation}\label{b136}
\sqrt{G}=R\sqrt{\frac{I}{m}}{\rm sh}\frac{r}{R}.
\end{equation}
It is seen that the spherically very special case $I=mR^{2}$ here,
in the pseudospherical case also leads to some simplification of
$\left[G_{ij}\right]$, but not so striking one as previously. This
fact has deep geometric reasons which will be explained in the
sequel.

Namely, in the spherical space an essential point is the natural
identification between the quotient manifold
SO$(3,\mathbb{R})/{\rm SO}(2,\mathbb{R})$ and the spheres
$S^{2}(0,R)$, $S^{2}(0,1)$. And this has to do with the formal
identification between two-dimen\-sional rigid body moving over
the spherical surface and the three-dimen\-sional symmetrical top
without translational degrees of freedom. The special case
$I=mR^{2}$ corresponds to the spherical top. In general the
kinetic energy is then invariant under SO$(3,\mathbb{R})\times
{\rm SO}(2,\mathbb{R})$. In the three-dimensional top analogy
SO$(3,\mathbb{R})$ is acting as left regular translations and
SO$(2,\mathbb{R})$ as right regular translations corresponding to
the group of rotations around the body-fixed $z$-axis. If
$I=mR^{2}$ we have the full invariance under
SO$(3,\mathbb{R})\times {\rm SO}(3, \mathbb{R})$.

In the hyperbolic pseudospherical  geometry the problem is
isomorphic with the three-dimensional Lorentzian (Minkowskian) top
on $\mathbb{R}^{3}$. The rotation group SO$(3, \mathbb{R})$ is
replaced by the three-dimensional Lorentz group SO$(1, 2)$. And
still an important role is played by SO$(2, \mathbb{R})$
interpreted again as the group of usual rotations in Euclidean
space of $(x,y)$-variables (thus, not affecting the
$z$-direction). The above kinetic energy (\ref{a135}),
(\ref{b135}) is invariant under SO$(1, 2)\times {\rm SO}(2,
\mathbb{R})$. But it is never invariant under SO$(1, 2)\times {\rm
SO}(1, 2)$, i.e., under left and right Lorentz regular
translations in the SO$(1,2)$-sense. The spherical special case
$I=mR^{2}$ does not help here. Indeed, the underlying metric $G$
(and the kinetic energy itself) is positively definite. But the
doubly-invariant (SO$(1,2)\times {\rm SO}(1,2)$-invariant) metric
on SO$(1,2)$, i.e., its Killing metric is not positively definite.
Instead it has the normal-hyperbolic signature $(++-)$. The reason
is that it is semi-simple (even simple) non-compact group. This
brings about the question about non-positive kinetic energies
(metric tensors) on our configuration space. As the negative
contribution to the Killing metric tensor on  SO$(1,2)$ comes from
its compact subgroup SO$(2,\mathbb{R})$ of $(x,y)$-rotations,
i.e., from the gyroscopic degree of freedom in the language of
$H^{2,2,+}(0,R)$, there is a natural suggestion to invert the sign
of the gyroscopic contribution to (\ref{a92}), (\ref{a93}), i.e.,
to make it negative. One is naturally reluctant to indefinite
kinetic energies but there are examples when they are just
convenient and very useful as tools for describing some kinds of
physical interactions
\cite{JJS02_2,JJS04,JJS04S,JJS-VK03,JJS-VK04,JJS-VK04S,all-book04,all04,all05},
just encoding them even without any use of potentials.

So, we can try to use, or at least mathematically analyze, the
"Lorentz-type kinetic energies" $T_{L}$ of the form
\begin{eqnarray}
 T_{L}&=&\frac{m}{2}\left(\left(\frac{dr}{dt}\right)^{2}+R^{2}{\rm sh}^{2}
  \frac{r}{R}\left(\frac{d\varphi}{dt}\right)^{2}\right)-\frac{I}{2}\left(\frac{d\psi}{dt}+
  {\rm ch}\frac{r}{R}\frac{d\varphi}{dt}\right)^{2}
  \nonumber\\
&=&\frac{mR^{2}}{2}\left(\left(\frac{d\vartheta}{dt}\right)^{2}+{\rm
sh}^{2}
  \vartheta\left(\frac{d\varphi}{dt}\right)^{2}\right)-\frac{I}{2}\left(\frac{d\psi}{dt}+
  {\rm ch} \vartheta \frac{d\varphi}{dt}\right)^{2}.\label{a138}
\end{eqnarray}
Thus, it is so as if the extra rotation diminished effectively the
kinetic energy of translational motion. If we write as usual that
\[
T_{L}=\frac{m}{2}{}_{L}G_{ij}(q)\frac{dq^{i}}{dt}\frac{dq^{j}}{dt},
\]
then, with the same as previously convention concerning the
ordering of coordinates $(r,\varphi,\psi)$, we have that
\[
\left[{}_{L}G_{ij}\right]=\left[
\begin{array}{ccc}
  1 & 0 & 0 \\
  & & \\
  0 & R^{2}\left({\rm sh}^{2}\frac{r}{R}-\frac{I}{mR^{2}}{\rm ch}^{2}\frac{r}{R}\right) &
   -\frac{I}{m} {\rm ch} \frac{r}{R} \\
   & & \\
  0 & -\frac{I}{m} {\rm ch} \frac{r}{R} & -\frac{I}{m} \\
\end{array}
\right]
\]
(compare this with (\ref{c135})). And now, obviously, the
remarkable simplification occurs in the very special case
$I=mR^{2}$ just as in the spherical symmetry. This has to do ``as
usual" with the enlarging of the symmetry group from SO$(1,
2)\times {\rm SO}(2, \mathbb{R})$ to SO$(1, 2)\times {\rm SO}(1,
2)$ (two additional parameters of symmetry). And namely, ${}_{L}G$
becomes then ${}_{L}\breve{G}$, i.e.,
\[
\left[{}_{L}\breve{G}_{ij}\right]=\left[
\begin{array}{ccc}
  1 & 0 & 0 \\
  0 & -R^{2} &
   -R^{2}{\rm ch} \frac{r}{R}\\
  0 & -R^{2}{\rm ch} \frac{r}{R} & -R^{2} \\
\end{array}
\right]
\]
(compare this with (\ref{a95}) and notice the characteristic sign
differences).

The weight-two scalar density built of $_{L}G$ has the form
\[
{\rm det}\left[{}_{L}G_{ij}\right]=-R^{2}\frac{I}{m}{\rm
sh}^{2}\frac{r}{R},
\]
so it differs in sign from (\ref{a136}). This difference obviously
does not influence the density of Riemannian volume element, i.e.,
\[
\sqrt{|\det\left[{}_{L}G_{ij}\right]|}=R\sqrt{\frac{I}{m}}{\rm sh}
\frac{r}{R},
\]
exactly as in (\ref{b136}). This fact is of some interest.

As usual, the geodetic Hamiltonian is given by the expression
\[
\mathcal{T}_{L}=\frac{1}{2m}{}_{L}G^{ij}(q)p_{i}p_{j},
\]
where ${}_{L}G^{ij}$ are components of the contravariant inverse
of ${}_{L}G_{ij}$. They are given as follows:
\[
\left[{}_{L}G^{ij}\right]=\left[
\begin{array}{ccc}
  1 & 0 & 0 \\
  & & \\
  0 & \frac{1}{R^{2}{\rm sh}^{2}\frac{r}{R}} &
   -\frac{{\rm ch}\frac{r}{R}}{R^{2}{\rm sh}^{2}\frac{r}{R}}\\
   & & \\
  0 & -\frac{{\rm ch}\frac{r}{R}}{R^{2}{\rm sh}^{2}\frac{r}{R}} &
   -\frac{m}{I}+\frac{1}{R^{2}}{\rm cth}^{2}\frac{r}{R} \\
\end{array}
\right].
\]
Obviously, if we use the above isomorphism between the
two-dimensional top sliding over the Lobatshevski plane with the
three-dimensional Lorentz top without translational motion in
$\mathbb{R}^{3}$, then it is clear that ${}_{L}G$ is, up to
normalization, identical with the Killing metric tensor of SO$(1,
2)$. Let us quote some formulas and concepts analogous to
three-dimensional angular velocities, i.e., to (\ref{b94}),
(\ref{c94}). And then the kinetic energy will be expressed like in
(\ref{d94}).

First of all we parameterize SO$(1,2)$ with the help of what we
call the ``pseudo-Euler angles". So, let us write that
\[
{\rm SO}(1, 2)\ni L(\varphi, \vartheta,
\psi)=U_{z}(\varphi)L_{x}(\vartheta)U_{z}(\psi),
\]
where the meaning of $U_{z}$ is like in (\ref{b93}) and $L_{x}$
denotes some Lorentz transformation in $\mathbb{R}^{3}$, namely,
the ``boost" along the $x$-axis, i.e.,
\[
L_{x}(\vartheta)=\left[
\begin{array}{ccc}
  1 & 0 & 0 \\
  0 & {\rm ch} \vartheta &
  {\rm sh} \vartheta\\
  0 & {\rm sh} \vartheta &
   {\rm ch} \vartheta \\
\end{array}
\right].
\]
During the motion all these quantities are functions of time and
we can calculate the corresponding Lie-algebraic element
\[
\widehat{\lambda}=L^{-1}\frac{dL}{dt},
\]
i.e., the co-moving pseudo-angular velocity. After some
calculations we obtain formulas analogous to (\ref{c94}),
(\ref{e94}), (\ref{f94}), (\ref{a94}), and namely,
\[
\widehat{\lambda}=\widehat{\lambda}_{1}\left[
\begin{array}{ccc}
  0 & 0 & 0 \\
  0 & 0 &
 1\\
  0 & 1 &
  0 \\
\end{array}
\right]+\widehat{\lambda}_{2}\left[
\begin{array}{ccc}
  0 & 0 & 1 \\
  0 & 0 &
 0\\
  1 & 0 &
  0 \\
\end{array}
\right]+\widehat{\lambda}_{3}\left[
\begin{array}{ccc}
  0 & -1 & 0 \\
  1 & 0 &
 0\\
  0 & 0 &
  0 \\
\end{array}
\right],
\]
where
\begin{eqnarray}
\widehat{\lambda}_{1}&=&{\rm sh} \vartheta \sin \psi
\frac{d\varphi}{dt}+\cos \psi \frac{d\vartheta}{dt},\nonumber\\
\widehat{\lambda}_{2}&=&-{\rm sh} \vartheta \cos
\psi\frac{d\varphi}{dt}+\sin \psi \frac{d\vartheta}{dt},\nonumber\\
\widehat{\lambda}_{3}&=&{\rm ch} \vartheta \frac{d\varphi}{dt}+
\frac{d\psi}{dt}.\nonumber
\end{eqnarray}
Both the similarities and differences in comparison with the
corresponding spherical formulas are easily seen.

And now we can write two formulas analogous to (\ref{d94}), i.e.,
\begin{eqnarray}
 T&=&\frac{K}{2}\left(\widehat{\lambda}_{1}\right)^{2}+
 \frac{K}{2}\left(\widehat{\lambda}_{2}\right)^{2}+
 \frac{I}{2}\left(\widehat{\lambda}_{3}\right)^{2},\label{a142}\\
 T&=&\frac{K}{2}\left(\widehat{\lambda}_{1}\right)^{2}+
 \frac{K}{2}\left(\widehat{\lambda}_{2}\right)^{2}-
 \frac{I}{2}\left(\widehat{\lambda}_{3}\right)^{2},\label{b142}
\end{eqnarray}
where $K>0$ and $I>0$. This is the symmetric SO$(1, 2)$-top in
$\mathbb{R}^{3}$. The indefinite expression (\ref{b142}) is
structurally suited to the normal-hyperbolic signature of SO$(1,
2)$. When $K=I$, then it becomes the spherical Lorentz top in
$\mathbb{R}^{3}$ in the indefinite version based on the Killing
metric.

It is easily seen that both expressions (\ref{a142}) and
(\ref{b142}) are invariant under SO$(1, 2)\times {\rm SO}(2,
\mathbb{R})$, where SO$(1, 2)$ and SO$(2, \mathbb{R})$ acts on
SO$(1, 2)$ through respectively the left and right regular
translations. The form (\ref{b142}) with $K=I$ is invariant under
all regular translations (both left and right), i.e., under SO$(1,
2)\times {\rm SO}(1, 2)$. And specifying $K=mR^{2}$ in
(\ref{a142}) and (\ref{b142}), we obtain respectively (\ref{a135})
and (\ref{a138}).

Let us now turn to the Hamilton-Jacobi equation and action-angle
variables. Just as previously, $\varphi, \psi$ are cyclic
variables in the kinetic energy term and we assume that the same
is true for the total Hamiltonian, i.e., that the potential energy
depends only on the variable $r$. Because of the non-diagonal
structure of $\left[G_{ij}\right]$ and $\left[{}_{L}G_{ij}\right]$
in the natural variables $(r,\varphi,\psi)$, it is rather
difficult to decide a priori what would be (if any!) the form of
the general separable potential. So, just as in (\ref{a100}) the
reduced action $S_{0}$ is sought (with the same meaning of all
symbols) in the form
\[
S_{0}(r, \varphi, \psi; E, \ell, s)=S_{r}(r, E)+\ell \varphi+s
\psi.
\]
Just as in (\ref{a101}) the genuine dynamics reduces to the radial
function which satisfies the equation
\begin{equation}\label{a143}
\left(\frac{dS_{r}}{dr}\right)^{2}=2m\left(E-V(r)\right)-\frac{\left(\ell-
s\,{\rm ch} \frac{r}{R}\right)^{2}}{R^{2}{\rm
sh}^{2}\frac{r}{R}}\pm \frac{m}{I}s^{2}.
\end{equation}
The $\pm$ signs in (\ref{a143}) refer respectively to the models
(\ref{a135}), (\ref{a138}). Just as in the spherical case
$p_{\varphi}$, $p_{\psi}$ are constants of motion and their
constant values on fixed trajectories equal
\[
p_{\varphi}=\frac{\partial S_{0}}{\partial \varphi}=\ell, \qquad
p_{\psi}=\frac{\partial S_{0}}{\partial \psi}=s.
\]
And, of course, their action variables are given as follows:
\begin{eqnarray}
J_{\varphi}&=&\oint p_{\varphi}d\varphi=\int\limits_{0}^{2
\pi}\ell d\varphi=2\pi \ell,\nonumber\\
J_{\psi}&=&\oint p_{\psi}d\psi=\int\limits_{0}^{2 \pi} s
d\psi=2\pi s.\nonumber
\end{eqnarray}
Substituting this to the radial action expression and to
(\ref{a143}) we obtain for
\[
J_{r}=\oint p_{r}dr
\]
the following formula:
\begin{equation}\label{a144}
J_{r}=\oint \sqrt{2m(E-V(r))-\frac{\left(J_{\varphi}-J_{\psi} {\rm
ch}\frac{r}{R}\right)^{2}}{4\pi^{2}R^{2}{\rm sh}^{2}\frac{r}{R}}
\pm \frac{m}{I}\frac{J^{2}_{\psi}}{4\pi^{2}}}dr
\end{equation}
(with the above-mentioned meaning of the $\pm$ signs).

From now on the procedure is just as previously. After
substituting here some particular form of $V$ we perform the above
contour integration and (at least in principle) determine the
dependence of $J_{r}$ on $E$, $J_{\varphi}$, $J_{\psi}$, i.e.,
\[
J_{r}=J_{r}(E, J_{\varphi}, J_{\psi}).
\]
Solving this equation with respect to $E$, one obtains the final
expression for Hamiltonian:
\[
E=\mathcal{H}(J_{r}, J_{\varphi}, J_{\psi}).
\]
Just as in the spherical case no degeneracy structure may be
directly deduced from (\ref{a144}) even in the special case of the
minus sign and $I=mR^{2}$.

An important point is that, because of the uncompactness of the
Lobatshevski space, the geodetic motion $(V=0)$ is unbounded, and
then obviously the action-angle formalism becomes meaningless.
Nevertheless, even for unbounded motion the isotropic models with
$V$ depending only on $r$ may be effectively studied with the use
of explicitly known constants of motion and reduced to first-order
ordinary differential equations.

Let us now consider a deformable top moving in Lobatshevski space.
All symbols concerning internal degrees of freedom are just those
used in spherical geometry. The metric tensor $G$ underlying the
kinetic energy expression, i.e.,
\[
T=\frac{m}{2}G_{ij}(q)\frac{dq^{i}}{dt}\frac{dq^{j}}{dt},
\]
has the form analogous to the spherical case with the
trigonometric functions simply replaced by the hyperbolic ones
without any change of sign. Thus, the matrix $\left[G_{ij}\right]$
consists of three blocks $M_{1}, M_{2}, M_{3}$, where
\[
M_{1}=[1],
\]
\[
M_{2}=\left[
\begin{array}{ccc}
  R^{2}{\rm sh}^{2}\frac{r}{R}+\frac{I}{m}\left(x^{2}+y^{2}\right){\rm ch}^{2}\frac{r}{R}
  & \frac{I}{m}x^{2}{\rm ch}\frac{r}{R} & \frac{I}{m}y^{2}{\rm ch}\frac{r}{R} \\
  & & \\
  \frac{I}{m}x^{2}{\rm ch}\frac{r}{R} & \frac{I}{m}x^{2}&
  0\\
  & & \\
  \frac{I}{m}y^{2}{\rm ch}\frac{r}{R} & 0 &
  \frac{I}{m}y^{2}  \\
\end{array}
\right],
\]
\[
M_{3}=\frac{I}{m} I_{2}= \left[
\begin{array}{cc}
 \frac{I}{m} & 0 \\
  0&
   \frac{I}{m} \\
\end{array}
\right].
\]

When deformations are admitted, then there is no particular
geometric motivation for discussing the ``Lorentzian" model with
the indefinite kinetic energy. This is nevertheless possible and
formally simply consists in replacing $I$ by $-I$ in all formulas;
in other words negative ``inertial moments" are admitted.

For the inverse contravariant metric $\left[G^{ij}\right]$
underlying the geodetic Hamilto\-nian, i.e.,
\[
\mathcal{T}=\frac{1}{2m}G^{ij}(q)p_{i}p_{j},
\]
we have the block structure also quite analogous to the spherical
formulas:
\[
M_{1}^{-1}=[1],
\]
\[
M_{2}^{-1}=\left[
\begin{array}{ccc}
  \frac{1}{R^{2}{\rm sh}^{2}\frac{r}{R}}
  & -\frac{{\rm ch} \frac{r}{R}}{R^{2}{\rm sh}^{2}\frac{r}{R}}&
   -\frac{{\rm ch} \frac{r}{R}}{R^{2}{\rm sh}^{2}\frac{r}{R}} \\
   & & \\
 -\frac{{\rm ch} \frac{r}{R}}{R^{2}{\rm sh}^{2}\frac{r}{R}} &
  \frac{m}{I}\frac{1}{x^{2}}+\frac{1}{R^{2}}{\rm cth}^{2}\frac{r}{R}&
  \frac{1}{R^{2}}{\rm cth}^{2}\frac{r}{R}\\
  & & \\
  -\frac{{\rm ch} \frac{r}{R}}{R^{2}{\rm sh}^{2}\frac{r}{R}} &
  \frac{1}{R^{2}}{\rm cth}^{2}\frac{r}{R} &
  \frac{m}{I}\frac{1}{y^{2}}+\frac{1}{R^{2}}{\rm cth}^{2}\frac{r}{R} \\
\end{array}
\right],
\]
\[
M_{3}^{-1}=\frac{m}{I} I_{2}= \left[
\begin{array}{cc}
 \frac{m}{I} & 0 \\
  0&
   \frac{m}{I} \\
\end{array}
\right].
\]
The corresponding scalar $G$-densities are given as follows:
\begin{eqnarray}
G=\det\left[G_{ij}\right]&=&R^{2}\left(\frac{I}{m}\right)^{4}x^{2}y^{2}{\rm
sh}^{2}\frac{r}{R},\nonumber \\
\sqrt{|G|}&=&R\left(\frac{I}{m}\right)^{2}\left|xy\right|{\rm
sh}\frac{r}{R}.\nonumber
\end{eqnarray}

Again we assume that the potential energy does not depend on the
angles $\varphi$, $\alpha$, $\beta$, i.e., they are cyclic
variables for the total Hamiltonian. So, just as previously, the
complete integral of
\[
\frac{1}{2m} G^{ij}(q)\frac{\partial S_{0}}{\partial
q^{i}}\frac{\partial S_{0}}{\partial q^{j}}+V(q)=E
\]
will be sought as a linear function of all angles, i.e.,
\begin{eqnarray}
S_{0}(q)&=& S_{r}(r)+S_{x}(x)+S_{y}(y)+\ell
\varphi+C_{\gamma}\gamma+C_{\delta}\delta\nonumber\\
&=& S_{r}(r)+S_{x}(x)+S_{y}(y)+\ell
\varphi+C_{\alpha}\alpha+C_{\beta}\beta,\nonumber
\end{eqnarray}
with the same as previously relationships between constants.
Coming back to the more familiar spin and vorticity symbols $s,
j$, we have that $C_{\alpha}=s$ and $C_{\beta}=j$. The action
variables corresponding to $J_{\alpha}$, $J_{\beta}$,
$J_{\varphi}$ are respectively given by the same formulas like in
the spherical geometry:
\begin{eqnarray}
J_{\alpha}&=&\oint p_{\alpha}d\alpha=C_{\alpha}\int\limits_{0}^{2
\pi}
d\alpha=2\pi C_{\alpha}=2\pi s,\nonumber\\
J_{\beta}&=&\oint p_{\beta}d\beta=C_{\beta}\int\limits_{0}^{2 \pi}
d\beta=2\pi C_{\beta}=2\pi j,\nonumber\\
J_{\varphi}&=&\oint p_{\varphi}d\varphi=\ell\int\limits_{0}^{2
\pi} d\varphi=2\pi \ell.\nonumber
\end{eqnarray}
Similarly, when the $(x,y)$-deformation invariants are used, the
most natural separable potentials have the explicitly separated
form:
\[
V(r, x, y)=V_{r}(r)+V_{x}(x)+V_{y}(y).
\]
Then, just as in the spherical case, denoting the separation
constants by $C_{r}$, $C_{x}$, and $C_{y}$, we have that
\[
C_{r}+ C_{x}+ C_{y}=E
\]
and the expressions for $J_{r}, J_{x}, J_{y}$ just analogous to
(\ref{b123}), (\ref{c123}), (\ref{d123}), namely,
\begin{eqnarray}
J_{x}&=&\oint
p_{x}dx=\oint\sqrt{2I(C_{x}-V_{x}(x))-\frac{(J_{\alpha}+
J_{\beta})^{2}}{16\pi^{2}x^{2}}} dx,\nonumber\\
J_{y}&=&\oint
p_{y}dy=\oint\sqrt{2I(C_{y}-V_{y}(y))-\frac{(J_{\alpha}-
J_{\beta})^{2}}{16\pi^{2}y^{2}}} dy,\nonumber\\
J_{r}&=&\oint
p_{r}dr=\oint\sqrt{2m(E-C_{x}-C_{y}-V_{r}(r))-\frac{(J_{\varphi}-
J_{\alpha}ch\frac{r}{R})^{2}}{4\pi^{2}R^{2}sh^{2}\frac{r}{R}}}
dr.\nonumber
\end{eqnarray}
The procedure of eliminating constants $C_{x}$, $C_{y}$ and
expressing $E$ through the action variables, i.e.,
\[
E=\mathcal{H}(J_{r}, J_{\varphi}, J_{\alpha}, J_{\beta}, J_{x},
J_{y}),
\]
proceeds exactly as in the spherical case.

Exactly as in the theory of deformable gyroscope in the spherical
space it is convenient and practically useful to parameterize
deformation invariants with the use of polar variables $\varrho$,
$\varepsilon$ (see (\ref{a125})). The only formal difference is
that the trigonometric functions of $r/R$ (but not those of
$\varepsilon !$) are replaced by the hyperbolic ones without the
change of sign, thus, (\ref{a126}) and (\ref{a127}) take on
respectively the forms
\begin{eqnarray}
K_{2}&=&\left[
\begin{array}{ccc}
  R^{2}{\rm sh}^{2}\frac{r}{R}+\frac{I}{m}\varrho^{2}{\rm ch} ^{2}\frac{r}{R}&
   \frac{I}{m}\varrho^{2}\sin ^{2} \varepsilon {\rm ch} \frac{r}{R} &
   \frac{I}{m}\varrho^{2}\cos ^{2} \varepsilon {\rm ch} \frac{r}{R}  \\
   & & \\
   \frac{I}{m}\varrho^{2}\sin ^{2} \varepsilon {\rm ch} \frac{r}{R} &
   \frac{I}{m}\varrho^{2}\sin ^{2} \varepsilon  & 0 \\
   & & \\
    \frac{I}{m}\varrho^{2}\cos ^{2} \varepsilon {\rm ch} \frac{r}{R} & 0 &
     \frac{I}{m}\varrho^{2}\cos ^{2} \varepsilon \\
\end{array}
\right],\nonumber\\
K_{2}^{-1}&=&\left[
\begin{array}{ccc}
  \frac{1}{R^{2}{\rm sh} ^{2}\frac{r}{R}}&
   -\frac{{\rm ch} \frac{r}{R}}{R^{2}{\rm sh} ^{2}\frac{r}{R}} &
   -\frac{{\rm ch} \frac{r}{R}}{R^{2}{\rm sh} ^{2}\frac{r}{R}}  \\
   & & \\
   -\frac{{\rm ch} \frac{r}{R}}{R^{2}{\rm sh} ^{2}\frac{r}{R}} &
   \frac{m}{I}\frac{1}{\varrho^{2}\sin ^{2} \varepsilon }+
   \frac{1}{R^{2}}{\rm cth} ^{2}\frac{r}{R} & \frac{1}{R^{2}}{\rm cth} ^{2}\frac{r}{R} \\
   & & \\
    -\frac{{\rm ch} \frac{r}{R}}{R^{2}{\rm sh} ^{2}\frac{r}{R}} &
    \frac{1}{R^{2}}{\rm cth} ^{2}\frac{r}{R} &
      \frac{m}{I}\frac{1}{\varrho^{2}\cos ^{2} \varepsilon }+
   \frac{1}{R^{2}}{\rm cth} ^{2}\frac{r}{R} \\
\end{array}
\right].\nonumber
\end{eqnarray}
And obviously the other blocks are identical with those from the
spherical geometry. The metrical scalar densities are given as
follows:
\begin{eqnarray}
G=\det\left[G_{ij}\right]&=&R^{2}\left(\frac{I}{m}\right)^{4}\varrho^{4}\sin^{2}\varepsilon
\cos^{2}\varepsilon\ {\rm sh}^{2}\frac{r}{R},\nonumber\\
\sqrt{|G|}&=&R\left(\frac{I}{m}\right)^{2}\varrho^{2}\left|\sin
\varepsilon \cos \varepsilon \right| {\rm sh}
\frac{r}{R}.\nonumber
\end{eqnarray}
There are also no essential changes with the integration
procedure, reasonable deformative potentials models, and the
action-angle variables. Just the trigonometric functions  of $r/R$
replaced by the hyperbolic ones.

\subsection{Toroidal case}

Let us now consider another interesting two-dimensional model,
namely, the motion of structured material points over the toroidal
manifold, so this time over the Riemann space of not constant
curvature. Obviously, we do not mean the torus
$\mathbb{R}^{2}/\mathbb{Z}^{2}$, i.e., the quotient of
$\mathbb{R}^{2}$ with respect to the ``crystal lattice"
$\mathbb{Z}^{2}$; this would be flat, although topologically not
trivial, space. Our torus is a ``tyre", or rather ``inner tube",
injected into $\mathbb{R}^{3}$ and endowed with the metric tensor
induced from $\mathbb{R}^{3}$ (from its flat Euclidean structure).

Let $R$ denote the ``small" radius, i.e., the radius of circles
obtained from transversal orthogonal cross-sections of the tube by
planes. The ``large" radius, i.e., the radius of the centrally
placed inner-tube circle will be denoted by $L$. The parametric
description of the torus is given as follows:
\begin{eqnarray}
x&=&(L+R\cos \vartheta)\cos \varphi,\nonumber\\
y&=&(L+R\cos \vartheta)\sin \varphi,\nonumber\\
z&=&R \sin \vartheta,\nonumber
\end{eqnarray}
where $\varphi$ is the angle measured along the central inner
circle (``along-tyre" angle) and $\vartheta$ denotes the polar
angle of cross-section circles (``around-tyre" angle) measured
from the external circle (maximally remote from the ``tyre"
centre). Both $\varphi$ and $\vartheta$ run over the usual angular
range $[0, 2\pi]$ with its obvious non-uniqueness. One can easily
show that such a torus, denoted by $T^{2}(0, L, R)$, consists of
points $(x, y, z)\in \mathbb{R}^{3}$ satisfying the fourth-degree
algebraic equation
\[
(x^{2}+y^{2}+z^{2}+L^{2}-R^{2})^{2}-4L^{2}(x^{2}+y^{2})=0.
\]
Therefore, it is an algebraic manifold of fourth-degree because it
is impossible to lower the degree of the polynomial on the
left-hand side.

After the easy calculation one obtains that the metric element
induced on the surface $T^{2}(0, L, R)$ has the form
\[
ds^{2}=R^{2}d\vartheta^{2}+\left(L+R\cos
\vartheta\right)^{2}d\varphi^{2}.
\]
Introducing as previously the geodetic length along the
"around-tyre" circles, i.e.,
\[
r=R\vartheta \in \left[0, 2\pi R\right],
\]
we can write that
\[
ds^{2}=dr^{2}+\left(L+R\cos \frac{r}{R}\right)^{2}d\varphi^{2}.
\]
Therefore, the kinetic energy of the structure-less material point
is given by the expression
\begin{eqnarray}
T_{\rm
tr}&=&\frac{m}{2}\left(\left(\frac{dr}{dt}\right)^{2}+\left(L+R\cos
\frac{r}{R}\right)^{2}\left(\frac{d\varphi}{dt}\right)^{2}\right)
\nonumber \\
&=&\frac{m}{2}\left(R^{2}\left(\frac{d\vartheta}{dt}\right)^{2}+\left(L+R\cos
\vartheta\right)^{2}\left(\frac{d\varphi}{dt}\right)^{2}\right).\nonumber
\end{eqnarray}

In analogy to spherical and pseudo-spherical formulas it is also
convenient to use the form
\[
T_{\rm
tr}=\frac{mR^{2}}{2}\left(\left(\frac{d\vartheta}{dt}\right)^{2}+\left(\frac{L}{R}+\cos
\vartheta\right)^{2}\left(\frac{d\varphi}{dt}\right)^{2}\right).
\]
It is seen that again $\varphi$ is the cyclic variable in $T_{\rm
tr}$. The along-$\varphi$ rotations, i.e., $\varphi \mapsto
\varphi+\alpha$, are isometries, thus, $\partial/\partial \varphi$
is the Killing vector field.

As previously, the coordinates on $T^{2}(0, L, R)$ will be ordered
as $(\vartheta, \varphi)$ or $(r, \varphi)$, thus,
\begin{eqnarray}
\left[g_{ij}\right]&=& \left[
\begin{array}{cc}
  R^{2} & 0 \\
  0 & (L+R\cos
\vartheta)^{2} \\
\end{array}
\right], \nonumber \\
\left[g^{ij}\right]&=&\left[
\begin{array}{cc}
  \frac{1}{R^{2}} & 0 \\
  0 & \frac{1}{(L+R\cos
\vartheta)^{2}} \\
\end{array}
\right], \nonumber
\end{eqnarray}
and the density of Riemannian volume (area) is given as follows:
\[
\sqrt{\left|g\right|}=R\left(L+R\cos \vartheta\right).
\]
Separable potentials have the form
\[
V(\vartheta, \varphi)=V_{\vartheta}(\vartheta)+
\frac{V_{\varphi}(\varphi)}{(L+R\cos \vartheta)^{2}}.
\]
The simplest and geometrically interesting is the class of models
invariant under $\varphi$-rotations, i.e., $V_{\varphi}=0$. The
angle $\varphi$ is then a cyclic variable. Unfortunately, even the
simplest case, i.e., the geodetic model $V=0$, is technically
rather complicated and leads to elliptic integrals (after an
appropriate change of coordinates).

One can easily show that the only not vanishing Christoffel
symbols are as follows:
\[
\Gamma^{\varphi}{}_{\varphi
\vartheta}=\Gamma^{\varphi}{}_{\vartheta \varphi}=-\frac{R \sin
\vartheta}{L+R\cos \vartheta},\qquad \Gamma^{\vartheta}{}_{\varphi
\varphi}=\left(\frac{L}{R}+\cos \vartheta\right)\sin
\vartheta.\nonumber
\]

As expected, the most natural choice of the auxiliary field of
frames $E$ is one suited to coordinate lines, i.e.,
\[
E_{\vartheta}=\frac{1}{R}\frac{\partial}{\partial
\vartheta}=\frac{1}{R} \mathcal{E}_{\vartheta}, \qquad
E_{\varphi}=\frac{1}{L+R\cos \vartheta}\frac{\partial}{\partial
\varphi}=\frac{1}{L+R\cos \vartheta} \mathcal{E}_{\varphi}.
\]
It is evidently orthogonal because
\[
g_{\vartheta \varphi}=g_{\varphi \vartheta}=0, \qquad g_{\vartheta
\vartheta}=R^{2}, \qquad g_{\varphi \varphi}=(L+R\cos
\vartheta)^{2}.
\]

Let us now take into account internal degrees of freedom. We begin
with the infinitesimal gyroscope. The angular velocities are then
analytically expressed by the expressions
\[
\omega_{\rm rl}=\frac{d\psi}{dt},\qquad \omega =\omega_{\rm
rl}+\omega_{\rm dr}=\frac{d\psi}{dt}+\sin
\vartheta\frac{d\varphi}{dt}.
\]
Therefore, the kinetic energy of the infinitesimal rotator is
given as follows:
\begin{eqnarray}
T&=&T_{\rm tr}+T_{\rm int}\nonumber\\
&=&\frac{m}{2}\left(R^{2}\left(\frac{d\vartheta}{dt}\right)^{2}+(L+R\cos
\vartheta)^{2}\left(\frac{d\varphi}{dt}\right)^{2}\right)+
\frac{I}{2}\left(\frac{d\psi}{dt}+\sin \vartheta
\frac{d\varphi}{dt}\right)^{2}.\nonumber
\end{eqnarray}
If we write it in the previous form, i.e.,
\[
T=\frac{m}{2}G_{ij}(q)\frac{dq^{i}}{dt}\frac{dq^{j}}{dt},
\]
with the ordering
\[
(q^{1}, q^{2}, q^{3})=(\vartheta, \varphi, \psi),
\]
then
\[
\left[G_{ij}\right]=\left[
\begin{array}{ccc}
  R^{2} & 0 & 0 \\
  & & \\
  0 & (L+R\cos
\vartheta)^{2}+\frac{I}{m}\sin^{2}\vartheta & \frac{I}{m}\sin \vartheta\\
& & \\
  0 & \frac{I}{m}\sin \vartheta & \frac{I}{m} \\
\end{array}
\right].
\]
The inverse metric underlying the geodetic Hamiltonian is given by
the expression
\[
 \left[G^{ij}\right]=\left[
\begin{array}{ccc}
  \frac{1}{R^{2}} & 0 & 0 \\
  & & \\
  0 & \frac{1}{(L+R\cos
\vartheta)^{2}}&- \frac{\sin \vartheta}{(L+R\cos
\vartheta)^{2}}\\
& & \\
  0 & - \frac{\sin \vartheta}{(L+R\cos
\vartheta)^{2}} &\frac{m}{I}+\frac{\sin^{2} \vartheta}{(L+R\cos
\vartheta)^{2}} \\
\end{array}
\right].
\]

Up to the $I$-dependent normalization, the Riemannian density
function is like that for the structure-less material point, i.e.,
\[
\sqrt{\left|G\right|} = \sqrt{\frac{I}{m}} \, R \left(L + R \cos
\vartheta \right).
\]
Just as in the spherical and pseudospherical spaces, the metric
$G$ is not diagonal in natural coordinates, therefore, without
rather complicated analysis it would be difficult to fix the class
of potentials compatible with the separation of variables
procedure. And again evidently integrable models are isotropic
ones when the variables $\varphi$, $\psi$ are cyclic, i.e.,
\[
H = {\mathcal{T}} + V(\vartheta) = \frac{1}{2} G^{ij} p_{i}p_{j} +
V(\vartheta).
\]
Obviously, the phase space coordinates are ordered in the usual
way, i.e.,
\[
\left(q^{1},q^{2},q^{3};p_{1},p_{2},p_{3}\right)=\left(\vartheta,
\varphi, \psi ;p_{\vartheta},p_{\varphi},p_{\psi}\right).
\]
For the action variables $J_{\varphi}$, $J_{\psi}$ we obtain that
\begin{eqnarray}
J_{\varphi} &=& \oint p_{\varphi} d \varphi = \int\limits_{0}^{2
\pi} \ell d \varphi = 2 \pi \ell, \label{a156}\\
J_{\psi} &=& \oint p_{\psi} d \psi = \int\limits_{0}^{2 \pi} s d
\psi = 2 \pi s, \label{b156}
\end{eqnarray}
with the same meaning of symbols as previously and
\[
S_{0}(\vartheta, \varphi, \psi; E, \ell, s)=
S_{\vartheta}(\vartheta, E)+ \ell \varphi + s \psi.
\]
The true dynamics is contained in $J_{\vartheta}$ which after the
substitution of (\ref{a156}) and (\ref{b156}) into
\[
\oint p_{\varphi} d \vartheta = \oint \frac{d S_{\vartheta}}{d
\vartheta} d \vartheta
\]
becomes as follows:
\[
J_{\vartheta}= R \oint\sqrt{2m \left(E - V(\vartheta)\right)-
\frac{\left(J_{\varphi}-J_{\psi}\sin
\vartheta\right)^{2}}{4\pi^{2}\left(L + R \cos
\vartheta\right)^{2}}+\frac{m}{I}\frac{J^{2}_{\psi}}{4\pi^{2}}}d\vartheta.
\]
Assuming some particular shape of $V(\vartheta)$, ``performing"
the integration, and solving the result with respect to $E$, one
obtains the final formula:
\[
E = {\mathcal{H}}\left(J_{\vartheta}, J_{\varphi},
J_{\psi}\right).
\]
No degeneracy is a priori seen without performing the integration,
even for the geodetic models, i.e., when $V=0$. An elementary
integrability (if any!) may be suspected, of course, only for
$V(\vartheta)$ built in a simple way from $\sin \vartheta,\cos
\vartheta$-expressions.

Let us now quote some formulas concerning an infinitesimal
deformable top moving over the toroidal surface. We restrict
ourselves to the doubly isotropic dynamical models of internal
degrees of freedom and use the two-polar decomposition of the
matrix $\varphi$. All symbols are just as in the spherical and
pseudospherical case, in particular, generalized coordinates
$q^{i}$, $i=\overline{1, 6}$, are ordered as follows:
\[
\vartheta, \varphi, \gamma, \delta, x, y.
\]
Angular velocities $\chi$, $\theta$ of the two-polar decomposition
are given by the expressions
\[
\chi = \chi_{\rm rl} + \chi_{\rm dr}= \frac{d \alpha}{dt}+ \sin
\vartheta \frac{d \varphi}{dt}, \qquad \theta = \frac{d
\beta}{dt},
\]
where obviously $\alpha$, $\beta$ are the primary angles of the
decomposition, i.e.,
\[
\alpha = \frac{1}{2} \left(\gamma + \delta \right), \qquad \beta =
\frac{1}{2} \left(\gamma - \delta \right).
\]

The general formula (\ref{a84}) implies then that
\[
T = T_{\rm tr} + T_{\rm int} = \frac{m}{2} G_{ij}(q) \frac{d
q^{i}}{dt} \frac{d q^{j}}{dt},
\]
where with the above ordering of generalized coordinates the
matrix $\left[G_{ij}\right]$ consists of three distinguished
blocks $M_{1}$, $M_{2}$, $M_{3}$, and namely,
\[
M_{1} = \left[R^{2}\right],
\]
\[
M_{2} = \left[
\begin{array}{ccc}
  \left(L+R \cos
\vartheta\right)^{2} + \frac{I}{m}\left(x^{2} + y^{2}\right)
\sin^{2}
\vartheta & \frac{I}{m}x^{2} \sin \vartheta & \frac{I}{m} y^{2} \sin^{2} \vartheta \\
& & \\
  \frac{I}{m}x^{2} \sin \vartheta & \frac{I}{m}x^{2} & 0 \\
  & & \\
  \frac{I}{m}y^{2} \sin^{2} \vartheta & 0 & \frac{I}{m}y^{2} \\
\end{array}
\right],
\]
\[
M_{3} = \left[
\begin{array}{cc}
   \frac{I}{m} & 0 \\
  0 &  \frac{I}{m} \\
\end{array}
\right]=  \frac{I}{m} I_{2}.
\]
The Riemann density has the form
\[
\sqrt{\left|G\right|}= R\left(\frac{I}{m}\right)^{2}
\left|xy\right| \left(L+R \cos \vartheta\right).
\]
The inverse matrix $\left[G^{ij}\right]$ consists of the blocks
\[
M_{1}{}^{-1}= \left[\frac{1}{R^{2}}\right],
\]
\[
M_{2}{}^{-1}= \left[
\begin{array}{ccc}
  \frac{1}{(L+R \cos \vartheta)^{2}} & -\frac{\sin \vartheta}{(L+R
  \cos \vartheta)^{2}} & -\frac{\sin \vartheta}{(L+R \cos \vartheta)^{2}} \\
  & & \\
  -\frac{\sin \vartheta}{(L+R \cos \vartheta)^{2}} & \frac{m}{I}
  \frac{1}{x^{2}}+\frac{\sin^{2} \vartheta}{(L+R \cos \vartheta)^{2}} &
  \frac{\sin^{2} \vartheta}{(L+R \cos \vartheta)^{2}} \\
  & & \\
  -\frac{\sin \vartheta}{(L+R \cos \vartheta)^{2}} & \frac{\sin^{2}
  \vartheta}{(L+R \cos \vartheta)^{2}} & \frac{m}{I} \frac{1}{y^{2}}+\frac{\sin^{2}
  \vartheta}{(L+R \cos \vartheta)^{2}} \\
\end{array}
\right],
\]
\[
M_{3}{}^{-1}= \left[
\begin{array}{cc}
  \frac{m}{I}  & 0 \\
  0 & \frac{m}{I} \\
\end{array}
\right] = \frac{m}{I}I_{2}.
\]
As in all previous examples, the $G$-non-orthogonality of the
natural coordinates is a serious difficulty in analysis of
integrability problems. And just as previously, one can do
something when the angles $\varphi$, $\alpha$, $\beta$ are cyclic
variables, i.e., when $V$ depends only on $\vartheta$ and
deformations invariants $x$, $y$. Namely, the system is separable
for Hamiltonians of the form
\begin{equation}\label{a160}
H=\mathcal{T}+V_{\vartheta}(\vartheta)+V_{x}(x)+V_{y}(y).
\end{equation}
When $V_{x}$, $V_{y}$ are chosen so that to prevent an unlimited
extension or contraction of the body, then they are capable to
encode something like the dynamics of nonlinear elastic
vibrations. This is even much more true when we use the polar
coordinates $(\varrho, \varepsilon)$, i.e., (\ref{a125}), on the
plane of deformation invariants:
\[
x= \varrho \sin \varepsilon, \qquad y = \varrho \cos \varepsilon.
\]
In these coordinates the matrix $\left[G_{ij}\right]$ consists
again of three blocks $K_{1}$, $K_{2}$, $K_{3}$, where $K_{1} =
M_{1}$, $K_{2}$ is just $M_{2}$ with formally substituted
(\ref{a125}), and $K_{3}$ has the form
\[
K_{3}= \left[
\begin{array}{cc}
  \frac{I}{m} & 0 \\
  0 & \frac{I}{m} \varrho^{2} \\
\end{array}
\right] = \frac{I}{m} \left[
\begin{array}{cc}
  1 & 0 \\
  0 & \varrho^{2} \\
\end{array}
\right].
\]
Just as in the spherical and pseudospherical case the problem is
separable for deformation potentials (\ref{a128}):
\[
V(\varrho, \varepsilon)= V_{\varrho}(\varrho)+
\frac{V_{\varepsilon}(\varepsilon)}{\varrho^{2}},
\]
i.e., for the total potentials
\[
V(\vartheta, \varrho, \varepsilon)= V_{\vartheta}(\vartheta)+
V_{\varrho}(\varrho)+
\frac{V_{\varepsilon}(\varepsilon)}{\varrho^{2}}.
\]
In particular, (\ref{a131}) is very interesting from the point of
view of nonlinear elasticity.

The general procedure of constructing the action $J$-variables and
discussing degeneracy problems is just like in the spherical and
pseudospherical cases. Let us write down some explicit formulas.

According to our assumption about independence of $V$ of the
angles $\varphi$, $\alpha$, $\beta$ we have the following forms of
the reduced action $S_{0}$:
\begin{eqnarray}
S_{0}&=& S_{\vartheta}(\vartheta)+S_{x}(x)+S_{y}(y)+ \ell \varphi
+ s \alpha + j \beta \nonumber \\
&=&
S_{\vartheta}(\vartheta)+S_{\varrho}(\varrho)+S_{\varepsilon}(\varepsilon)+
\ell \varphi + s \alpha + j \beta, \nonumber
\end{eqnarray}
depending on whether we use the $(x,y)$- or $(\varrho,
\varepsilon)$-variables in the plane of deformations invariants.
Therefore, just as previously,
\begin{eqnarray}
J_{\alpha}&=& \oint p_{\alpha}d \alpha =
\int\limits_{0}^{2 \pi} s d \alpha = 2 \pi s,\nonumber \\
J_{\beta}&=& \oint p_{\beta}d \beta = \int\limits_{0}^{2 \pi} j d
\beta =
2 \pi j,\nonumber \\
J_{\varphi}&=& \oint p_{\varphi}d \varphi = \int\limits_{0}^{2
\pi} \ell d \varphi = 2 \pi \ell, \nonumber
\end{eqnarray}
where $s$, $j$, $\ell$ denote respectively the constant fixed
values of spin, vorticity, and ``orbital" angular momenta.

If Hamiltonian has the form (\ref{a160}), then the separation of
variables procedure tells us that
\begin{equation}
J_{x} = \oint \sqrt{2I \left(C_{x} - V_{x}(x)\right)-
\frac{\left(J_{\alpha}+J_{\beta}\right)^{2}}{16 \pi^{2} x^{2}}}
dx,\label{a162}
\end{equation}
\begin{equation}
J_{y} = \oint \sqrt{2I \left(C_{y} - V_{y}(y)\right)-
\frac{\left(J_{\alpha}-J_{\beta}\right)^{2}}{16 \pi^{2} y^{2}}}
dy, \label{b162}
\end{equation}
\begin{equation}
J_{\vartheta} = R \oint \sqrt{2m \left(E - C_{x}- C_{y} -
V_{\vartheta}(\vartheta)\right)- \frac{\left(J_{\varphi}-
J_{\alpha} \sin \vartheta \right)^{2}}{4 \pi^{2} \left(L+R \cos
\vartheta\right)^{2}}} d\vartheta. \label{a163}
\end{equation}
And again the procedure is like previously. When some explicit
forms of $V_{x}$, $V_{y}$, $V_{\vartheta}$ are assumed, the
integrals (\ref{a162}), (\ref{b162}) are ``calculated", and then
$C_{x}$, $C_{y}$ are expressed as functions of $J_{x}$,
$J_{\alpha}$, $J_{\beta}$, i.e.,
\[
C_{x}=C_{x}\left(J_{x}, J_{\alpha} + J_{\beta}\right), \qquad
C_{y}=C_{y} \left(J_{y}, J_{\alpha} - J_{\beta}\right).
\]
Substituting this to (\ref{a163}) and ``calculating" the integral,
we express $J_{\vartheta}$ through $E$, $J_{\alpha}$, $J_{\beta}$,
$J_{\varphi}$, $J_{x}$, $J_{y}$. And finally, solving this
expression with respect to $E$, we determine $E$ as a function of
action variables, i.e.,
\[
E = {\mathcal{H}}\left(J_{\vartheta}, J_{\varphi}, J_{\alpha},
J_{\beta}, J_{x}, J_{y}\right).
\]
And unfortunately again without explicit calculations nothing may
be decided a priori concerning even some partial degeneracy.

\section{Three-dimensional problems}

Now we are going to discuss the motion of structured material
points in three dimensions. We concentrate on the very interesting
special case of the Einstein Universe, i.e., compact
constant-curvature space metrically diffeomorphic with the
three-dimensional sphere $S^{3}(0, R) \subset {\mathbb{R}}^{4}$.
This three-dimensional model is quite different from the
previously studied two-dimen\-sional ones. The point is that the
spheres $S^{3}(0,R)$ are parallelizable, moreover, $S^{3}(0,1)$
may be identified in a natural way with the group SU$(2)$. This
means that the most natural choice of the auxiliary reference
frame $E$ is that of basic generators of left or right regular
translations on SU$(2)$, i.e., respectively, the basic system of
right- or left-invariant vector fields on SU$(2)$.  This frame may
be easily chosen orthonormal. And all this means that the
principal bundle of orthonormal frames over SU$(2)$, $FM({\rm
SU}(2),g)$ may be naturally identified with the Cartesian product
SU$(2) \times {\rm SO}(3,{\mathbb{R}})$. The metric $g$ here is
obtained by the natural restriction of the Euclidean metric in
${\mathbb{R}}^{4}$ to $S^{3}(0,R)$ and in the SU$(2)$-terms it is
identical up to a constant factor with the Killing-Cartan metric
on SU$(2)$. The latter is invariant under left and right regular
translations on the group SU$(2)$.

Before going any further we return temporarily to the more general
$n$-dimensional formulation. This makes our description more
lucid. So, we return to the general formula (\ref{a74}) for the
co-moving affine velocity $\widehat{\Omega}$, i.e,
\begin{equation}\label{b165}
\widehat{\Omega}^{B}{}_{A}= \varphi^{-1 B}{}_{F}\Gamma^{F}{}_{DC}
\varphi^{D}{}_{A}\varphi^{C}{}_{E}\widehat{V}^{E}+ \varphi^{-1
B}{}_{C} \frac{d \varphi^{C}{}_{A}}{dt}.
\end{equation}
Let us remind that $\widehat{\Omega}$ is defined by the formula
\[
\frac{D e_{A}}{D t}= e_{B}\widehat{\Omega}^{B}{}_{A},
\]
i.e.,
\[
\widehat{\Omega}^{B}{}_{A}= \left\langle e^{B}, \frac{D e_{A}}{D
t} \right\rangle = e^{B}{}_{i} \frac{D e^{i}{}_{A}}{D t}.
\]
When the internal motion is gyroscopic, i.e., $\varphi \in {\rm
SO}(n, \mathbb{R})$, then obviously $\widehat{\Omega}$ is
skew-symmetric in the Kronecker-delta sense, i.e.,
\[
{\widehat{\Omega}}^{B}{}_{A}= -{\widehat{\Omega}}_{A}{}^{B} = -
\delta_{AC} \delta^{BD}{\widehat{\Omega}}^{C}{}_{D}.
\]
In the above formula (\ref{b165}) $\widehat{V}^{A}$ denote the
co-moving components of the translational velocity, i.e.,
\[
\widehat{V}^{A}= e^{A}{}_{i}V^{i}= e^{A}{}_{i}\frac{dx^{i}}{dt}.
\]
And we shall use the abbreviations (\ref{a81}), i.e.,
\begin{eqnarray}
{\widehat{\Omega}}_{\rm dr}{}^{A}{}_{B}&=& \varphi^{-1 A}{}_{F}
\Gamma^{F}{}_{DC}\varphi^{D}{}_{B}\varphi^{C}{}_{E}\widehat{V}^{E},\label{a165a}\\
{\widehat{\Omega}}_{\rm rl}{}^{A}{}_{B}&=& \varphi^{-1 A}{}_{C}
\frac{d \varphi^{C}{}_{B}}{dt}, \qquad \Omega_{\rm
rl}{}^{A}{}_{B}= \frac{d \varphi^{A}{}_{C}}{dt} \varphi^{-1
C}{}_{B},\label{a165b}
\end{eqnarray}
thus,
\[
\widehat{\Omega}= \widehat{\Omega}_{\rm dr}+\widehat{\Omega}_{\rm
rl} = \widehat{\Omega}_{\rm dr} + \varphi^{-1} \Omega_{\rm rl}
\varphi.
\]

If, as we always assume, the auxiliary frame $E$ is
$g$-orthonormal and $\Gamma$ is the $g$-Levi-Civita connection, or
more generally it is some Riemann-Cartan connection, then the
aholonomic components $\Gamma^{A}{}_{BC}$ are
Kronecker-skew-symmetric in the first pair of indices, i.e.,
\[
\Gamma^{A}{}_{BC} = - \Gamma_{B}{}^{A}{}_{C}=
-\delta_{BK}\delta^{AL} \Gamma^{K}{}_{LC}.
\]
It is not true for the holonomic components $\Gamma^{i}{}_{jk}$
because they are not $g$-skew-symmetric,
\[
\Gamma^{i}{}_{jk} \neq - \Gamma_{j}{}^{i}{}_{k}= -
g_{ja}g^{ib}\Gamma^{a}{}_{bk}.
\]

Let us now assume that the ``legs" $E_{A}$ of the auxiliary field
of frames $E$ form a Lie algebra, i.e.,
\[
\left[E_{A}, E_{B}\right]= C^{K}{}_{AB} E_{K},
\]
where $C^{K}{}_{AB}$ are some constants, just the structure
constants of the algebra with respect to the basis $(\ldots,E_{A},
\ldots)$. Lie bracket of vector fields in the above formula is
meant in the convention
\[
\left[X,Y\right]^{i}= X^{j}Y^{i}{}_{,j}- Y^{j}X^{i}{}_{,j},
\]
i.e., as the commutator of $X$, $Y$ when the vector fields are
identified with first-order differential operators:
\[
Xf = X^{i} f_{,i}.
\]

If some point $x_{0} \in M$ is fixed, then the manifold $M$ may be
identified with some Lie group and $x_{0}$ with its identity.
Without distinguished $x_{0}$ the manifold $M$ is a homogeneous
space of the corresponding Lie group with trivial isotropy groups.
In the Abelian case with the $\mathbb{R}^{n}$-topology of $M$ this
is just the difference between the linear and affine spaces.

The algebraic Killing metric of the Lie algebra with structure
constants $C^{K}{}_{LM}$ has with respect to the basis
$(\ldots,E_{A},\ldots)$ the components
\[
\gamma_{AB}= C^{K}{}_{LA} C^{L}{}_{KB}.
\]
The corresponding Killing tensor field on $M$ is given in local
coordinates as follows:
\[
\gamma_{ij} = \gamma_{AB}E^{A}{}_{i}E^{B}{}_{j},
\]
i.e., in the absolute tensor notation we have that
\[
\gamma = \gamma_{AB}E^{A} \otimes E^{B}.
\]
Obviously, it is not singular if and only if the Lie algebra given
by structure constants $C$ is semi-simple. It has the elliptic
signature, i.e., is essentially Riemannian when the underlying Lie
group is compact. Then strictly speaking the metric is negatively
definite, but changing its over-all sign we obtain the usual
positive metric. The inverse algebraic metric $\gamma^{AB}$
satisfies the condition
\[
\gamma^{AC}\gamma_{CB}= \delta^{A}{}_{B},
\]
thus, the contravariant metric field on $M$ is given as follows:
\[
\gamma^{ij} = E^{i}{}_{A}E^{j}{}_{B}\gamma^{AB},
\]
or in the index-free form:
\[
\widetilde{\gamma} = \gamma^{AB}E_{A} \otimes E_{B}.
\]

While $\gamma_{ij}E^{i}{}_{A}E^{j}{}_{B}=\gamma_{AB}={\rm const}$,
then the metric tensors $\gamma$, $g$ on $M$ define practically
"the same" geometries. And from now on we assume that $\gamma$ and
$g$ just coincide at least up to a constant multiplier. Let us
mention by the way that this multiplier does not affect the
Levi-Civita connection.

Let $\Gamma[E]$ denote the teleparallelism affine connection
(\ref{a72}), (\ref{b72}) built of $E$ and $S[E]$ is its torsion
tensor (\ref{c72}). It follows from (\ref{a73}), (\ref{a74}) that
$S[E]^{A}{}_{BC}$, i.e., the aholonomic coefficients of $S[E]$
with respect to $E$, are given as follows:
\[
S\left[E\right]^{A}{}_{BC}= \frac{1}{2}C^{A}{}_{BC}.
\]
We have just assumed that the metric tensors $\gamma$, $g$ differ
by an overall constant multiplier $\lambda$, thus,
\begin{equation}\label{a168}
\gamma_{AB}=\lambda \delta_{AB},\qquad \gamma_{ij}=\lambda g_{ij}.
\end{equation}
Their common Levi-Civita connection will be denoted by $\left\{
\right\}$ or componentwisely by $\left\{\begin{array}{c}
  i \\
  jk
\end{array}\right\}$. According to the assumed relationships between $g$, $E$, $\gamma$,
the metric tensors $\gamma$, $g$ are parallel under the connection
$\Gamma[E]$ (i.e., the teleparallelism connection is metrical with
respect to $\gamma$, $g$). Therefore, following (\ref{a42}),
\[
\Gamma \left[E\right]^{i}{}_{jk} = \left\{\begin{array}{c}
  i \\
  jk
\end{array}\right\} + S^{i}{}_{jk}+S_{jk}{}^{i} + S_{kj}{}^{i},
\]
i.e., in the aholonomic terms:
\[
\Gamma \left[E\right]^{K}{}_{LM} = \left\{\begin{array}{c}
  K \\
  LM
\end{array}\right\} + \frac{1}{2} C^{K}{}_{LM}+\frac{1}{2}
C_{LM}{}^{K}- \frac{1}{2} C_{M}{}^{K}{}_{L}.
\]
The shift of ``small" holonomic indices is meant here in the sense
of $g_{ij}$, whereas the shift of aholonomic ``capital" indices is
meant in the Kro\-necker-delta sense. But the $E$-aholonomic
components of the $E$-tele\-pa\-rallelism connection evidently
vanish, i.e., $\Gamma [E]^{K}{}_{LM} = 0$, so we obtain that
\[
\left\{\begin{array}{c}
  K \\
  LM
\end{array}\right\} =- \frac{1}{2} C^{K}{}_{LM}-\frac{1}{2}
C_{LM}{}^{K}+ \frac{1}{2} C_{M}{}^{K}{}_{L}.
\]
In semi-simple Lie algebras the structure constants are
skew-symmetric in the first pair of indices with respect to the
Killing metric. As we have assumed the latter to be proportional
to the Kronecker delta, then we finally obtain that
\[
\left\{\begin{array}{c}
  K \\
  LM
\end{array}\right\} = -\frac{1}{2} C_{M}{}^{K}{}_{L} = -\frac{1}{2}
C^{K}{}_{LM}.
\]
So, finally (\ref{a165a}), (\ref{a165b}) imply that
\begin{eqnarray}
\widehat{\Omega}^{A}{}_{B}&=& \widehat{\Omega}_{\rm
dr}{}^{A}{}_{B}+
\widehat{\Omega}_{\rm rl}{}^{A}{}_{B} \nonumber \\
&=& -
\frac{1}{2}\varphi^{-1A}{}_{F}C^{F}{}_{DC}\varphi^{D}{}_{B}\varphi^{C}{}_{E}{\widehat{V}}^{E}
+ \varphi^{-1A}{}_{C} \frac{d \varphi^{C}{}_{B}}{dt}.\nonumber
\end{eqnarray}
It turns out that in certain formulas for kinetic energy it may be
convenient to use other representations for the relative
(internal) aholonomic velocity. Namely, from the point of view of
GL$(n, \mathbb{R})$-objects $\widehat{\Omega}_{\rm rl}$ is the
"co-moving" affine/angular velocity. One can also use the
"spatial" representation in ${\mathbb{R}}^{n}$, i.e.,
\[
\Omega_{\rm rl}{}^{A}{}_{B}= \frac{d \varphi^{A}{}_{C}}{dt}
\varphi^{-1C}{}_{B} = \varphi^{A}{}_{C}\widehat{\Omega}_{\rm
rl}{}^{C}{}_{D}\varphi^{-1D}{}_{B}.
\]
There are also some additional possibilities concerning the
$M$-representa\-tion, i.e., literally understood spatial
representation. Obviously, the most geometric one is that based on
the $e$-injection, i.e.,
\begin{eqnarray}
\Omega^{i}{}_{j}&=& e^{i}{}_{A} {\widehat{\Omega}}^{A}{}_{B}
e^{B}{}_{j}=
\frac{De^{i}{}_{A}}{Dt} e^{A}{}_{j}, \nonumber \\
\Omega_{\rm rl}{}^{i}{}_{j}&=& e^{i}{}_{A} {\widehat{\Omega}}_{\rm
rl}{}^{A}{}_{B} e^{B}{}_{j}, \qquad \varphi^{i}{}_{j} =
e^{i}{}_{A} \varphi^{A}{}_{B}e^{B}{}_{j}, \nonumber
\end{eqnarray}
and so on. But one can also use the $E$-injection of
${\mathbb{R}}^{n}$ into the tangent spaces of $M$, i.e.,
\[
\widetilde{\Omega}_{\rm rl}{}^{i}{}_{j}= E^{i}{}_{A}
\widehat{\Omega}_{\rm rl}{}^{A}{}_{B} E^{B}{}_{j}, \qquad
\widetilde{V}^{i}= E^{i}{}_{A}{\widehat{V}}^{A}, \qquad
\widetilde{\varphi}^{i}{}_{j}= E^{i}{}_{A} \varphi^{A}{}_{B}
E^{B}{}_{j},
\]
and so on.

Let us now go back to the three-dimensional simple compact groups
SU$(2)$, SO$(3, \mathbb{R})$. In the standard representation
structure constants are given by the three-dimensional Ricci
symbol:
\[
C^{K}{}_{LM} = \varepsilon^{K}{}_{LM}, \qquad C_{KLM} =
\varepsilon_{KLM}
\]
(again the shift of indices is meant in the Kronecker sense).

Some analytical expressions must be quoted now. Lie algebra of
SU$(2)$ consists of trace-less anti-hermitian $2 \times 2$
matrices and it is customary to use the standard basis $a \in {\rm
SU}(2)^{\prime}$, $j = 1,2,3$, where
\[
a_{B} = \frac{1}{2i} \sigma_{B},
\]
and $\sigma_{B}$ denote the Pauli matrices:
\[
\sigma_{1} =\left[
\begin{array}{cc}
  0 & 1 \\
  1 & 0 \\
\end{array}
\right] , \qquad \sigma_{2} =\left[
\begin{array}{cc}
  0 & -i \\
  i & 0 \\
\end{array}
\right] , \qquad \sigma_{3} = \left[
\begin{array}{cc}
  1 & 0 \\
  0 & -1 \\
\end{array}
\right].
\]
Using canonical coordinates of the first kind $\overline{k} \in
{\mathbb{R}}^{3}$, we have the expressions for the
SU$(2)$-elements
\begin{equation} \label{a171}
u\left(\overline{k}\right) = \exp \left(k^{B} a_{B} \right) =
I_{2}\cos \frac{k}{2}-i\sigma_{B}\frac{k^{B}}{k}\sin \frac{k}{2},
\end{equation}
where $I_{2}= \left[
\begin{array}{cc}
  1 & 0 \\
  0 & 1 \\
\end{array}
\right]$ is the $2 \times 2$ identity matrix and
\[
k=\sqrt{\overline{k} \cdot \overline{k}}=
\sqrt{\left(k^{1}\right)^{2}+ \left(k^{2}\right)^{2}+
\left(k^{3}\right)^{2}}
\]
is the Euclidean length of $\overline{k}$. The direction
$\overline{k}/k \in S^{2}(0, 1)$ of $\overline{k}$ runs over the
total indicated range, whereas the modulus $k$ changes between $0$
and $2 \pi$. All vectors $\overline{k}$ with $k = 2 \pi$ refer to
the same group element $-I_{2} \in {\rm SU}(2)$. For the values of
$k$ larger than $2 \pi$ the formula (\ref{a171}) would repeat the
former elements of SU$(2)$. In this way $k=0$ and $k=2 \pi$ are
singularities of the above coordinate system. Obviously, $a_{J}$
obey the standard commutation rules, i.e.,
\[
\left[a_{J}, a_{K}\right] = \varepsilon^{M}{}_{JK} a_{M}.
\]
The Lie algebra of SO$(3, \mathbb{R})$, i.e., SO$(3,
\mathbb{R})^{\prime}$, consists of real $3 \times 3$
skew-symmetric matrices. The standard basis is given by $A_{K}$,
$K= 1,2,3$, where
\[
\left(A_{K}\right)^{L}{}_{M} = -\varepsilon_{K}{}^{L}{}_{M},
\]
and the trivial (purely cosmetic) Kronecker shift of indices is
meant. With this convention we have the same commutation rules,
i.e.,
\[
\left[A_{J}, A_{K}\right] = \varepsilon^{M}{}_{JK} A_{M}.
\]
The exponential mapping
\begin{equation} \label{a172}
R\left(\overline{k}\right) = \exp\left(k^{J} A_{J}\right)
\end{equation}
leads to the explicit form of finite rotation matrices
$R\left(\overline{k}\right)$:
\[
R\left(\overline{k}\right) \cdot \overline{x} = \cos k \cdot
\overline{x} +(1 - \cos k) \left(\frac{\overline{k}}{k}\cdot
\overline{x}\right) \frac{\overline{k}}{k} + \sin k \;
\frac{\overline{k}}{k} \times \overline{x},
\]
with the standard notation $\overline{a} \cdot \overline{b}$ and
$\overline{a} \times \overline{b}$ respectively for the scalar and
vector product in ${\mathbb{R}}^{3}$.

Now $k$ runs over the range $[0, \pi]$ and antipodal points on the
coordinate sphere $k= \pi$ in ${\mathbb{R}}^{3}$ are pairwise
identified because
\[
R(\pi \overline{n}) = R(- \pi \overline{n})
\]
for any versor $\overline{n} \in {\mathbb{R}}^{3}$, $\overline{n}
\cdot \overline{n} =1$. In this way $\overline{k}$ is the usual
rotation vector, $k$ is the rotation angle (in radians), and
$\overline{n} = \overline{k}/k$ is the rotation axis in the
right-hand-screw sense, $R(\overline{k})\overline{k}=
\overline{k}$. In particular, we have the intuitive formula
\[
R(\overline{k}) \overline{x} = \overline{x} +\overline{k} \times
\overline{x} +\frac{1}{2!} \overline{k} \times (\overline{k}
\times \overline{x}) + \cdots + \frac{1}{n!} \overline{k} \times
(\overline{k} \times \cdots (\overline{k} \times \overline{x})
\ldots ) + \cdots,
\]
where the term with the $1/n!$-factor contains $n$ copies of
$\overline{k}$. The natural homomorphism
\[
{\rm pr}:{\rm SU}(2) \rightarrow {\rm SO}(3, \mathbb{R})
\]
is given by $v \mapsto R$, where
\[
vu(\overline{k})v^{-1} = u(R \overline{k}).
\]
Therefore,
\[
{\rm pr}^{-1}\left(R(\overline{k})\right)= \left\{
u(\overline{k}), u \left(-(2 \pi - k)
\frac{\overline{k}}{k}\right)\right\} = \left\{u(\overline{k}),
-u(\overline{k}) \right\}.
\]

The formula (\ref{a171}) injects the group SU$(2)$ into the real
four-dimensional space spanned by matrices $I_{2}$, $-i
\sigma_{B}$, $B=1,2,3$. Obviously, this space may be identified
with ${\mathbb{R}}^{4}$. It is seen that the sum of squared
coefficients at the mentioned matrices equals one. This is just
the mentioned identification between SU$(2)$ and three-dimensional
unit sphere $S^{3}(0,1) \subset {\mathbb{R}}^{4}$.

The Lie-algebraic Killing metric built of the structure constants
$\varepsilon^{A}{}_{BC}$ has the components
\[
\gamma_{AB} = -2 \delta_{AB},
\]
therefore, in (\ref{a168}) we have that
\[
\lambda= -2.
\]

The metric field on SU$(2)$ or SO$(3, \mathbb{R})$ is analytically
given as follows:
\begin{equation} \label{a174}
g_{ij} = \frac{4}{k^{2}}\sin^{2}\frac{k}{2}\delta_{ij} + \left( 1
- \frac{4}{k^{2}}\sin^{2}\frac{k}{2}\right) \frac{k^{i}}{k}
\frac{k^{j}}{k},
\end{equation}
i.e., the corresponding arc element has the form
\begin{equation} \label{b174}
ds^{2} = dk^{2} + 4 \sin^{2} \frac{k}{2} \left( d \vartheta^{2}+
\sin^{2} \vartheta d\varphi^{2}  \right),
\end{equation}
where $k$, $\vartheta$, $\varphi$ are spherical variables in the
space of rotation vectors $\overline{k}$. The angular variables
$\vartheta$, $\varphi$ parameterize the manifold of rotation
versors $\overline{n}(\vartheta, \varphi)= \overline{k}/k$ and we
can write that
\[
ds^{2} = dk^{2} + 4 \sin^{2} \frac{k}{2} d\overline{n} \cdot
d\overline{n},
\]
or using more sophisticated tensorial terms:
\[
g = dk \otimes dk +4 \sin^{2} \frac{k}{2} \delta_{AB} dn^{A}
\otimes dn^{B}.
\]

The reference frame $E$ may be chosen either as ${}^{l}E$ or
${}^{r}E$ respectively consisting of vector fields generating left
or right translations on SU$(2)$, SO$(3, \mathbb{R})$. The vector
fields ${}^{l}E_{A}$, ${}^{r}E_{A}$ are respectively right- and
left-invariant. Let ${}^{l}{\widetilde{E}}$,
${}^{r}{\widetilde{E}}$ be their dual co-frames. The covector
fields (differential one-forms) ${}^{l}E^{A}$, ${}^{r}E^{A}$ are
respectively right- and left-invariant.

After some calculations one obtains the analytical expressions in
terms of coordinates $\overline{k}$:
\begin{eqnarray}
{}^{l}E_{A} &=& \frac{k}{2} {\rm ctg} \frac{k}{2}
\frac{\partial}{\partial k^{A}} + \left(1 -   \frac{k}{2} {\rm
ctg} \frac{k}{2} \right) \frac{k_{A}}{k} \frac{k^{J}}{k}
\frac{\partial}{\partial k^{J}} + \frac{1}{2}
\varepsilon_{AM}{}^{J} k^{M}
\frac{\partial}{\partial k^{J}}, \nonumber \\
{}^{r}E_{A} &=& \frac{k}{2} {\rm ctg} \frac{k}{2}
\frac{\partial}{\partial k^{A}} + \left(1 -   \frac{k}{2} {\rm
ctg} \frac{k}{2} \right) \frac{k_{A}}{k} \frac{k^{J}}{k}
\frac{\partial}{\partial k^{J}} - \frac{1}{2}
\varepsilon_{AM}{}^{J} k^{M}
\frac{\partial}{\partial k^{J}}, \nonumber \\
{}^{l}E^{A} &=& \frac{\sin k}{k} dk^{A} + \left( 1- \frac{\sin
k}{k} \right) \frac{k^{A}}{k} \frac{k_{B}}{k} dk^{B} + \frac{2
}{k^{2}}\sin^{2} \frac{k}{2}\
\varepsilon^{A}{}_{BC}k^{B} dk^{C}, \nonumber \\
{}^{r}E^{A} &=& \frac{\sin k}{k} dk^{A} + \left( 1- \frac{\sin
k}{k} \right) \frac{k^{A}}{k} \frac{k_{B}}{k} dk^{B} - \frac{2
}{k^{2}}\sin^{2} \frac{k}{2}\ \varepsilon^{A}{}_{BC}k^{B} dk^{C}.
\nonumber
\end{eqnarray}
The vector fields
\[
D_{A} := {}^{l}E_{A} - {}^{r}E_{A} = \varepsilon_{AB}{}^{C} k^{B}
\frac{\partial}{\partial k^{C}}
\]
are infinitesimal generators of inner automorphisms, i.e.,
\[
u \mapsto vuv^{-1}.
\]

It is convenient to separate explicitly the ``radial" variable $k$
and the angles $\vartheta$, $\varphi$. To do this in a symmetric
way, we use differential operators (vector fields) $D_{A}$ which
depend only on $\vartheta$, $\varphi$ and act only on these
variables, i.e.,
\[
D_{A} f = 0,
\]
if $f$ is a function of the variable $k$. Similarly, the redundant
system of quantities $\overline{n} =\overline{k}/k$ are used.
Obviously, $n^{A}$ depend only on $\vartheta$, $\varphi$, but not
on $k$.

Then one can show that
\begin{eqnarray}
{}^{l}E_{A} &=& n_{A} \frac{\partial}{\partial k} - \frac{1}{2}
{\rm ctg}
\frac{k}{2}\ \varepsilon_{ABC} n^{B} D^{C} + \frac{1}{2} D_{A}, \nonumber \\
{}^{r}E_{A} &=& n_{A} \frac{\partial}{\partial k} - \frac{1}{2}
{\rm ctg}
\frac{k}{2}\ \varepsilon_{ABC} n^{B} D^{C} - \frac{1}{2} D_{A}, \nonumber \\
{}^{l}E^{A} &=& n^{A} d k + 2 \sin^{2} \frac{k}{2}\
\varepsilon^{ABC} n_{B} dn_{C} + \sin k\ dn^{A}, \nonumber \\
{}^{r}E^{A} &=& n^{A} d k - 2 \sin^{2} \frac{k}{2}\
\varepsilon^{ABC} n_{B} dn_{C} + \sin k\ dn^{A}, \nonumber
\end{eqnarray}
or in a brief, little symbolic way:
\begin{eqnarray}
{}^{l}\overline{E} &=& \overline{n} \frac{\partial}{\partial k} -
\frac{1}{2} {\rm ctg} \frac{k}{2}\ \overline{n} \times
\overline{D} +
\frac{1}{2} \overline{D}, \nonumber \\
{}^{r}\overline{E} &=& \overline{n} \frac{\partial}{\partial k} -
\frac{1}{2} {\rm ctg} \frac{k}{2}\ \overline{n} \times
\overline{D} -
\frac{1}{2} \overline{D}, \nonumber \\
{}^{l}\widetilde{E} &=& \overline{n} d k + 2 \sin^{2} \frac{k}{2}\
\overline{n} \times d\overline{n} + \sin k\ d\overline{n}, \nonumber \\
{}^{r}\widetilde{E} &=& \overline{n} d k - 2 \sin^{2} \frac{k}{2}\
\overline{n} \times d\overline{n} + \sin k\ d\overline{n}.
\nonumber
\end{eqnarray}
Obviously,
\begin{eqnarray}
\left\langle dk, D_{A}\right\rangle = D_{A} k = 0, &\qquad&
\left\langle dk, \frac{\partial}{\partial k}\right\rangle = 1,
\nonumber\\
\left\langle dn_{A}, \frac{\partial}{\partial k} \right\rangle =
\frac{\partial n_{A}}{\partial k}= 0, &\qquad& \left\langle
dn_{A}, D_{B}\right\rangle = D_{B} n_{A} = \varepsilon_{ABC}
n^{C}.\nonumber
\end{eqnarray}
Commutation relations between the above vector fields read as
follows:
\[
\left[{}^{l}E_{A}, {}^{l}E_{B} \right]= -\varepsilon_{AB}{}^{C}\
{}^{l}E_{C}, \qquad \left[{}^{r}E_{A}, {}^{r}E_{B} \right]=
\varepsilon_{AB}{}^{C}\ {}^{r}E_{C},
\]
\[
\left[{}^{l}E_{A}, {}^{r}E_{B} \right]=0.
\]
The vanishing of the last Lie bracket is due to the fact that the
left and right group translations mutually commute.

The metric tensor $g$ (see (\ref{a174}), (\ref{b174})), i.e., the
$(-2)$-multiple of the Killing tensor, may be written down as
follows:
\[
g = \delta_{AB}{}^{l}E^{A} \otimes {}^{l}E^{B} =
\delta_{AB}{}^{r}E^{A} \otimes {}^{r}E^{B}.
\]
Similarly, the contravariant inverse metric
\[
g^{ij} = \frac{k^{2}}{4 \sin^{2} \frac{k}{2}} \delta^{ij} + \left(
1 - \frac{k^{2}}{4 \sin^{2} \frac{k}{2}}  \right) n^{i} n^{j}
\]
may be expressed as follows:
\[
\widetilde{g} = \delta^{AB}{}^{l}E_{A} \otimes {}^{l}E_{B} =
\delta^{AB}{}^{r}E_{A} \otimes {}^{r}E_{B},
\]
or using the $k,\overline{n}$-variables:
\[
\widetilde{g} = \frac{\partial}{\partial k} \otimes
\frac{\partial}{\partial k} + \frac{1}{4 \sin^{2} \frac{k}{2}}
\delta^{AB} D_{A} \otimes D_{B}.
\]

Let us now pass over from the standard unit sphere $S^{3}(0,1)
\simeq {\rm SU}(2)$ to the sphere $S^{3}(0,R) \subset
{\mathbb{R}}^{4}$ of the radius $R$. We introduce the new
``radial" variable $r=Rk/2$, i.e., $k=2r/R$. Obviously, the range
of $r$ is $[0,  \pi R]$. The coordinate vector $\overline{k}$ is
replaced by $\overline{r}$ with the length $r$ and such that
\[
\frac{\overline{r}}{r} = \frac{\overline{k}}{k} = \overline{n}.
\]
Expressions for the auxiliary reference frames $E$ and co-frames
$\widetilde{E}$ take on the forms
\begin{eqnarray}
{}^{l}E_{A} = \frac{R}{2}{}^{l}E(R)_{A}, &\qquad& {}^{r}E_{A} =
\frac{R}{2}{}^{r}E(R)_{A},\nonumber\\
{}^{l}E^{A} = \frac{2}{R}{}^{l}E(R)^{A}, &\qquad& {}^{r}E^{A} =
\frac{2}{R}{}^{r}E(R)^{A},\nonumber
\end{eqnarray}
where
\begin{eqnarray}
{}^{l}E(R)_{A} &=& n_{A} \frac{\partial}{\partial r} -
\frac{1}{R} {\rm ctg} \frac{r}{R} \varepsilon_{ABC} n^{B} D^{C} +\frac{1}{R} D_{A}, \label{a178a} \\
{}^{r}E(R)_{A} &=& n_{A} \frac{\partial}{\partial r} -
\frac{1}{R} {\rm ctg} \frac{r}{R} \varepsilon_{ABC} n^{B} D^{C} -\frac{1}{R} D_{A}, \label{a178b} \\
{}^{l}E(R)^{A} &=& n^{A} dr+ R\sin^{2} \frac{r}{R}
\varepsilon^{ABC}n_{B} dn_{C} + \frac{R}{2} \sin \frac{2r}{R}dn^{A}, \label{a178c} \\
{}^{r}E(R)^{A} &=& n^{A} dr- R\sin^{2} \frac{r}{R}
\varepsilon^{ABC}n_{B} dn_{C} + \frac{R}{2} \sin
\frac{2r}{R}dn^{A}. \label{a178d}
\end{eqnarray}
The basic Lie brackets are as follows:
\begin{eqnarray}
\left[{}^{l}E(R)_{A}, {}^{l}E(R)_{B} \right] &=& - \frac{2}{R}
\varepsilon_{AB}{}^{Cl}E(R)_{C}, \label{b178a} \\
\left[{}^{r}E(R)_{A}, {}^{r}E(R)_{B} \right] &=&  \frac{2}{R}
\varepsilon_{AB}{}^{Cr}E(R)_{C}, \label{b178b} \\
\left[{}^{l}E(R)_{A}, {}^{r}E(R)_{B} \right] &=& 0. \label{b178c}
\end{eqnarray}
Therefore, the corresponding structure constants are given as
follows:
\[
{}^{l}C(R)^{A}{}_{BC} = - \frac{2}{R} \varepsilon^{A}{}_{BC},
\qquad {}^{r}C(R)^{A}{}_{BC} = \frac{2}{R} \varepsilon^{A}{}_{BC},
\]
and the trivial (cosmetic) Kronecker shift of indices is meant.

The resulting metric
\[
g(R) = \delta_{AB}{}^{l}E(R)^{A} \otimes {}^{l}E(R)^{B}=
\delta_{AB}{}^{r}E(R)^{A} \otimes {}^{r}E(R)^{B}
\]
has the coordinate form
\[
g(R)_{ij} = \left(\frac{R}{r}\right)^{2}\sin^{2}\frac{r}{R}
\delta_{ij}+ \left( 1 - \left(\frac{R}{r}\right)^{2} \sin^{2}
\frac{r}{R}\right) n_{i}n_{j},
\]
i.e.,
\begin{eqnarray}
ds^{2} &=& dr^{2} + R^{2} \sin^{2}\frac{r}{R} \left(d
\vartheta^{2} + \sin^{2} \vartheta d \varphi^{2}\right)
\nonumber\\
&=&  dr^{2} + R^{2} \sin^{2}\frac{r}{R} d\overline{n} \cdot
d\overline{n},\label{a179}
\end{eqnarray}
or in more sophisticated terms:
\[
g(R) = dr \otimes dr + R^{2} \sin^{2}\frac{r}{R} \delta_{AB}
dn^{A} \otimes dn^{B}.
\]
The algebraic Killing metric $\gamma_{AB}$ is now given as
follows:
\[
\gamma_{AB} = -\frac{8}{R^{2}}\delta_{AB},
\]
therefore,
\[
\lambda = - \frac{8}{R^{2}}.
\]

The above analytical formulas may be geometrically interpreted in
such a way that the unit sphere in ${\mathbb{R}}^{4}$ is subject
to the $R$-factor dilatation resulting in $S^{3}(0,R)$, i.e., the
sphere of radius $R$ and origin $0$ in ${\mathbb{R}}^{4}$. The
metric $g(R)$ may be obtained by restriction of the Kronecker
metric in ${\mathbb{R}}^{4}$ to $S^{3}(0,R)$.

Parametrizing $S^{3}(0,R) \subset {\mathbb{R}}^{4}$ by $(r,
\vartheta, \varphi)$-coordinates, i.e.,
\begin{eqnarray}
x^{1} = R \sin \frac{r}{R} \sin \vartheta \cos \varphi,& \qquad&
x^{2} = R \sin \frac{r}{R} \sin \vartheta \sin \varphi, \nonumber \\
x^{3} = R \sin \frac{r}{R} \cos \vartheta,& \qquad& x^{4} = R \cos
\frac{r}{R}, \nonumber
\end{eqnarray}
and expressing
\[
ds^{2} = \left(dx^{1}\right)^{2} + \left(dx^{2}\right)^{2} +
\left(dx^{3}\right)^{2} + \left(dx^{4}\right)^{2}
\]
through $(r, \vartheta, \varphi)$, one obtains exactly
(\ref{a179}).

It is seen from (\ref{a178a}), (\ref{a178b}), (\ref{a178c}),
(\ref{a178d}), (\ref{b178a}), (\ref{b178b}), and (\ref{b178c})
that when the limit transition $R \rightarrow \infty$ is
performed, then ${}^{l}E(R)_{A}$ and ${}^{r}E(R)_{A}$ become
asymptotically commutative, then
\[
\lim_{R \rightarrow \infty} {}^{l}E(R)_{A}= \lim_{R \rightarrow
\infty} {}^{r}E(R)_{A}= \frac{\partial}{\partial x^{A}}.
\]
Similarly,
\begin{eqnarray}
\lim_{R \rightarrow \infty} {}^{l}E(R)^{A}=\lim_{R \rightarrow
\infty} {}^{r}E(R)^{A}&=&  dx^{A}, \nonumber \\
\lim_{R \rightarrow \infty} g(R)_{ij} &=& \delta_{ij}, \nonumber
\end{eqnarray}
i.e., as expected, one obtains the $\mathbb{R}^{3}$-relationships
(flat space).

Now we make use of the peculiarity of the dimension three, i.e.,
the identification between skew-symmetric second-order tensors and
axial vectors. Therefore, the matrices of angular velocity will be
represented as follows:
\begin{eqnarray}
{\widehat{\Omega}}(R)^{A}{}_{B} =-
\varepsilon^{A}{}_{BC}{\widehat{\Omega}}(R)^{C}, &\quad&
{\widehat{\Omega}}_{\rm rl}{}^{A}{}_{B} =-
\varepsilon^{A}{}_{BC}{\widehat{\Omega}}_{\rm rl}{}^{C}, \nonumber \\
{\widehat{\Omega}}_{\rm dr}(R)^{A}{}_{B} =-
\varepsilon^{A}{}_{BC}{\widehat{\Omega}}_{\rm dr}(R)^{C}, &\quad&
\Omega_{\rm rl}{}^{A}{}_{B} = \frac{d \varphi^{A}{}_{C}}{dt}
\varphi^{-1 C}{}_{B}=- \varepsilon^{A}{}_{BC}\Omega_{\rm
rl}{}^{C}, \nonumber
\end{eqnarray}
and conversely,
\begin{eqnarray}
{\widehat{\Omega}}(R)^{A} =- \frac{1}{2}
\varepsilon^{A}{}_{B}{}^{C}
 {\widehat{\Omega}}(R)^{B}{}_{C}, &\qquad&
{\widehat{\Omega}}_{\rm rl}{}^{A} =- \frac{1}{2}
\varepsilon^{A}{}_{B}{}^{C}
{\widehat{\Omega}}_{\rm rl}{}^{B}{}_{C}, \nonumber \\
{\widehat{\Omega}}_{\rm dr}(R)^{A} =- \frac{1}{2}
\varepsilon^{A}{}_{B}{}^{C} {\widehat{\Omega}}_{\rm
dr}(R)^{B}{}_{C}, &\qquad& \Omega_{\rm rl}{}^{A} =- \frac{1}{2}
\varepsilon^{A}{}_{B}{}^{C} \Omega_{\rm rl}{}^{B}{}_{C}. \nonumber
\end{eqnarray}
As always, the capital indices are moved in the ``cosmetic"
Kronecker sense. There is no radius label $R$ at
${\widehat{\Omega}}_{\rm rl}$, $\Omega_{\rm rl}$ because these
quantities are independent of $R$.

After some calculations one obtains the  expressions for the basic
aholonomic velocities (``angular velocities") for the motion on
the sphere $S^{3}(0,R)$:
\begin{eqnarray}
{}^{l} \Omega_{\rm dr}(R)^{A} = {}^{l}E^{A}{}_{i}(R; \overline{r})
\frac{dr^{i}}{dt},& \qquad& {}^{r} \Omega_{\rm dr} (R)^{A} =
 {}^{r}E^{A}{}_{i}(R; \overline{r}) \frac{dr^{i}}{dt}, \nonumber \\
{}^{l} \Omega_{\rm rl}{}^{A}=
{}^{l}E^{A}{}_{i}(\overline{\varkappa}) \frac{d
\varkappa^{i}}{dt},& \qquad& {}^{r} \Omega_{\rm rl}{}^{A} =
{}^{r}E^{A}{}_{i}(\overline{\varkappa}) \frac{d
\varkappa^{i}}{dt}, \nonumber
\end{eqnarray}
where $\overline{\varkappa}$ is the rotation vector on SO$(3,
\mathbb{R})$ which parameterizes the manifold of internal degrees
of freedom in the sense of (\ref{a172}) with
$\overline{\varkappa}$ substituted instead of $\overline{k}$:
\[
{\rm SO}(3, \mathbb{R}) \ni \varphi (\overline{\varkappa}) = \exp
\left(\varkappa^{J} A_{J}\right).
\]
Denoting the canonical momenta conjugate to $r^{i}$,
$\varkappa^{i}$ respectively by $p_{i}$, $\pi_{i}$, we have the
 expressions for the aholonomic momenta (canonical
angular momenta) conjugate to $\Omega_{\rm dr}(R)^{A}$,
$\Omega_{\rm rl}{}^{A}$:
\begin{eqnarray}
{}^{l} S_{\rm dr} (R)_{A} = {}^{l}E^{i}{}_{A}(R;
\overline{r})p_{i},& \qquad&
{}^{r}S_{\rm dr}(R)_{A} = {}^{r}E^{i}{}_{A}(R; \overline{r}) p_{i}, \nonumber \\
{}^{l} S_{\rm rl}{}_{A} = {}^{l}E^{i}{}_{A}(\overline{\varkappa})
\pi_{i},& \qquad& {}^{r}S_{\rm rl}{}_{A} =
{}^{r}E^{i}{}_{A}(\overline{\varkappa}) \pi_{i}. \nonumber
\end{eqnarray}
Their Poisson brackets are given as follows:
\begin{eqnarray}
\left\{ {}^{l} S_{\rm dr} (R)_{A}, {}^{l} S_{\rm dr} (R)_{B}
\right\} &=&
\frac{2}{R} \varepsilon_{AB}{}^{C}\ {}^{l} S_{\rm dr} (R)_{C}, \nonumber \\
\left\{ {}^{r} S_{\rm dr} (R)_{A}, {}^{r} S_{\rm dr} (R)_{B}
\right\} &=&
-\frac{2}{R} \varepsilon_{AB}{}^{C}\ {}^{r} S_{\rm dr} (R)_{C}, \nonumber \\
\left\{ {}^{l} S_{\rm dr} (R)_{A}, {}^{r} S_{\rm dr} (R)_{B}
\right\} &=&0, \nonumber \\
\left\{ {}^{l} S_{\rm rl}{}_{A}, {}^{l} S_{\rm rl}{}_{B} \right\}
&=&\varepsilon_{AB}{}^{C}\ {}^{l} S_{\rm rl}{}_{C}, \nonumber \\
\left\{ {}^{r} S_{\rm rl}{}_{A}, {}^{r} S_{\rm rl}{}_{B} \right\}
&=&- \varepsilon_{AB}{}^{C}\ {}^{r} S_{\rm rl}{}_{C}, \nonumber \\
\left\{ {}^{l} S_{\rm rl}{}_{A}, {}^{r} S_{\rm rl}{}_{B} \right\}
&=& 0. \nonumber
\end{eqnarray}
The drive and relative quantities are mutually in involution,
e.g.,
\[
\left\{ {}^{l} S_{\rm dr} (R)_{A}, {}^{l} S_{\rm rl}{}_{B}
\right\} = 0,
\]
and so on. After some calculations we obtain the expression for
the kinetic energy
\begin{eqnarray}
T &=& \frac{1}{2} \left( m + \frac{I}{R^{2}}\right) \delta_{AB}
{}^{l}\Omega (R)^{A} {}^{l}\Omega (R)^{B} \nonumber \\
&&- \frac{I}{R} \delta_{AB} {}^{l} \Omega_{\rm rl}{}^{A}\
{}^{l}\Omega(R)^{B} +  \frac{I}{2} \delta_{AB} {}^{l} \Omega_{\rm
rl}{}^{A}\ {}^{l}\Omega_{\rm rl}{}^{B}. \nonumber
\end{eqnarray}
There are a few interesting features of this formula, namely,
existing of the mass-modifying term $I/R^{2}$ in the purely
translational part and the characteristic gyroscopic term of
coupling between two angular velocities.

The most convenient procedure of deriving equations of motion is
one based on the Hamiltonian formalism and Poisson brackets. We do
almost exclusively with the potential models when Lagrangian has
the form
\[
L = T - V\left(\overline{r}, \overline{\varkappa} \right).
\]
The Legendre transformation expressed in aholonomic terms is as
follows:
\begin{eqnarray}
{}^{l}S(R)_{A} &=& \frac{\partial L}{\partial {}^{l}\Omega(R)^{A}}
=
\frac{\partial T}{\partial {}^{l}\Omega(R)^{A}}, \nonumber \\
{}^{l}S_{\rm rl}{}_{A} &=& \frac{\partial L}{\partial
{}^{l}\Omega_{\rm rl}{}^{A}} = \frac{\partial T}{\partial
{}^{l}\Omega_{\rm rl}{}^{A}}, \nonumber
\end{eqnarray}
i.e., explicitly,
\begin{eqnarray}
{}^{l}S(R)_{A} &=& \left( m+ \frac{I}{R^{2}} \right)
{}^{l}\Omega(R)_{A} -
\frac{I}{R} {}^{l}\Omega_{\rm rl}{}_{A}, \nonumber \\
{}^{l}S_{\rm rl}{}_{A} &=&- \frac{I}{R} {}^{l}\Omega(R)_{A}+ I
{}^{l}\Omega_{\rm rl}{}_{A}. \nonumber
\end{eqnarray}
The corresponding inverse rule reads
\begin{eqnarray}
{}^{l}\Omega(R)^{A} &=& \frac{1}{m} {}^{l}S(R)^{A} +\frac{1}{mR}
{}^{l}S_{\rm rl}{}^{A}, \nonumber \\
{}^{l}\Omega_{\rm rl}{}^{A} &=& \frac{1}{mR} {}^{l}S(R)^{A} +
\frac{I+m R^{2}}{ImR^{2}} {}^{l}S_{\rm rl}{}^{A}, \nonumber
\end{eqnarray}
where the raising and lowering of capital indices are meant in the
"cosmetic" Kronecker-delta sense.

After substituting all this to the kinetic energy formula, we
obtain the expression for the kinetic Hamiltonian
\begin{eqnarray}
{\mathcal{T}} &=& \frac{1}{2m} \delta^{AB}\  {}^{l}S(R)_{A}
{}^{l}S(R)_{B} +
\frac{1}{mR} \delta^{AB}\  {}^{l}S(R)_{A} {}^{l}S_{\rm rl}{}_{B} \nonumber \\
&&+\frac{I+mR^{2}}{2ImR^{2}}\delta^{AB} {}^{l}S_{\rm
rl}{}_{A}{}^{l}S_{\rm rl}{}_{B}.  \label{a184}
\end{eqnarray}
When some potential term is admitted, then the Hamiltonian has the
form
\[
H= {\mathcal{T}} + {V}(\overline{r}, \overline{\varkappa}).
\]
For the purely geodetic models, when ${V}=0$, the Poisson-bracket
form of equations of motion, i.e.,
\[
\frac{dF}{dt} = \left\{F, H \right\},
\]
leads to the dynamical system for canonical angular momenta
\begin{eqnarray}
\frac{d {}^{l}S(R)_{A} }{dt} &=& \frac{2}{mR^{2}}
\varepsilon_{A}{}^{BC} {}^{l}S_{\rm rl}{}_{B} {}^{l}S(R)_{C}, \label{a185a} \\
\frac{d {}^{l}S_{\rm rl}{}_{A} }{dt} &=& \frac{1}{mR}
\varepsilon_{A}{}^{BC} {}^{l}S(R)_{B} {}^{l}S_{\rm
rl}{}_{C}.\label{a185b}
\end{eqnarray}
It is interesting that only the interference term in (\ref{a184})
contributes to these equations because the first and the third
terms have vanishing Poisson brackets with ${}^{l}S(R)_{A}$,
${}^{l}S_{\rm rl}{}_{A}$ (they are Casimir invariants). Using the
three-dimensional vector notation in $\mathbb{R}^{3}$, we can
write the above equations as follows:
\begin{eqnarray}
\frac{d {}^{l}\overline{S(R)}}{dt} &=& \frac{2}{mR^{2}}
{}^{l}\overline{S_{\rm rl}} \times {}^{l}\overline{S(R)},  \label{b185a} \\
\frac{d {}^{l}\overline{S_{\rm rl}}}{dt} &=& \frac{1}{mR}
{}^{l}\overline{S(R)} \times {}^{l}\overline{S_{\rm
rl}},\label{b185b}
\end{eqnarray}
where there is meant the standard vector product in
$\mathbb{R}^{3}$.

The Poisson rules quoted above imply that the squared magnitudes
of angular momenta, i.e.,
\begin{eqnarray}
\left|{}^{l}S(R)\right|^{2} &=& \delta^{AB} {}^{l}S(R)_{A} {}^{l}S(R)_{B}, \nonumber \\
\left|{}^{l}S_{\rm rl} \right|^{2} &=& \delta^{AB} {}^{l}S_{\rm
rl}{}_{A} {}^{l}S_{\rm rl}{}_{B}, \nonumber
\end{eqnarray}
are constants of motion in virtue of the equations above
(\ref{a185a}), (\ref{a185b}), (\ref{b185a}), (\ref{b185b}). And
moreover, these equations imply that the $\mathbb{R}^{3}$-vector
\begin{equation}\label{a186}
\frac{R}{2}{}^{l}S(R)_{A} + {}^{l}S_{\rm rl}{}_{A}
\end{equation}
is a constant of motion. This implies in particular that the
$\mathbb{R}^{3}$-scalar product of angular momenta
\[
\delta^{AB}\ {}^{l}S(R)_{A} {}^{l}S_{\rm rl}{}_{B}
\]
and the angle between ${}^{l}\overline{S(R)}$,
${}^{l}\overline{S_{\rm rl}}$ are constants of motion functionally
dependent on the previous ones.

In this way the six-dimensional dynamical system (\ref{a185a}),
(\ref{a185b}), (\ref{b185a}), (\ref{b185b}) has five independent
constants of motion and there is only one time-dependent degree of
freedom in the
$\left({}^{l}\overline{S(R)},{}^{l}\overline{S}\right)$-space. The
only essentially active degree of freedom is the orientation
(polar angle) of the two-dimensional plane in $\mathbb{R}^{3}$
spanned by the vectors ${}^{l}\overline{S(R)}$,
${}^{l}\overline{S_{\rm rl}}$ and rotating about the axis given by
the vector (\ref{a186}).

{\bf Remark:} it has been told that the motion of vectors ${}^{l}\overline{S(R)}$,
${}^{l}\overline{S_{\rm rl}}$ is independent of the first and second terms in (\ref{a184}), in
particular, of the inertial moment $I$. But of course these terms and the quantity $I$ are
relevant for the total description of motion. Namely, $I$ appears in the Legendre
transformation and its inverse. Therefore, the time evolution of configuration variables
$(\overline{r}, \overline{\varkappa})$ depends explicitly on $I$ and on all terms in the
kinetic energy expression.

\end{document}